\documentclass[12pt]{article}
\newlength{\pubnumber} 

\catcode`\@=11
\@addtoreset{equation}{section}

\def\section{\@startsection{section}{1}{\z@}{3.5ex plus 1ex minus .2ex}
 {2.3ex plus .2ex}{\large\bf}}
\def\subsection{\@startsection{subsection}{2}{\z@}{2.3ex plus .2ex}
 {2.3ex plus .2ex}{\bf}}

\begin{document}

\begin{titlepage}
\samepage{
\setcounter{page}{0}
\rightline{\tt hep-th/0602286}
\rightline{February 2006}
\vfill
\begin{center}
    {\Large \bf  Statistics on the Heterotic Landscape: \\
    {\bf Gauge Groups and Cosmological Constants
    of Four-Dimensional Heterotic Strings}\\}
\vfill
\vspace{.10in}
   {\large
      Keith R. Dienes\footnote{
     E-mail address:  dienes@physics.arizona.edu}
    \\}
\vspace{.10in}
 {\it  Department of Physics, University of Arizona, Tucson, AZ  85721  USA\\}
\end{center}
\vfill
\begin{abstract}
  {\rm 
    Recent developments in string theory have reinforced 
    the notion that the space of stable supersymmetric and non-supersymmetric
    string vacua fills out a ``landscape'' 
    whose features are largely unknown. 
    It is then hoped that progress in extracting phenomenological predictions 
    from string theory ---
    such as correlations between gauge groups, matter representations,
    potential values of the cosmological constant, and so forth ---
    can be achieved through statistical studies of these vacua.  
    To date, most of the efforts in these directions have focused on 
    Type~I vacua.
    In this note, we present the first results of a statistical 
    study of the {\it heterotic}\/ landscape, focusing on more than 
    $10^5$ explicit non-supersymmetric 
    tachyon-free heterotic string vacua and their associated gauge 
    groups and one-loop cosmological constants.  
    Although this study has several important limitations, we find a number of
    intriguing features which may be relevant for the heterotic landscape
    as a whole.  These features include
    different probabilities and correlations for different possible gauge groups
    as functions of the number of orbifold twists.
    We also find a vast degeneracy amongst 
    non-supersymmetric string models, leading to a severe reduction
    in the number of realizable values of the cosmological constant as compared
    with naive expectations.  Finally, we also find strong correlations
    between cosmological constants and gauge groups which suggest that
    heterotic string models with extremely small cosmological constants 
    are overwhelmingly more likely to 
    exhibit the Standard-Model gauge group 
    at the string scale than any of its grand-unified extensions.
    In all cases, heterotic worldsheet symmetries such as modular
    invariance provide important constraints that do not appear
    in corresponding studies of Type~I vacua.  
   }
\end{abstract}
\vfill
\smallskip}
\end{titlepage}

\setcounter{footnote}{0}

% ========================= DEFINITIONS ===================================
\def\beq{\begin{equation}}
\def\eeq{\end{equation}}
\def\beqn{\begin{eqnarray}}
\def\eeqn{\end{eqnarray}}
\def\half{{\textstyle{1\over 2}}}

\def\calO{{\cal O}}
\def\calE{{\cal E}}
\def\calT{{\cal T}}
\def\calM{{\cal M}}
\def\calF{{\cal F}}
\def\calY{{\cal Y}}
\def\calV{{\cal V}}
\def\calN{{\cal N}}
\def\ibar{{\overline{\i}}}
\def\qbar{{\overline{q}}}
\def\mm{{\tilde m}}
\def\ahat{{\hat a}}
\def\nn{{\tilde n}}
\def\rep#1{{\bf {#1}}}
\def\ie{{\it i.e.}\/}
\def\eg{{\it e.g.}\/}

\def\Str{{{\rm Str}\,}}
\def\bone{{\bf 1}}

\def\thetai{{\vartheta_i}}
\def\thetaj{{\vartheta_j}}
\def\thetak{{\vartheta_k}}
\def\thetaibar{\overline{\vartheta_i}}
\def\thetajbar{\overline{\vartheta_j}}
\def\thetakbar{\overline{\vartheta_k}}
\def\etainv{{\overline{\eta}}}

\def\modinvmeasure{{  {{{\rm d}^2\tau}\over{\tautwo^2} }}}
\def\qbar{{  \overline{q} }}
\def\ahat{{ \hat a }}

\newcommand{\newc}{\newcommand}
\newc{\gsim}{\lower.7ex\hbox{$\;\stackrel{\textstyle>}{\sim}\;$}}
\newc{\lsim}{\lower.7ex\hbox{$\;\stackrel{\textstyle<}{\sim}\;$}}

%==============================================================================
\hyphenation{su-per-sym-met-ric non-su-per-sym-met-ric}
\hyphenation{space-time-super-sym-met-ric}
\hyphenation{mod-u-lar mod-u-lar--in-var-i-ant}
%==============================================================================

%================== BLACKBOARD BOLD CHARACTERS ==============================

\def\inbar{\,\vrule height1.5ex width.4pt depth0pt}

\def\IC{\relax\hbox{$\inbar\kern-.3em{\rm C}$}}
\def\IQ{\relax\hbox{$\inbar\kern-.3em{\rm Q}$}}
\def\IR{\relax{\rm I\kern-.18em R}}
 \font\cmss=cmss10 \font\cmsss=cmss10 at 7pt
\def\IZ{\relax\ifmmode\mathchoice
 {\hbox{\cmss Z\kern-.4em Z}}{\hbox{\cmss Z\kern-.4em Z}}
 {\lower.9pt\hbox{\cmsss Z\kern-.4em Z}}
 {\lower1.2pt\hbox{\cmsss Z\kern-.4em Z}}\else{\cmss Z\kern-.4em Z}\fi}

% Redefine caption to put text and formulas in smaller font
\long\def\@caption#1[#2]#3{\par\addcontentsline{\csname
  ext@#1\endcsname}{#1}{\protect\numberline{\csname
  the#1\endcsname}{\ignorespaces #2}}\begingroup
    \small
    \@parboxrestore
    \@makecaption{\csname fnum@#1\endcsname}{\ignorespaces #3}\par
  \endgroup}
\catcode`@=12

\input epsf
%============================== TEXT BEGINS HERE ============================

%=============================================================================
\section{Introduction}
\setcounter{footnote}{0}

One of the most serious problems faced by 
practitioners of string phenomenology is the multitude
of possible, self-consistent string vacua.
That there exist large numbers of potential string solutions
has been known since the earliest
days of string theory;  these result from the 
large numbers of possible ways in which one may choose an appropriate compactification
manifold (or orbifold), an appropriate set of background fields and fluxes, 
and appropriate expectation values for the plethora of additional moduli
to which string theories generically give rise.
Although historically these string solutions were not completely stabilized,
it was tacitly anticipated for many years that some unknown vacuum stabilization
mechanism would ultimately lead to a unique vacuum state.
Unfortunately, recent developments suggest that there continue
to exist huge numers of self-consistent string solutions
(\ie, string ``models'' or ``vacua'') even after stabilization.
Thus, a picture emerges in which there exist huge numbers of possible string vacua,
all potentially stable (or sufficiently metastable),
with apparently no dynamical principle to select amongst them.
Indeed, each of these potential vacua can be viewed
as sitting at the local minimum of a complex terrain
of possible string solutions
dominated by hills and valleys.
This terrain has come to be known as the ``string-theory 
landscape''~\cite{landscape}. 

The existence of such a landscape 
has tremendous practical significance because
the specific low-energy phenomenology that 
can be expected to emerge from string theory depends
critically on the particular choice of vacuum state.
Detailed quantities such as particle masses and mixings,
and even more general quantities and structures such as the
choice of gauge group, number of chiral particle generations,
magnitude of the supersymmetry-breaking scale,
and even the cosmological constant can be expected to vary significantly
from one vacuum solution to the next.
Thus, in the absence of some sort of vacuum selection principle,
it is natural to tackle a secondary but perhaps more tractible question
concerning whether there might exist generic string-derived {\it correlations}\/ 
between different phenomenological features.
In this way, one can still hope to extract phenomenological predictions
from string theory. 

This idea has triggered
a recent surge of activity concerning the {\it statistical}\/ properties
of the landscape~[2--13].
%  \cite{douglas,SUSYbreaking,splitSUSY,blumenhagen,schellekens,Weinberg,BP,
%  lambdapapers,kane,fieldtheory,anthropics,banks,review}.
Investigations along these lines have
focused on diverse phenomenological issues,
including 
 the value of the supersymmetry-breaking scale~\cite{douglas,SUSYbreaking},
  the value of the cosmological constant~\cite{Weinberg,BP,lambdapapers},
  and the preferred rank of the corresponding gauge groups,
  the prevalence of the Standard-Model gauge group, 
  and possible numbers of chiral generations~\cite{douglas,blumenhagen,schellekens}.
Discussions of the landscape have also led to various theoretical paradigm shifts,
ranging from
  alternative landscape-based notions of naturalness~\cite{SUSYbreaking,splitSUSY}
   and novel cosmological inflationary scenarios~\cite{Weinberg,BP,lambdapapers}
  to the use of anthropic arguments to constrain 
   the set of viable string vacua~\cite{Weinberg,lambdapapers,anthropics}.
  There have even been proposals for field-theoretic analogues of 
   the string-theory landscape~\cite{fieldtheory}
   as well as discussions concerning whether there 
   truly exist effective field theories that can describe it~\cite{banks}.
Collectively, these developments have even given birth to a large, 
ambitious, organized  
effort dubbed the ``String Vacuum Project (SVP)''~\cite{SVP},
one of whose purposes is to map out the properties of this
landscape of string vacua.  It is envisioned that this will happen
not only through direct enumeration/construction
of viable string vacua, but also through planned large-scale statistical 
studies across the landscape as a whole.  

Unfortunately, although there have been many abstract theoretical 
discussions of such vacua and their statistical properties, 
there have been relatively few direct statistical examinations of 
actual string vacua. 
Despite considerable effort,
there have been relatively few pieces of
actual data gleaned from direct studies of the string landscape
and the vacua which populate it.
This is because, in spite of recent progress, the construction and analysis
of completely stable string vacua remains a rather complicated affair~\cite{fluxes,constructions}.
Surveying whole classes of such vacua and doing a proper statistical
analysis thus remains a formidable task.
 
There are exceptions, however.
For example, one recent computer analysis examined millions of supersymmetric
intersecting D-brane models on a particular orientifold background~\cite{blumenhagen}.
Although the models which were constructed for such analyses
are not completely stable (since they continue to have flat directions),
the analysis reported in Ref.~\cite{blumenhagen} examined important questions 
such as the statistical occurrences of various gauge groups, chirality,
numbers of generations, and so forth.  
A similar statistical study focusing on Gepner-type orientifolds exhibiting 
chiral supersymmetric Standard-Model spectra was performed in Ref.~\cite{schellekens}. 
By means of such studies, a number of interesting statistical correlations were uncovered.

To date, however, there has been almost no discussion of the {\it heterotic}\/
landscape.
This is somewhat ironic, especially since perturbative heterotic strings were 
the framework in which most of the original work in string phenomenology 
was performed in the late 1980's and early 1990's. 

In this paper, we shall present the results of the first statistical study
of the heterotic string landscape. 
Thus, in some sense, this work can be viewed as providing a heterotic analogue
of the work reported in Refs.~\cite{blumenhagen,schellekens}. 
In this paper, we shall focus on a sample of approximately $1.2\times 10^5$ distinct 
four-dimensional perturbative heterotic string models, all randomly generated, 
and we shall analyze statistical information concerning their gauge groups and one-loop 
cosmological constants. 

As we shall see, the statistical properties of perturbative heterotic strings 
are substantially different from those of Type~I strings.
This is already apparent at the level of gauge groups:  while the gauge
groups of Type~I strings are constrained
only by allowed D-brane configurations and anomaly-cancellation constraints, 
those of perturbative heterotic strings necessarily
have a maximum rank.  Moreover, as we shall repeatedly see, 
modular invariance shall also prove to play an important role in constraining
the features of the heterotic landscape.
This too is a feature that is lacking for Type~I landscape. 
On the other hand, there will be certain similarities.
For example, one of our results will concern a probability for randomly
obtaining the Standard-Model gauge group from perturbative heterotic strings.  
Surprisingly, this probability shall be very close to what is obtained for 
Type~I strings.

For various technical and historical reasons,
our statistical study will necessarily have certain limitations.
There will be discussed more completely below and in Sect.~2.
However, three limitations are critical and deserve immediate mention. 

First, as mentioned, our sample size is relatively small, consisting of 
only $\sim 10^5$ distinct models.   However, although this number is 
miniscule compared with the numbers of string models that are currently 
quoted in most landscape discussions, we believe that the statistical results 
we shall obtain have already achieved saturation --- \ie, we do not believe that
they will change as more models are added.
We shall discuss this feature in more detail in Sect.~2.

Second, for historical reasons to be discussed below, our statistical
study in this paper shall be limited to only two phenomenological
properties of these models:  their low-energy gauge groups, and their
one-loop vacuum amplitudes (cosmological constants).  Nevertheless,
as we shall see, this represents a considerable wealth of data.
Further studies are currently underway to investigate other properties
of these models and their resulting spacetime spectra, and we hope to
report those results in a later publication.

Perhaps most importantly, however, all of the models 
we shall be analyzing are non-supersymmetric.
Therefore, even though they are all tachyon-free,
they have non-zero dilaton tadpoles and thus 
are not stable beyond tree level.
Indeed, the models we shall be examining can be viewed as four-dimensional 
analogues of the $SO(16)\times SO(16)$ heterotic string in ten dimensions~\cite{SOsixteen}.  
Such models certainly satisfy all of the necessary
string self-consistency constraints --- 
they have worldsheet conformal/superconformal invariance,
they have one-loop and multi-loop modular-invariant amplitudes,
they exhibit proper spin-statistics relations, 
and they contain physically sensible GSO projections and orbifold twists.
However, they are not stable beyond tree level.

Clearly, such models do not represent
the sorts of truly stable vacua that we would ideally like to be studying.
Again invoking landscape imagery, such models 
do not sit at local minima in the landscape --- they sit on
hillsides and mountain passes, valleys and even mountaintops.
Thus, in this paper, we shall in some sense be 
surveying the entire {\it profile}\/ of
the landscape rather than merely the properties of its local minima.
Indeed, we can call this a ``raindrop'' study:
we shall let the rain fall randomly over the perturbative heterotic landscape
and collect statistical data where each raindrop hits the surface.
Clearly this is different in spirit from a study in which
our attention is restricted to the locations of the puddles which remain
after the rain has stopped and the sun comes out.
 
Despite these limitations, we believe that such a study
can be of considerable value.
First, such models do represent valid string solutions at tree level,
and it is therefore important to understand their properties as a first
step towards understanding the full phenomenology of non-supersymmetric
strings and their contributions to the overall architecture of the landscape.
Indeed, since no stable perturbative non-supersymmetric heterotic strings have 
yet been constructed, our study represents the current state of the art
in the statistical analysis of perturbative non-supersymmetric heterotic strings. 

Second, as we shall discuss further in Sect.~2,
the models we shall be examining range from the extremely simple,
involving a single set of sectors, to the extraordinarily complex,
involving many convoluted layers of overlapping orbifold twists and Wilson lines.
In all cases, these sets of orbifolds twists and Wilson lines were randomly generated,
yet each satisfies all necessary self-consistency constraints. 
These models thus exhibit an unusual degree of intricacy and complexity,
just as we expect for models which might eventually exhibit low-energy
phenomenologies resembling that of the real world.

Third, an important question for any landscape study is to understand
the phenomenological roles played by supersymmetry and by the need for
vacuum stability.  However, the only way in which we might develop an understanding
of the statistical significance of the effects that spacetime supersymmetry
might have on other phenomenological properties (such as gauge groups, numbers
of chiral generations, {\it etc}\/.) is to already have the results of a study
of strings in which supersymmetry is absent.

But most importantly, we know as an experimental fact that the 
low-energy world is non-supersymmetric.  Therefore, if we believe that
perturbative heterotic strings are relevant to its description,
it behooves us to understand the properties of non-supersymmetric strings.
Although no such strings have yet been found which are stable beyond tree level,
analyses of these unstable vacua may prove useful in pointing the way towards their
eventual constructions.
Indeed, as we shall see, some of our results shall suggest some of the likely
phenomenological properties that such string might ultimately have.

This paper is organized as follows.
In Sect.~2, we shall provide an overview of the models that we will
be analyzing in this paper.  We shall also discuss, in more detail,
the limitations and methodologies of our statistical study.
In Sect.~3, we shall then provide a warm-up discussion that focuses
on the better-known properties of the {\it ten}\/-dimensional heterotic 
landscape.
We will then turn our attention to heterotic strings in four dimensions
for the remainder of the paper.
In Sect.~4, we shall focus on the gauge groups of such strings,
and in Sect.~5 we shall focus on their one-loop vacuum energies (cosmological
constants).  Finally, in Sect.~6, we shall analyze the statistical
 {\it correlations}\/ between the gauge groups and cosmological constants.
A short concluding section will then outline some future directions.
Note that even though these string models are unstable beyond tree level, 
we shall use the terms ``string models'' and ``string vacua''
interchangeably in this paper to refer 
to these non-supersymmetric, tachyon-free string solutions.

%===========
\bigskip
\noindent{\it Historical note}
\medskip

This paper has a somewhat unusual provenance.
Therefore, before beginning, we provide a brief historical note.

In the late 1980's, soon after the development of the free-fermionic
construction~\cite{KLT}, a number of string theorists 
undertook various computer-automated randomized searches through the space
of perturbative four-dimensional heterotic string models. 
The most detailed and extensive of such searches was described
in Ref.~\cite{Senechal};  to the best of our knowledge, this represents
the earliest automated search through the space of heterotic string models.
Soon afterwards, other searches were also performed (see, {\it e.g.}\/, Ref.~\cite{PRL}).

At that time, the goals of such studies were to find
string models with certain favorable phenomenological properties. 
In other words, these investigations were viewed as searches rather
than as broad statistical studies.

One such search at that time~\cite{PRL}
was aimed at finding four-dimensional
perturbative non-supersymmetric tachyon-free heterotic string models
which nevertheless have zero one-loop cosmological constants.
Inspired by Atkin-Lehner symmetry and its possible extensions~\cite{Moore},
we conducted a 
search using the techniques (and indeed some of the software) first described
in Ref.~\cite{Senechal}.
At that time, our interest was purely on 
the values of the cosmological constant.  However, along the way, 
the corresponding gauge groups of these models were also determined and recorded.

In this paper, we shall report on the results of 
a new, comprehensive, statistical analysis of this ``data'' 
which was originally collected in the late 1980's.
As a consequence of the limited scope of our original search,
our statistical analysis here shall therefore be focused on non-supersymmetric tachyon-free models.
Likewise, in this paper we shall concentrate on only the two phenomenological
properties of such models (gauge groups and cosmological constants) for which such data 
already existed.  As mentioned above, a more exhaustive statistical study using 
modern software and a significantly larger data set is currently underway:
this will include both supersymmetric and non-supersymmetric heterotic string models,
and will involve many additional properties of the physical
spectra of the associated models  (including their gauge groups, 
numbers of generations, chirality properties, and so forth).
However, the study described in this paper shall be limited to the 
data set that was generated as part of the investigations of Ref.~\cite{PRL}. 
Although this data was generated over fifteen years ago,
we point out that almost all of statistical results of this paper 
were obtained recently and have not been published or reported elsewhere 
in the string literature.

%=============================================================================
\section{The string vacua examined}
\setcounter{footnote}{0}

In this section we shall describe the class of string vacua
which are included in our statistical analysis.

Each of these vacua represents a weakly coupled critical heterotic string
compactified to, or otherwise constructed directly in, four 
large (flat) spacetime dimensions.
In general, such a string may be described in terms of its
left- and right-moving worldsheet conformal field theories (CFT's);
in four dimensions, 
in addition to the spacetime coordinates
and their right-moving worldsheet superpartners,
these internal CFT's must have central charges
$(c_R,c_L)=(9,22)$ in order to enforce worldsheet conformal anomaly 
cancellation.
While the left-moving internal CFT must merely exhibit conformal invariance,
the right-moving internal CFT must actually exhibit superconformal invariance.  
While any CFT's with these 
central charges may be considered, in this paper we shall
focus on those string models for which these internal worldsheet CFT's may 
be taken to consist of tensor products of free, non-interacting,
complex (chiral) bosonic or fermionic fields. 

This is a huge class of models which has been discussed and analyzed
in many different ways in the string literature.
On the one hand, taking these worldsheet fields as fermionic
leads to the so-called   
``free-fermionic'' construction~\cite{KLT}
which will be our primary tool throughout
this paper.
In the language of this construction, 
different models are achieved by varying (or ``twisting'') the boundary conditions of
these fermions around the two non-contractible loops of the 
worldsheet torus while simultaneously varying 
the phases according to which the contributions of each such 
spin-structure sector are summed in producing the one-loop
partition function.  
However, alternative but equivalent languages for constructing such models exist.
For example, we may bosonize these worldsheet fermions and
construct ``Narain'' models~\cite{Narain,Lerche} in which the resulting complex worldsheet
bosons are compactified on internal lattices of appropriate dimensionality
with appropriate self-duality properties.
Furthermore, many of these models have additional geometric realizations 
as orbifold compactifications with randomly chosen 
Wilson lines;  in general, the process of orbifolding
is quite complicated in these models, involving many sequential layers of projections
and twists.
Note that all of these constructions generally overlap to a large degree, 
and all are capable of producing models in which the corresponding 
gauge groups and particle
contents are quite intricate.
Nevertheless, in all cases, we must ensure that all required self-consistency
constraints are satisfied.  These include modular invariance, physically 
sensible GSO projections, proper spin-statistics identifications, and so forth.
Thus, each of these vacua represents a fully self-consistent string
solution at tree level.

In order to efficiently survey the space
of such non-supersymmetric four-dimensional string-theoretic vacua,
we implemented a computer search based on 
the free-fermionic spin-structure construction, as originally developed
in Ref.~\cite{KLT}.
Recall that in this light-cone gauge construction,
each of the six compactified bosonic spacetime coordinates is fermionized
to become two left-moving and two right-moving internal free real fermions,
and consequently
our four-dimensional heterotic strings
consist of the following fields on the worldsheet:
20 right-moving free real
fermions (the eight original supersymmetric partners of the eight
transverse bosonic coordinates of the ten-dimensional string, along
with twelve additional internal fermions resulting from compactification);  
44 left-moving free real
fermions (the original 32 in ten dimensions plus the additional twelve
resulting from compactification);  and
of course the two transverse bosonic (coordinate) fields $X^\mu$.
Of these 20 right-moving
real fermions, only two (the supersymmetric partners of the
two remaining transverse coordinates) carry Lorentz indices.  In our 
analysis, we restricted our
attention to those models for which our real fermions can
always be uniformly paired to form complex fermions, and therefore
it was possible to specify the boundary conditions (or spin-structures)
of these real fermions
in terms of the complex fermions directly.
We also restricted our attention to 
cases in which the worldsheet fermions 
exhibited either antiperiodic (Neveu-Schwarz) or periodic (Ramond)
boundary conditions.
Of course, in order to build a self-consistent
string model in this framework, these boundary conditions 
must satisfy tight constraints.  These constraints are
necessary in order to ensure that the  
one-loop partition function is modular invariant and that the resulting
Fock space of states can be 
interpreted as arising from a physically sensible projection
from the space of all worldsheet states onto the subspace of physical
states with proper spacetime spin-statistics.  Thus, within a given
string model, it is necessary
to sum over appropriate sets of untwisted and twisted sectors 
with different boundary conditions and projection phases.

Our statistical analysis consisted of an examination of 
$123,573$ distinct vacua,
each randomly generated through the free-fermionic construction.
(In equivalent orbifold language, each vacuum was constructed from randomly chosen
sets of orbifold twists and Wilson lines, subject to the constraints
described above.)
Details of this study are similar to those of the earlier study described 
in Ref.~\cite{Senechal},
and made use of model-generating software borrowed from that earlier study.
Essentially, each set of boundary conditions was chosen randomly in each sector,
subject only to the required self-consistency constraints.
However, in our statistical sampling, we placed no limits on the complexity of 
the orbifold twisting (\ie, on the number of basis vectors in the free-fermionic
language).  
Thus, our statistical analysis included models of arbitrary intricacy and
sophistication.

As discussed above, for the purpose of this search, 
we demanded that supersymmetry be broken without introducing tachyons. 
Thus, these vacua are all non-supersymmetric but tachyon-free, and
can be considered as four-dimensional analogues of
the ten-dimensional
$SO(16)\times SO(16)$ heterotic string~\cite{SOsixteen} 
which is also non-supersymmetric but tachyon-free.
As a result, these models all have non-vanishing but finite
one-loop cosmological constants/vacuum energies $\Lambda$,
and we shall examine these values of $\Lambda$ in Sect.~5. 
However, other than demanding that supersymmetry
be broken in a tachyon-free manner,
we  placed no requirements on other possible 
phenomenological properties of these vacua
such as the possible gauge groups, numbers of chiral generations, 
or other aspects of the particle content.
We did, however, require that our string construction 
begin with a supersymmetric theory in which the supersymmetry
is broken only through subsequent orbifold twists.  
(In the language of the free-fermionic construction,
    this is tantamount to demanding that our 
    fermionic boundary conditions include a superpartner
    sector, typically denoted ${\bf W}_1$ or ${\bf V}_1$.)
This is to be distinguished from a potentially more general
class of models in which supersymmetry does not appear at
any stage of the construction.
This is merely a technical detail in our construction,
and we do not believe that this ultimately affects our
results. 

Because of the tremendous redundancy inherent in the free-fermionic
construction, string vacua were judged to be distinct based on
their spacetime characteristics --- {\it i.e.}\/, their 
low-energy gauge groups and massless particle content.
Thus, as a minimum condition,
distinct string vacua necessarily exhibit different massless spacetime spectra.\footnote{
    As a result of conformal invariance and modular invariance (both of which simultaneously relate 
    states at all mass levels), it is extremely difficult for two string models to share the same 
    massless spectrum (as well as the same off-shall tachyonic structure) 
    and yet differ in their massive spectra.  
    Thus, for all practical purposes, our requirement
    that two models must have different massless spectra is not likely to eliminate
    potential models whose spectra might differ only at the massive level.} 
As we shall discuss further below, such a requirement about the distinctness of 
the spacetime spectrum must be an 
important component of any statistical study of string models.
Since the same string model may have a plethora of different worldsheet
realizations, one cannot verify that one is accurately surveying  
the space of distinct, independent string models based on their 
worldsheet realizations alone. 
This ``redundancy'' issue becomes increasingly pressing as larger
and larger numbers of models are considered.

Clearly, this class of string models is not all-encompassing.
By its very nature, the free-fermionic construction
reaches only certain specific points in the full space of self-consistent
string models.  For example, since each worldsheet fermion is nothing but a worldsheet
boson compactified at a specific radius, a larger (infinite) class of models 
can immediately be realized through a bosonic formulation 
by varying these radii away from their free-fermionic values.  
However, this larger class of models will typically have 
only abelian gauge groups and consequently uninteresting particle representations.
Indeed, the free-fermionic points typically represent precisely those points
at which additional (non-Cartan) gauge-boson states become massless,
thereby enhancing the gauge symmetries to become non-abelian.
Thus, the free-fermionic construction naturally leads to 
precisely the set of models which are likely to 
be of direct phenomenological relevance.

Similarly, it is possible to go beyond the class of free-field string
models altogether, and consider models built from more complicated worldsheet 
CFT's (\eg, Gepner models).
We may even transcend the realm of critical string theories, and consider
non-critical strings and/or strings with non-trivial background fields.
Likewise, we may consider heterotic strings beyond the usual perturbative
limit.
However, although such models may well give rise to phenomenologies
very different from those that emerge in free-field constructions,
their spectra are typically very difficult to analyze and are thus not amenable
to an automated statistical investigation.
Finally, even within the specific construction we are employing in this paper,
we may drop our requirement that our models be non-supersymmetric,
and consider models with varying degrees of unbroken supersymmetry.
This will be done in future work.

Finally, we should point out that strictly speaking, the class of models we are considering 
is only finite in size.
Because of the tight worldsheet self-consistency constraints
arising from modular invariance and the requirement of physically sensible GSO projections,
there are only a finite number of distinct boundary
condition vectors and GSO phases which may be chosen in our construction
as long as we restrict our attention to complex worldsheet fermions with 
only periodic (Ramond) or antiperiodic (Neveu-Schwarz) boundary conditions.
For example, in four dimensions there are a maximum of only  
$32$ boundary-condition vectors which can possibly be linearly independent,
even before we impose other dot-product modular-invariance
constraints. 

This is, nevertheless, a very broad and general class of theories.
Indeed, models which have been constructed using such techniques span
almost the entire spectrum of closed-string models,
including MSSM-like models, models with
and without extra exotic matter, and so forth.
Moreover,  worldsheet bosonic and fermionic constructions
can produce models which have an intricacy and complexity
which is hard to duplicate purely through geometric considerations
--- indeed, these are often models for which no geometric compactification space
is readily apparent.
It is for this reason that while most of our geometric insights about
string models have historically
come from Calabi-Yau and general orbifold analyses, much of the serious
work at realistic closed-string model-building over the past two
decades has been through the more algebraic bosonic or fermionic
formulations.
It is therefore within this class of string models that our analysis
will be focused.
Moreover, as we shall see, this set of models is still sufficiently
large to enable various striking statistical correlations to appear.

Finally, we provide some general comments 
about the statistical analysis we will be performing
and the interpretation of our results.

As with any statistical landscape study,
it is important to consider whether the properties we shall find are rigorously 
true for the landscape as a whole,
or are merely artifacts of having considered only a
finite statistical sample of models 
or a sample which is itself not representative of the landscape at large
because it is statistically biased or skewed in some way.
Clearly, without detailed analytical
knowledge of the entire landscape of models in the category
under investigation,  one can never definitively answer this question.
Thus, in each case,
it is necessary to judge which properties or statistical correlations
are likely to exist because of some deeper, identifiable string consistency constraint,
and which are not.
In this paper, we shall try to indicate in every circumstance what we
believe are the appropriate causes of each statistical correlation we find.

The issue concerning the finite size of our sample is particularly relevant
in our case, since we will be examining the properties of only
$\sim 10^5$ distinct models in this paper.
Although this is certainly a large number of string models on an absolute
scale, this number is extremely small compared with the current estimated  
size of the entire string landscape.
However, one way to judge the underlying validity of a particular
statistical correlation is to test whether it persists 
without significant modification
as our sample size increases.
If so, then
it is likely that 
we have already reached the ``continuum limit''  
(borrowing a phrase from our lattice gauge theory colleagues)
as far as the particular statistical correlation is concerned.
This can be verified by testing the numerical stability 
of a given statistical correlation as more and more string models
are added to our sample set, and can be checked {\it a posteriori}\/
by examining whether the correlation persists even if the 
final sample set is partitioned or subdivided.
All correlations that we will present in this paper
are stable in the ``continuum limit'' unless otherwise indicated.  

Finally, we point out that all correlations in this paper
will ultimately depend on a particular assumed measure across the landscape.
For example, when we plot a correlation between two quantities,
the averaging for these quantities is calculated across all models in
our data set, with each physically distinct string model weighted equally.
However, we expect that such averages would change significantly
if models were weighted in a different manner.  For example, as we
shall see, many of our results would be altered if we were to weight 
our probabilities equally across the set of distinct gauge groups
rather than across the set of distinct string models.  
This sensitivity to the underlying string landscape measure is, of course, well known.
In this paper, we shall employ a measure 
in which each distinct string model is 
weighted equally across our sample set.

%====================================================
\section{A preliminary example:\\
    The ten-dimensional heterotic landscape}

Before plunging into the four-dimensional case of interest,
let us first consider the ``landscape'' of {\it ten}\/-dimensional heterotic
string models.  Recall that in ten dimensions, such models have maximal gauge-group rank 
$16$, corresponding to sixteen left-moving worldsheet bosons (or complex fermions). 

It turns out that we can examine the resulting ``landscape'' of such models
by arranging them in the form of a family ``tree''.
First, at the root of the tree, we have what is literally the simplest 
ten-dimensional heterotic string model we can construct:
this is the supersymmetric $SO(32)$ heterotic string model in which 
our worldsheet fermionic fields all have identical boundary conditions in each
spin-structure sector of the theory.
Indeed, the internal $SO(32)$ rotational invariance amongst these fermions is nothing
but the spacetime gauge group of the resulting model.

Starting from this model, there are then a number of ways in which we may ``twist''
the boundary conditions of these fields (or, in orbifold language, mod out by discrete  
symmetries).  First, we might seek to twist the boundary conditions of these
sixteen complex fermions into two blocks of eight complex fermions each.  If   
we do this in a way that also breaks spacetime supersymmetry, we obtain a non-supersymmetric,
tachyon free model with gauge group $SO(16)\times SO(16)$;  in orbifold language, we 
have essentially chosen
a SUSY-breaking orbifold which projects out the non-Cartan gauge bosons 
in the coset $SO(32)/[SO(16)\times SO(16)]$.
However, if we try to do this in a way which simultaneously preserves 
spacetime supersymmetry, we find that we cannot obtain $SO(16)\times SO(16)$;  instead,
modular invariance requires 
that our SUSY-preserving orbifold twist simultaneously 
come with a twisted sector which supplies new gauge-boson states, enhancing this
gauge group to $E_8\times E_8$.  This produces
the well-known $E_8\times E_8$ heterotic string.
The $SO(16)\times SO(16)$ and $E_8\times E_8$ heterotic strings 
may thus be taken to sit on the second branch of our family tree. 

Continuing from these three heterotic strings, we may 
continue to perform subsequent orbifold
twists and thereby generate additional models.
For example,
we may act with other configurations of $\IZ_2$ twists 
on the supersymmetric $SO(32)$ string model: 
the three other possible self-consistent models that can be obtained
this way are
the non-supersymmetric $SO(32)$ heterotic string model,
the $SO(8)\times SO(24)$ string model, and a heterotic string
model with gauge group $U(16)=SU(16)\times U(1)$.  All have physical (on-shell) tachyons
in their spectrum.  Likewise, we may perform
various $\IZ_2$ orbifolds of the $E_8\times E_8$ string model:
self-consistent choices produce additional non-supersymmetric, tachyonic
models with gauge groups $SO(16)\times E_8$ and $(E_7)^2\times SU(2)^2$.
Finally, we may also orbifold the $E_8\times E_8$ model by a 
discrete symmetry (outer automorphism)
which exchanges the two $E_8$ gauge groups, producing a final
non-supersymmetric, tachyonic (rank-reduced) model with a single $E_8$ gauge group 
realized at affine level $k=2$~\cite{KLTclassification}. 
By its very nature, this last model is beyond the class of models with complex 
worldsheet fields that we will be considering, since modding out by the 
outer automorphism cannot be achieved on
the worldsheet except through the use of {\it real}\/ worldsheet fermions,
or by employing non-abelian orbifold techniques.

In this manner, 
we have therefore generated the nine self-consistent heterotic
string models in ten dimensions which are known to completely fill out the ten-dimensional
heterotic ``landscape''~\cite{KLTclassification}.    
However, the description we have provided above
represents only one possible route towards reaching these nine models;
other routes along different branches of the tree are possible.
For example, the non-supersymmetric, tachyon-free  $SO(16)\times SO(16)$ heterotic string   
can be realized either as a $\IZ_2$ orbifold of the supersymmetric $SO(32)$ string
or as a different $\IZ_2$ orbifold of the $E_8\times E_8$ string.
Thus, rather than a direct tree of ancestors and descendants,
what we really have are deeply interlocking webs of orbifold relations.

A more potent example of this fact
is provided by the single-$E_8$ heterotic string model.
This model can be constructed through several entirely different
constructions:  as a free-fermionic model involving necessarily real
fermions;  as an abelian orbifold of the $E_8\times E_8$ heterotic
string model in which the discrete symmetry is taken to be an
outer automorphism (exchange) of the two $E_8$ gauge symmetries;
and as a {\it non}\/-abelian orbifold model in which the non-abelian
discrete group is $D_4$~\cite{nonabelian}.  Moreover, as noted above,
even within a given construction
numerous unrelated combinations of parameter choices can yield
exactly the same string model.  These sorts of redundancy issues become
increasingly relevant as larger and larger sets of models
are generated and analyzed,
and must be addressed in order to allow efficient
progress in the task of enumerating models.

One way to categorize different branches of the ``tree'' of models 
is according to total numbers of irreducible gauge-group 
factors that these models contain.
As we have seen,
the ten-dimensional heterotic landscape 
contains exactly three models with only one irreducible gauge group:
these are the $SO(32)$ models, both supersymmetric and non-supersymmetric,
and the single-$E_8$ model.  
By contrast, there are five models with two gauge-group factors:
these are the models with gauge groups
$E_8\times E_8$,
$SO(16)\times SO(16)$,
$SO(24)\times SO(8)$,
$SU(16)\times U(1)$,
and $SO(16)\times E_8$.
Finally, there is one model with four gauge-group factors:
this is the $(E_7)^2\times SU(2)^2$ model.
Note that no other models with other numbers of gauge-group factors
appear in ten dimensions.
Alternatively, we may classify our models into groups depending
on their spacetime supersymmetry properties:  there are two models with
unbroken spacetime supersymmetry, one with broken supersymmetry but
without tachyons, and six with both broken supersymmetry and tachyons.  

Clearly, this ten-dimensional heterotic ``landscape'' is very 
restricted, consisting of only nine discrete models.
Nevertheless, many of the features we shall  
find for the four-dimensional heterotic landscape are already present here:
\begin{itemize}
\item First, we observe that 
    not all gauge groups can be realized.
    For example, we do not find any ten-dimensional heterotic
    string models with gauge group $SO(20)\times SO(12)$,
    even though this gauge group would have the appropriate total rank $16$.
    We also find no models with three gauge-group factors, even though we
    have models with one, two, and four such factors.
    Indeed, of all possible gauge groups with total rank $16$
    that may be constructed from 
    the simply-laced factors $SO(2n)$, $SU(n)$ and $E_{6,7,8}$, we see that
    only eight distinct combinations
    emerge from the complete set of self-consistent string models.
    Likewise, if we allow for the possibility of a broader class of models which
    incorporate rank-cutting, then we must also allow
    for the possibility that our gauge group can be composed of factors which 
    also include the non-simply laced gauge groups $SO(2n+1)$, $Sp(2n)$, $F_4$, and $G_2$.
    However, even from this broader set, only one additional gauge group (a single $E_8$) 
    is actually realized.
\item  Second, we see that certain phenomenological features are
     {\it correlated} in such strings.  For example, although there exist models
    with gauge groups $SO(16)\times SO(16)$ and $E_8\times E_8$, 
    the first gauge group is possible only in the non-supersymmetric case,
    while the second is possible only in the supersymmetric case.
    These two features (the presence/absence of spacetime supersymmetry and 
    the emergence of different possible gauge groups) are features that would be completely
    disconnected in quantum field theory, and thus represent intrinsically {\it stringy correlations}\/. 
    As such, these may be taken to represent 
    statistical predictions from string theory, manifestations
    of the deeper worldsheet self-consistency constraints that string theory imposes.
\item  Third, we have seen that a given string model can be realized in many different 
    ways on the worldsheet, none of which is necessarily special or preferred.
    This is part of the huge redundancy of string
    constructions that we discussed in Sect.~2.
    Thus, all of the string models that we shall
    discuss will be defined to be physically distinct on the basis of their {\it spacetime}\/ properties
    (\eg, on the basis of differing
    spacetime gauge groups or particle content).
\item  Fourth, we see that our ten-dimensional ``landscape'' contains 
    models with varying amounts of supersymmetry:
    in ten dimensions, we found $\calN=1$ supersymmetric models,
    non-supersymmetric tachyon-free models, and 
    non-supersymmetric models with tachyons.
    These are also features which will survive into the heterotic landscapes
    in lower dimensions,
    where larger numbers of unbroken supersymmetries are also possible.
    Since other phenomenological properties of these models may be correlated
    with their degrees of supersymmetry, it is undoubtedly useful
    to separate models according to this primary feature before undertaking
    further statistical analyses.
\item  Finally, we observe that a heterotic string
    model with the single rank-eight gauge group $E_8$ 
    is already present in the ten-dimensional heterotic landscape.
    This is a striking illustration of the fact
    that not all string models can be realized through
    orbifold techniques of the sort we will be 
    utilizing, and that our landscape
    studies will necessarily be limited in both class and scope.
\end{itemize}

%=============================================================================
\section{Gauge groups:  Statistical results} 
\setcounter{footnote}{0}

We now turn our attention to our main interest,
the landscape of heterotic string models in four dimensions.
As we discussed in Sect.~2, the string models we are examining 
are free-field models (\ie, models in which our worldsheet fields are free
and non-interacting).
As such, the gauge sector of such
four-dimensional heterotic string models can be described by 
even self-dual\footnote{
    For strings with spacetime supersymmetry,
    modular invariance requires these gauge lattices to be even and self-dual.
    In other cases, the self-duality properties actually apply
    to the full lattice corresponding to the internal gauge group as
    well as the spacetime Lorentz group.} 
Lorentzian lattices
of dimensionality $(6,22)$, as is directly evident in a bosonic (Narain) 
construction~\cite{Narain,Lerche}.   
[This is the four-dimensional analogue of the sixteen-dimensional lattice
that underlies the $SO(32)$ or $E_8\times E_8$ heterotic string models
in ten dimensions;  the remaining $(6,6)$ components arise internally 
through compactification from ten to four dimensions.]
In general, the right-moving (worldsheet supersymmetric)
six-dimensional components of these gauge lattices correspond at best to very 
small right-moving gauge groups composed of products of  
$U(1)$'s, or $SU(2)$'s realized at affine level $k=2$.
We shall therefore disregard these right-moving gauge groups and focus
exclusively on the left-moving gauge groups of these models.  
Moreover, because we are focusing on free-field constructions involving
only complex bosonic or fermionic worldsheet fields, the possibility
of rank-cutting is not available in these models.  Consequently, these
models all have left-moving simply-laced gauge groups with total rank 22, 
realized at affine level $k=1$.    

As we shall see, the twin requirements of modular invariance and physically
sensible projections impose powerful self-consistency constraints on such models
and their possible gauge groups.    
As such, these are features that are not present for open strings, but they are
ultimately responsible for most of the statistical features we shall observe.
Moreover, as discussed previously, in this paper we shall restrict our attention to models
which are non-supersymmetric but tachyon-free.
Such models are therefore stable at tree level, but have finite, 
non-zero one-loop vacuum energies (one-loop cosmological constants).  

In this section, we shall focus on statistical properties of the gauge groups of these models.
We believe that these properties are largely independent of the fact that our models
are non-supersymmetric.  In Sect.~5, we shall  then focus on the statistical
distributions of the values of their cosmological constants, and in Sect.~6
we shall discuss correlations between the gauge groups and the cosmological constants. 

In comparison with the situation for heterotic strings in ten dimensions,
the four-dimensional situation is vastly more complex, with literally billions and billions
of distinct, self-consistent heterotic string models exhibiting arbitrary degrees of
intricacy and complexity.   
These models are generated randomly, with 
increasingly many randomly chosen twists and 
overlapping orbifold projections.
Each time a self-consistent model is obtained, 
it is compared with all other models that have already been obtained.
It is deemed to represent a new model only if it has a massless spacetime spectrum
which differs from all models previously obtained.
Because of the tremendous worldsheet redundancy inherent 
in the free-fermionic approach, it becomes increasingly more difficult (both
algorithmically and in terms of computer time)
to find a ``new'' model as more and more models are constructed and stored.  
Nevertheless, through random choices for all possible twists 
and GSO projection phases, we 
have generated a set of more than $10^5$ distinct four-dimensional
heterotic string models which we believe provide a 
statistically representative sample of the heterotic landscape. 
Indeed, in many cases we believe that our sample is essentially complete, 
especially for relatively simple models involving relatively few 
orbifold twists.

Just as for ten-dimensional heterotic strings,
we can visualize a ``tree'' structure, grouping our models on higher and higher branches
of the tree in terms of their numbers of gauge-group factors. 
For this purpose, we shall decompose our gauge groups into irreducible factors;
\eg, $SO(4)\sim SU(2)\times SU(2)$ will be considered to have two factors.
We shall generally let $f$ denote the number of irreducible gauge-group factors.
Clearly, as $f$ increases, our gauge group of total rank 22 becomes increasingly
``shattered'' into smaller and smaller pieces.

The following is a description of some salient features of the tree that emerges
from our study.  Again we emphasize that our focus is restricted to models which
are non-supersymmetric but tachyon-free.  

\begin{itemize}
\item  {$f=1$}:\/  We find that there is only one 
heterotic string model whose gauge group contains only one factor:  this is an $SO(44)$ string
model which functions as the ``root'' of our subsequent tree.
This is a model in which all left-moving worldsheet fermions have 
identical boundary conditions, but in which all tachyons are GSO-projected
out of the spectrum.
This model is the four-dimensional analogue of the ten-dimensional $SO(32)$ heterotic string
model.

\item  {$f=2$}:\/  On the next branch, we find $34$ distinct string models whose
gauge groups contain 
two simple factors.  As might be expected, in all cases these gauge groups are 
of the form $SO(44-n)\times SO(n)$ for $n=8,12,16,20$.  Each of these models 
is constructed from the above $SO(44)$ string model by implementing a single twist.
Modular invariance, together with the requirement of worldsheet superconformal invariance
and a single-valued worldsheet supercurrent, are ultimately responsible for 
restricting the possible twists
to those with $n= 8,12,16,20$.   Note that $n=36,32,28,24$
are implicitly included in this set, yielding the same gauge groups as those with 
$n= 8,12,16,20$ respectively.   
Finally, note that cases with $n=4$ (or equivalently $n=40$) are not absent;
they are instead listed amongst the $f=3$ models since 
$SO(4)\sim SU(2)\times SU(2)$.

\item  {$f=3$}:\/  On the next branch, we find that $186$ distinct models with 
eight distinct gauge groups emerge.  Notably, this branch contains 
the first instance of
an exceptional group.  The eight relevant gauge groups, each with total rank 22,
are:
$SO(28)\times SO(8)^2$,
$SO(24)\times SO(12)\times SO(8)$,
$SO(20)\times SO(16)\times SO(8)$,
$SO(20)\times SO(12)^2$,
$SO(16)^2\times SO(12)$,
$SO(40)^2\times SU(2)^2$,
$E_8\times SO(20)\times SO(8)$,
and $E_8\times SO(16)\times SO(12)$.

\item $f=4$:  
This level gives rise to $34$ distinct gauge groups,
and is notable for the 
first appearance of $E_7$ as well as the first appearance of non-trivial
$SU(n)$ groups with $n=4,8,12$.
This is also the first level at which gauge groups begin to contain $U(1)$ factors.
This correlation between $SU(n)$ and $U(1)$ gauge-group factors will be discussed further below. 
Note that other `SU' groups are still excluded, presumably on the basis of modular-invariance
constraints.

\item $f=5$:  This is the first level at which $E_6$ appears, as well as gauge
groups $SU(n)$ with $n=6,10,14$.  We find $49$ distinct gauge groups at this level.

\end{itemize}

As we continue towards larger values of $f$, our gauge groups 
become increasingly ``shattered'' into 
more and more irreducible factors.  
As a result, it becomes more and more difficult to find models 
with relatively large gauge-group factors.  Thus, as $f$ increases, we 
begin to witness the disappearance of relatively large groups and a predominance
of relatively small groups.

\begin{itemize}

\item $f=6$:  At this level, $SU(14)$ disappears while $SU(7)$ makes
its first appearance.  We find $70$ distinct gauge groups at this level.

\item $f=7$:  We find $75$ distinct gauge groups at this level.
Only one of these gauge groups contains an $E_8$ factor. 
This is also the first level at which an $SU(5)$ gauge-group factor appears.

\item $f=8$:  We find $89$ distinct gauge groups at this level.
There are no $E_8$ gauge-group factors at this level,
and we do not find an $E_8$ gauge-group factor again at any higher level.
This also the first level at which an $SU(3)$ factor appears.
Thus, assuming these properties persist for the full heterotic landscape,
we obtain an interesting stringy correlation:
no string models in this class can have a gauge group containing $E_8\times SU(3)$.
These sorts of constraints emerge from modular invariance --- \ie, our inability
to construct a self-dual 22-dimensional lattice with these 
two factors.  Indeed, such a gauge group would have been possible based
on all other considerations (\eg, CFT central charges, total rank constraints, {\it etc.}\/).
 
\item $f=9$:  Here we find that our string models give rise to
$86$ distinct gauge groups.
This level also witnesses the permanent disappearance of $SU(12)$.

\end{itemize}

This tree ends, of course, at $f=22$,
where our gauge groups contain only $U(1)$ and $SU(2)$
factors, each with rank 1. 
It is also interesting to trace the properties of this tree backwards from
the $f=22$ endpoint.

\begin{itemize}

\item $f=22$:  Here we find only $16$ gauge groups, all of the form
$U(1)^n\times SU(2)^{22-n}$ for all values $0\leq n\leq 22$  
except for $n=1,2,3,5,7,9,11$.
Clearly no larger gauge-group factors are possible at this ``maximally shattered'' endpoint.

\item $f=21$:  Moving backwards one level, we find $10$ distinct 
gauge groups, all of the form $U(1)^n\times SU(2)^{20-n}\times SU(3)$ for
$n=9,11,12,13,14,15,16,17,18,19$.
Note that at this level, an $SU(3)$ factor must {\it always}\/ exist  
in the total gauge group since there are no simply-laced irreducible 
rank-two groups other than $SU(3)$.

\item $f=20$:  Moving backwards again, we find $24$ distinct gauge groups
at this level.  Each of these models contains either $SU(3)^2$ or $SU(4)\sim SO(6)$.

\item $f=19$:  Moving backwards one further level, we find $37$ distinct gauge groups,
each of which contains either $SU(3)^3$ or $SU(3)\times SU(4)$ or $SU(5)$ or $SO(8)$.

\end{itemize}

Clearly, this process continues for all branches of our ``tree'' over 
the range $1\leq f\leq 22$.
Combining these results from all branches,
we find a total of $1301$ distinct gauge groups from amongst 
over $120,000$ distinct models.

As we discussed at the end of Sect.~2,
it is important in such a statistical landscape study
to consider whether the properties we find are rigorously 
true or are merely artifacts of having considered 
a finite statistical sample of models 
or a sample 
which is itself not representative of the landscape at large.
One way to do this is to determine 
which properties or statistical correlations
are likely to persist because of some deeper string consistency constraint,
and which are likely to reflect merely a finite sample size.

For example, the fact that the $f=21,22$ levels give rise to gauge groups
of the forms listed above with only particular values of $n$ is likely to
reflect the finite size of our statistical sample.  Clearly, it is extremely
difficult to randomly find a sequence of overlapping sequential orbifold twists 
which breaks the gauge group down to such forms for any arbitrary values of $n$, all
without introducing tachyons, so it is
entirely possible that our random search through the space of models has
simply not happened to find them all.  Thus, such restrictions on the values of
$n$ in these cases are not likely to be particularly meaningful.  

However, the broader fact that $E_8$ 
gauge-group factors are not realized beyond a certain critical value of $f$,
or that $SU(3)$ gauge-group factors
are realized only beyond a different critical value of $f$,
are likely to be properties that relate directly back to the required
self-duality of the underlying charge lattices.  Such properties are therefore
likely to be statistically meaningful and representative of the perturbative
heterotic landscape as a whole.  Of course, the quoted values of these critical
values of $f$ should be viewed as approximate, as statistical results
describing probability distributions.  Nevertheless,
the relative appearance and disappearance
of different gauge-group factors are likely to be meaningful.  
Throughout this paper, we shall focus on only those statistical
correlations which we believe represent
the perturbative heterotic landscape at large.
Indeed, these correlations are
stable in the ``continuum limit''
(in the sense defined at the end of Sect.~2). 

Having outlined the basic tree structure of our heterotic mini-landscape, 
let us now examine its overall statistics and correlations.
The first question we might ask pertains to the {\it overall distribution}\/
of models across bins with different values of $f$.
Just how ``shattered'' are the typical gauge groups in our heterotic landscape?

%================== FIGURE ============================================
\begin{figure}[ht]
\centerline{
   \epsfxsize 4.0 truein \epsfbox {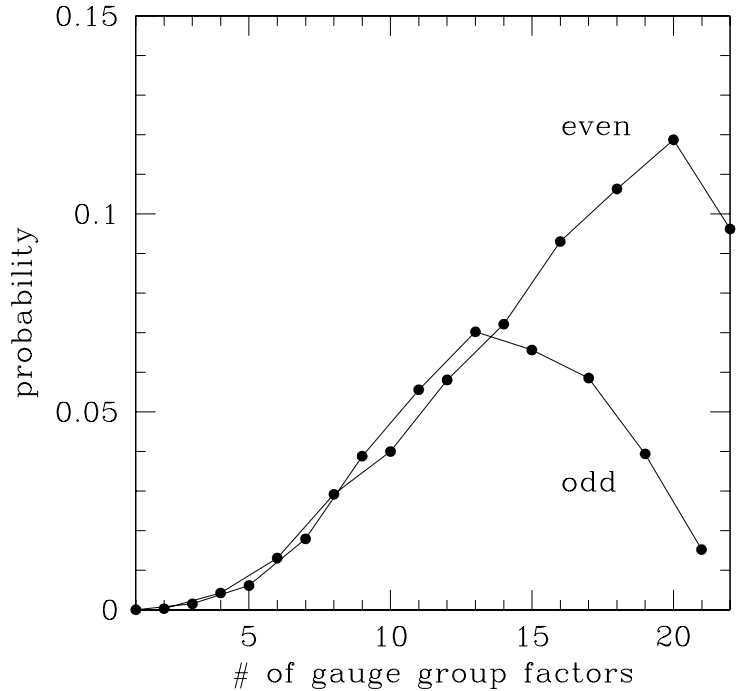}
    }
\caption{The absolute probabilities of obtaining distinct four-dimensional
   heterotic string models as a function of the degree to which their gauge
   groups are ``shattered'' into separate irreducible factors,
   stretching from a unique model with 
   the irreducible rank-22 gauge group $SO(44)$ to models with 
    only rank-one $U(1)$ and $SU(2)$ gauge-group factors.
   The total value of the points (the ``area under the curve'') is 1.  
   As the number of gauge-group factors increases,
    the behavior of the probability distribution bifurcates
    according to whether this number is  even or odd.
   Indeed, as this number approaches its upper limit $22$,
    models with even numbers of gauge-group factors become approximately
    ten times more numerous than those with odd numbers of gauge-group
    factors.}
\label{Fig1} 
\end{figure}
%================== END OF INSERTED FIGURE ============================

The results are shown in Fig.~\ref{Fig1}, where we plot
the absolute probabilities 
of obtaining distinct four-dimensional
   heterotic string models as a function of $f$, 
the number of factors in their spacetime gauge groups.
These probabilities are calculated by dividing the total number of distinct
models that we have found for each value of $f$ by the total number of models
we have generated.
(Note that we plot relative
probabilities rather than raw numbers 
of models 
because it is only these relative probabilities 
which are stable in the continuum limit discussed above.
Indeed, we have explicitly verified that restricting our sample size  
in any random manner
does not significantly affect the overall shape of the curves in Fig.~\ref{Fig1}.)
The average number of gauge-group factors in our sample
set is $\langle f\rangle \approx 13.75$. 

It is easy to understand the properties of these curves.
For $f=1$, we have only one string model, with gauge group
$SO(44)$.  However, as $f$ increases beyond $1$, the 
models grow in complexity, each with increasingly intricate
patterns of overlapping orbifold twists and Wilson lines, 
and consequently the number
of such distinct models grows considerably. 

For $f\gsim 14$, we find that the behavior of this
probability as a function of $f$ bifurcates   
to whether $f$ is even or odd.
Indeed, as $f\to 22$,
we find that models with even numbers of gauge-group factors 
become approximately ten times more numerous than those with 
odd numbers of gauge-group factors.
Of course, this behavior might be an artifact of our statistical
sampling methodology for randomly generating string models.  However,
we believe that this is actually a reflection of the underlying
modular-invariance constraints that impose severe self-duality restrictions
on the charge lattices corresponding to these models.

%================== FIGURE ============================================
\begin{figure}[t]
\centerline{
   \epsfxsize 4.0 truein \epsfbox {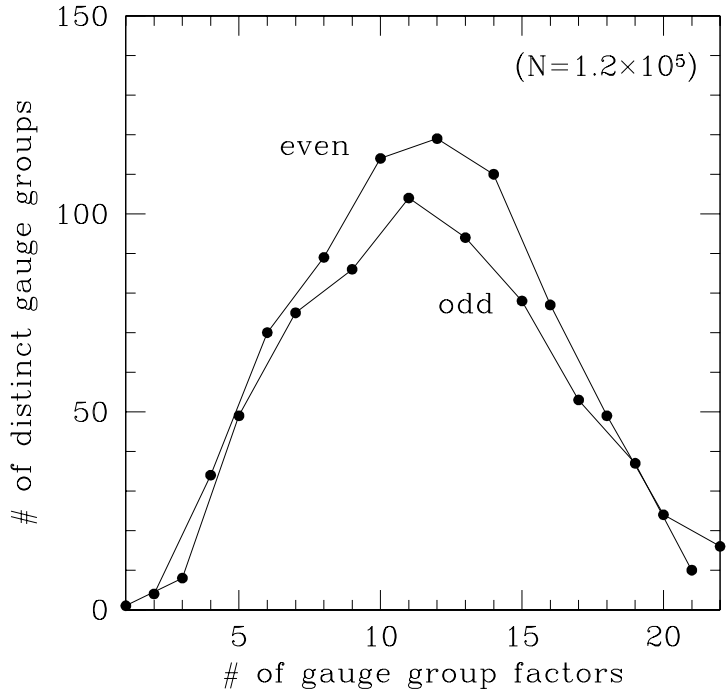}
    }
\caption{The number of distinct gauge groups realized from heterotic
   string models with $f$ gauge-group factors, plotted as a function of $f$.
   Only $1301$ distinct gauge groups are realized from $\sim 10^5$ distinct
   heterotic string models.}
\label{Fig2} 
\end{figure}
%================== FIGURE ============================================

Fig.~\ref{Fig1} essentially represents  the total number of distinct
 {\it models}\/ found as a function of $f$.
However, we can also examine the number of distinct {\it gauge groups}\/
found as a function of $f$.  This data appears in Fig.~\ref{Fig2}.
Once again, although we expect the raw number of distinct gauge groups
for each $f$ to continue to grow as our sample size increases, we do not
expect the overall shape of these curves to change significantly. 
For small values of $f$, the number of distinct realizable gauge groups
is relatively small, as discussed earlier.  For example, for $f=1$
we have a single realizable gauge group $SO(44)$, while
for $f=2$ we have the four groups $SO(44-n)\times SO(n)$ for $n=8,12,16,20$.
Clearly, as $f$ increases, the number of potential gauge group combinations
increases significantly, reaching a maximum for $f\approx 12$.
Beyond this, however, the relative paucity of Lie groups with small rank
becomes the dominant limiting factor, ultimately leading to very few 
distinct realizable gauge groups as $f\to 22$.

\begin{figure}[ht]
\centerline{
   \epsfxsize 4.0 truein \epsfbox {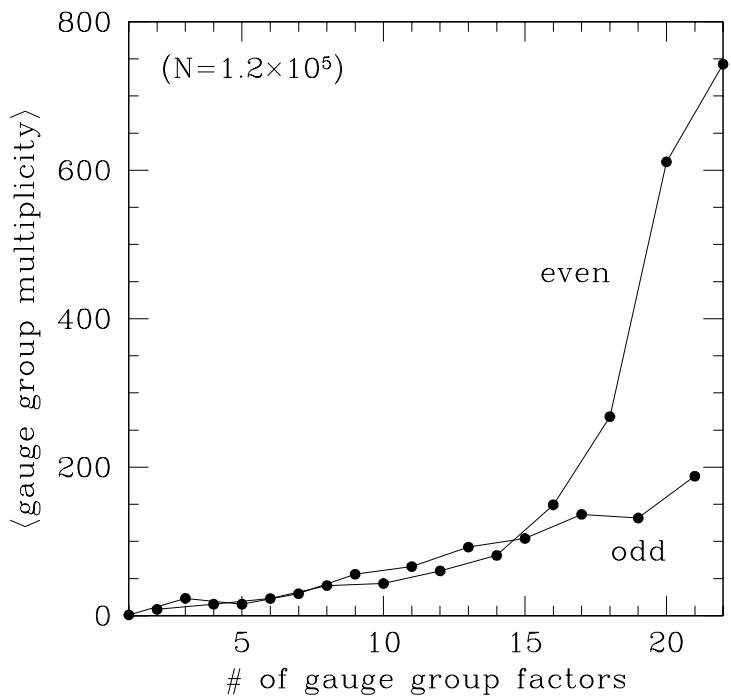}
    }
\caption{Average gauge-group multiplicity (defined as the number of distinct
   heterotic string models divided by the number of distinct gauge groups),
   plotted as a function of $f$, the number of gauge-group factors in the total
   gauge group.
   As the number of factors increases, we see that there are
   indeed more ways of producing a distinct string model with a
   given gauge group.
   Note that the greatest multiplicities occur for models with relatively
   large, {\it even}\/ numbers of gauge-group factors.}
\label{Fig3} 
\end{figure}

Since we have found that $N\approx 1.2\times 10^5$ distinct heterotic string models
yield only $1301$ distinct gauge groups, this number of models
yields an average gauge-group multiplicity factor $\approx 95$.
As we shall discuss later ({\it c.f.}\/ Fig.~\ref{Fig10}), 
we expect that this 
multiplicity will only increase as more models are added 
to our sample set.
However, it is also interesting to examine how this average multiplicity
is distributed across the set of distinct gauge groups.
This can be calculated by dividing
the absolute probabilities of obtaining distinct 
 heterotic string models (plotted as a function of $f$
in Fig.~\ref{Fig1})
by the number of distinct gauge groups 
(plotted as a function of $f$ in Fig.~\ref{Fig2}). 
The resulting average multiplicity factor,
distributed as a function of $f$, is
shown in Fig.~\ref{Fig3}. 
As we see, this average redundancy factor is relatively small for 
small $f$, but grows dramatically as $f$ increases.
This makes sense:  as the heterotic gauge group accrues more
factors, there are more combinations 
of allowed representations for our matter content,
thereby leading to more possibilities for distinct models
with distinct massless spectra.

%================== FIGURE ============================================
%================== END OF INSERTED FIGURE ============================

%================== FIGURE ============================================
\begin{figure}[ht]
\centerline{
   \epsfxsize 4.0 truein \epsfbox {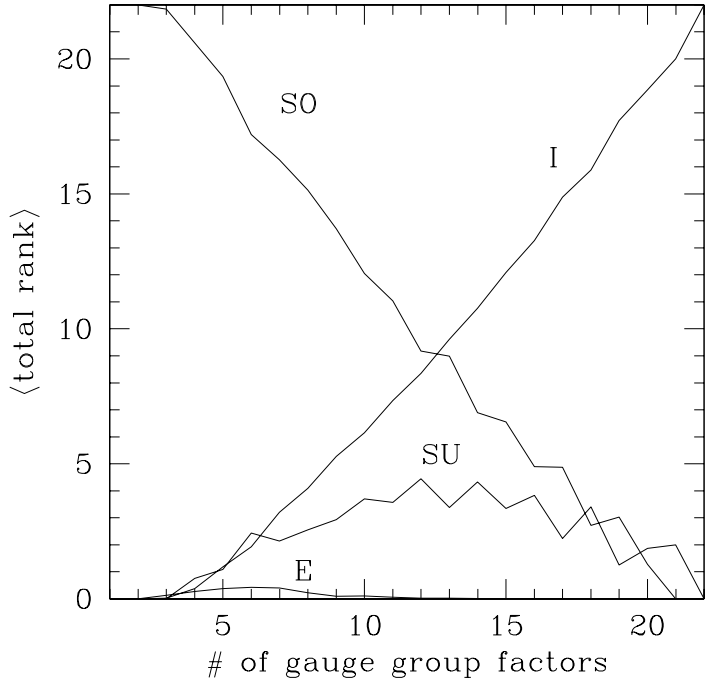}
    }
\caption{
   The composition of heterotic gauge groups, showing
  the average contributions to the total allowed rank
   from $SO(2n\geq 6)$ factors (denoted `SO'), $SU(n\geq 3)$ factors (denoted `SU'),
    exceptional group factors $E_{6,7,8}$ (denoted `E'),
    and rank-one factors $U(1)$ and $SU(2)$ (denoted `I').
    In each case, these contributions are plotted as functions of the number
    of gauge-group factors in the string model and averaged over all
    string models found with that number of factors.
    In the case of $SU(4)\sim SO(6)$ factors,
    the corresponding rank contribution
    is apportioned equally between the `SO' and `SU' categories.
    The total of all four lines is 22, as required.} 
\label{Fig4} 
\end{figure}
%================== END OF INSERTED FIGURE ============================

Another important statistical question we may consider concerns 
the relative abundances of `SO', `SU', and exceptional gauge groups.
Regardless of the value of $f$, the total rank of the full gauge group
is $22$;  thus, the interesting question concerns how this total rank
is apportioned amongst these different classes of Lie groups. 
This information is shown in Fig.~\ref{Fig4}.  For the purposes of this plot,
contributions from $SU(4) \sim SO(6)$ factors, when they occur, 
are equally shared between `SO' and `SU' categories.
The total of all four lines is 22, as required. 
Once again, we observe several important features which are consistent with
our previous discussions.
First, we see that all of the allowed rank is found to reside in `SO' groups
for $f=1,2$;  for $f=1$, this is because the unique realizable
gauge group in such models
is $SO(44)$, while for $f=2$, this is because 
the only realizable gauge-group breaking 
in such models is of the form 
$SO(44)\to SO(44-n)\times SO(n)$ for $n=8,12,16,20$.  
We also observe that the `E' groups do not contribute any net rank 
unless $f\geq 3$, while 
the `SU' groups do not contribute any net rank unless $f\geq 4$.
It is worth noting that the exceptional groups $E_{6,7,8}$ are exceedingly
rare for all values of $f$, especially given that their share of the total rank 
in a given string model must be 
at least six whenever they appear.
As $f$ grows increasingly large, however, the bulk of the rank is to be found
in $U(1)$ and $SU(2)$ gauge factors.
Of course, for $f=21$, the `SU' groups have an average rank 
which is exactly equal to $2$.
This reflects the fact that each of the realizable gauge groups for
$f=21$ necessarily contains a single $SU(3)$ factor, as previously discussed.

Across all of our models, we find that
\begin{itemize}
\item  {\bf 85.79\%} of our heterotic string models
contain $SO(2n\geq 6)$ factors;  amongst these models,
the average number of such factors is $\approx 2.5$.
\item  {\bf 74.35\%} of our heterotic string models
contain $SU(n\geq 3)$ factors;  amongst these models,
the average number of such factors is $\approx 2.05$.
\item  {\bf 0.57\%} of our heterotic string models
contain $E_{6,7,8}$ factors;  amongst these models,
the average number of such factors is $\approx 1.01$.
\item  {\bf 99.30\%} of our heterotic string models
contain $U(1)$ or $SU(2)$ factors;  amongst these models,
the average number of such factors is $\approx 13.04$.
\end{itemize}
In the above statistics, an $SU(4)\sim SO(6)$ factor
is considered to be a member of whichever category (`SU' or `SO') 
is under discussion.

Note that these statistics are calculated across distinct
heterotic string models.  However, as we have seen, there is
a tremendous redundancy of gauge groups amongst these string
models, with only 1301 distinct gauge groups appearing for these
$\approx 1.2\times 10^5$ models.
Evaluated across the set of distinct {\it gauge groups}\/ which emerge from
these string models 
(or equivalently, employing a different measure
which assigns statistical weights to models according to the distinctness of 
their gauge groups),
these results change somewhat:
\begin{itemize}
\item  {\bf 88.55\%} of our heterotic gauge groups 
contain $SO(2n\geq 6)$ factors;  amongst these groups,
the average number of such factors is $\approx 2.30$.
\item  {\bf 76.79\%} of our heterotic gauge groups 
contain $SU(n\geq 3)$ factors;  amongst these groups,
the average number of such factors is $\approx 2.39$.
\item  {\bf 8.38\%} of our heterotic gauge groups 
contain $E_{6,7,8}$ factors;  amongst these groups,
the average number of such factors is $\approx 1.06$.
\item  {\bf 97.62\%} of our heterotic gauge groups 
contain $U(1)$ or $SU(2)$ factors;  amongst these groups,
the average number of such factors is $\approx 8.83$.
\end{itemize}
Note that the biggest relative change occurs for the exceptional groups, with
over $8\%$ of our gauge groups containing exceptional factors.
Thus, we see that while exceptional gauge-group factors appear somewhat
frequently within the set of allowed distinct heterotic gauge 
groups, the gauge groups containing exceptional factors
emerge relatively rarely from our underlying 
heterotic string models.

%================== FIGURE ============================================
\begin{figure}[ht]
\centerline{
   \epsfxsize 4.0 truein \epsfbox {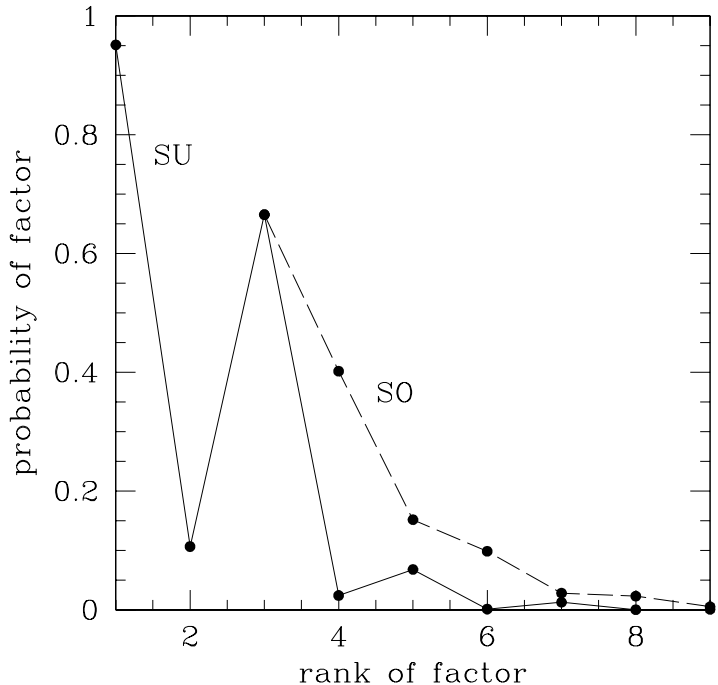}
    }
\caption{The probability that a given $SO(2n)$ or $SU(n+1)$ gauge-group 
    factor appears at least once in the gauge group of a randomly
    chosen heterotic string model, plotted as a function of the rank $n$
    of the factor.  
    While the `SU' curve (solid line) is plotted for all ranks $\geq 1$,
    the `SO' curve (dashed line) is only plotted for ranks $\geq 3$
    since $SO(2)\sim U(1)$ and $SO(4)\sim SU(2)^2$.
    These curves necessarily share a common point for rank 3,
    where $SU(4)\sim SO(6)$.    
    }  
\label{Fig5} 
\end{figure}
%================== END OF INSERTED FIGURE ============================

Of course, this information does not indicate the probabilities
of individual `SO' or `SU' groups. 
Such individual probabilities are shown in Fig.~\ref{Fig5}, where we indicate
the probabilities of individual `SO' or `SU' groups 
as functions of the ranks of these groups.  
We observe that for all ranks~$\geq 3$, 
the `SU' groups are significantly less common
than the corresponding `SO' groups.
This pattern exists even for ranks up to $22$, where
the probabilities for `SU' groups continue to be significantly
less than those of their corresponding `SO' counterparts.
This helps to explain why, despite the information itemized
above, the `SO' groups are able to make a significantly larger contribution to 
the total rank than do the `SU' groups, 
as illustrated in Fig.~\ref{Fig4}.

It is natural to wonder why the `SO' groups tend to dominate
over the `SU' groups in this way, especially since
ordinary quantum field theory would lead to no such preferences.
Of course, it is entirely possible that these results may indicate some sort
of bias in our sample of free-field string models.
However, in a heterotic string framework, we recall that gauge 
symmetries ultimately have their origins as 
internal symmetries amongst worldsheet fields.  
Indeed, within the free-field constructions we are examining, 
`SO' groups tend to be the most
natural since they represent rotational symmetries amongst
identical worldsheet fields.  
By contrast, `SU' groups are necessarily more difficult to construct,
especially as the rank of the `SU' group becomes relatively large.

We can illustrate this fact directly in the case of
of free-field worldsheet constructions by considering the relevant charge
lattices for the gauge-boson states.  These charges are nothing but 
the weights of the adjoint representations of these gauge groups, where
each direction of the charge lattice corresponds to a different worldsheet
field.  It is then straightforward to consider how the different
gauge groups are embedded in such a lattice, \ie, how these gauge-boson
states can be represented in terms of the underlying string
degrees of freedom.  For example, in a string formulation
based upon complex worldsheet bosons $\phi_\ell$ or fermions $\psi_\ell$,
each lattice direction $\hat e_\ell$ --- and consequently each
generator $U_\ell$ --- corresponds to a different worldsheet boson
or fermion:  $U_\ell\equiv i\partial \phi_\ell=\overline{\psi}_\ell \psi_\ell$.
Given such a construction,
we simply need to understand how the simple roots of each gauge group are
oriented with respect
to these lattice directions.

Disregarding irrelevant overall lattice permutations and inversions,
the `SO' groups have a natural lattice embedding.
For any group $SO(2n)$, the roots $\lbrace \vec \alpha\rbrace$
can be represented in an $n$-dimensional lattice
as $\lbrace \pm \hat e_i \pm \hat e_j\rbrace$, with the simple roots
given by $\vec\alpha_i = \hat e_i - \hat e_{i+1}$ for $1\leq i\leq n-1$,
and $\vec \alpha_n = \hat e_{n-1}+ \hat e_{n}$.
As we see, all coefficients for these embeddings are integers, which means
that these charge vectors  can be easily realized 
through excitations of Neveu-Schwarz worldsheet fermions.

By contrast, the group $SU(n)$ contains roots with necessarily
non-integer coefficients if embedded in an $(n-1)$-dimensional lattice
[as appropriate for the rank of $SU(n)$].
For example, $SU(3)$ has two simple roots whose relative angle is $2\pi /3$,
ensuring that no two-dimensional orthogonal coordinate system can be found with       
respect to which both roots have integer coordinates.
In free-field string constructions, this problem is circumvented by
embedding our $SU(n)$ groups into an $n$-dimensional lattice
rather than an $(n-1)$-dimensional lattice.
One can then represent the $n-1$ simple roots
as $\vec \alpha_i =  \hat e_i - \hat e_{i+1}$
for $1\leq i\leq n-1$, using only integer coefficients.
However, this requires the use of an {\it extra}\/ lattice direction
in the construction --- \ie, this requires the coordinated participation
of an additional worldsheet degree of freedom.
Indeed, the $SU(n)$ groups are realized non-trivially
only along diagonal hyperplanes within a higher-dimensional charge lattice. 
Such groups are consequently more difficult to achieve than their $SO(2n)$ cousins.

This also explains why the appearance of $SU(n)$ gauge groups is 
strongly correlated with the appearance of $U(1)$ factors in such free-field string
models.
As the above embedding indicates, in order to realize $SU(n)$ 
what we are really doing 
is first realizing $U(n)\equiv SU(n)\times U(1)$ in an $n$-dimensional
lattice.  In this $n$-dimensional lattice,
the $U(1)$ group factor amounts to the trace
of the $U(n)$ symmetry, and corresponds to the lattice
direction ${\vec E}\equiv \sum_{\ell=1}^n \hat e_\ell$.
The $(n-1)$-dimensional hyperplane
orthogonal to ${\vec E}$ then corresponds to the $SU(n)$ gauge group.
Thus, in such free-field string models, 
we see that the appearance of $SU(n)$ gauge groups is naturally
correlated with the appearance of $U(1)$ gauge groups.
Indeed, within our statistical sample of heterotic string
models, we find that
\begin{itemize}
\item
 {\bf 99.81\% of all heterotic string models which contain one or more 
$SU(n)$ factors also exhibit 
an equal or greater number of $U(1)$ factors.}   
[In the remaining 0.19\% of models, 
one or more of these $U(1)$ factors 
is absorbed to become part of another non-abelian group.]
By contrast, the same can be said for only $74.62\%$ of models
with $SO(2n\geq 6)$ factors
and only $61.07\%$ of models with $E_{6,7,8}$ factors.
Given that the average number of $U(1)$ factors
per model across our entire statistical sample
is $\approx 6.75$, these results for the `SO' and `E' groups
are essentially random and do not reflect 
any underlying correlations.  
\end{itemize}
Note that these last statements only apply to $SU(n)$ gauge-group
factors with $n=3$ or $n\geq 5$;  the special case $SU(4)$
shares a root system with the orthogonal group $SO(6)$ and consequently
does not require such an embedding.

\begin{figure}[ht]
\centerline{
   \epsfxsize 3.1 truein \epsfbox {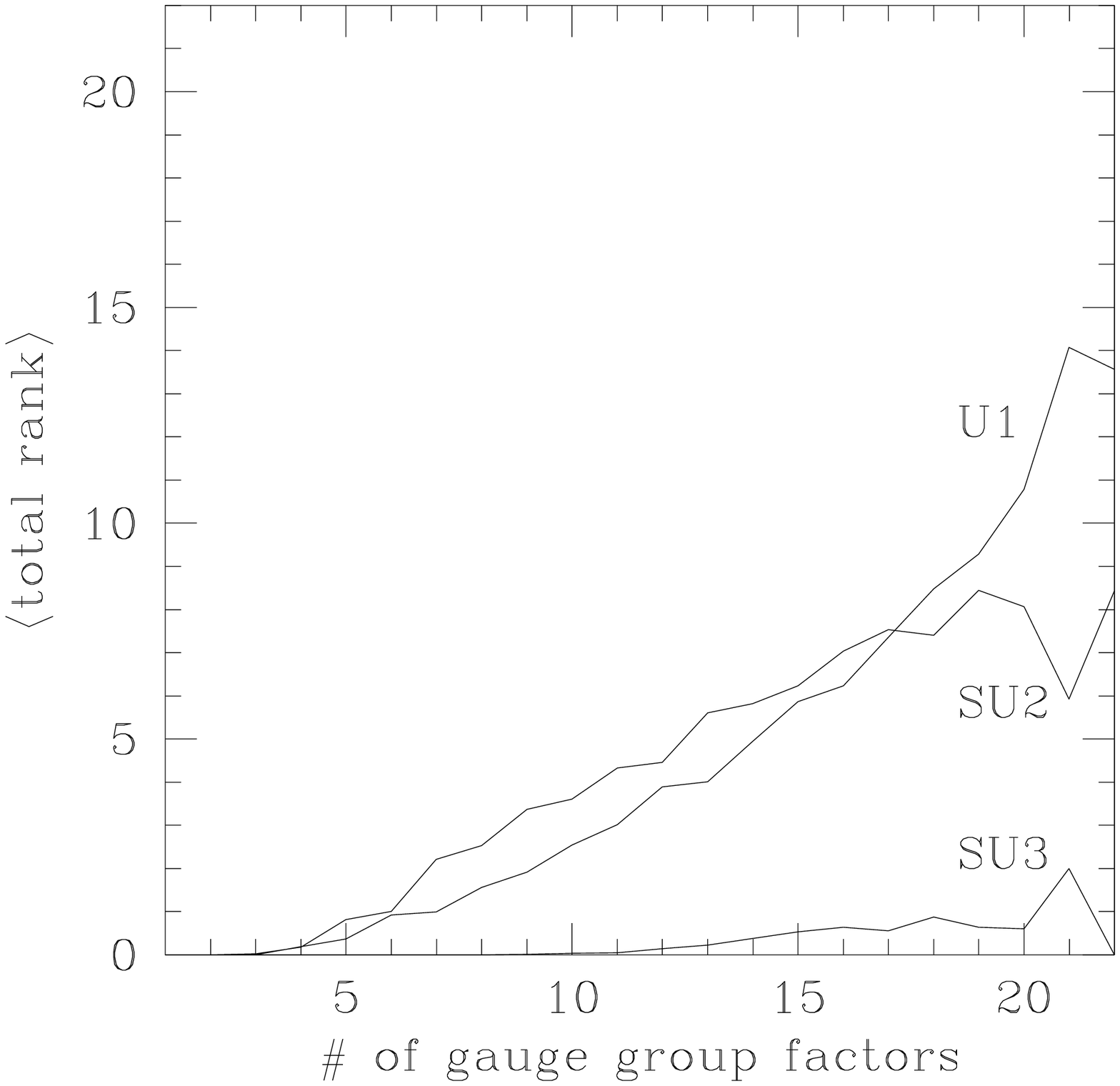}
   \epsfxsize 3.1 truein \epsfbox {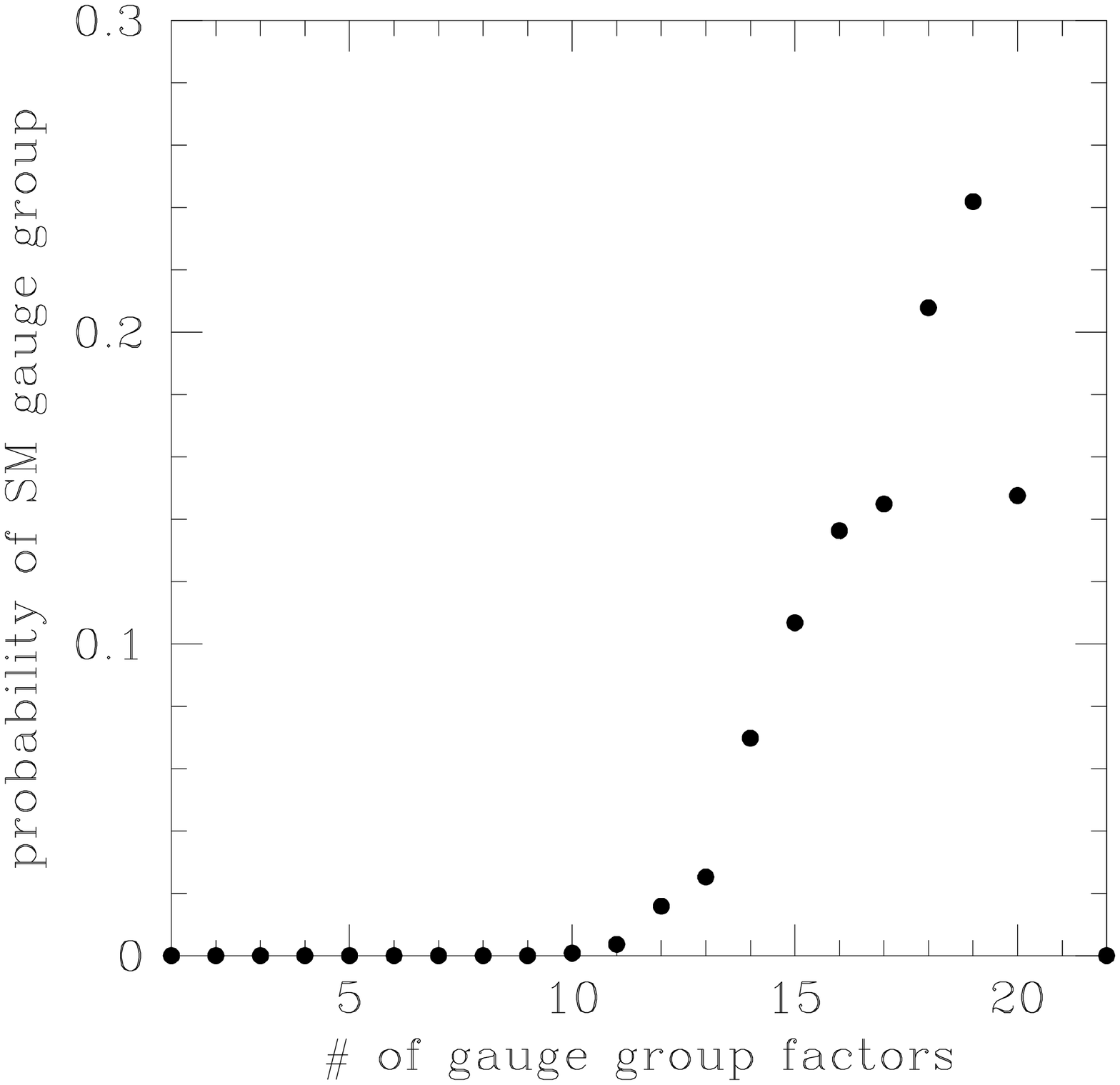}
    }
\caption{(a)  (left) 
    The distribution of total rank amongst $U(1)$, $SU(2)$, and $SU(3)$ 
    gauge-group factors as a function of the total number of gauge-group
    factors.  The sum of the $U(1)$ and $SU(2)$ lines reproduces
    the `I' line in Fig.~\protect\ref{Fig4}, while the $SU(3)$ line is a 
    subset of the `SU' line in Fig.~\protect\ref{Fig4}.
    (b) (right)  Absolute joint probability of obtaining a string model
    with the Standard-Model gauge group $G_{\rm SM}\equiv SU(3)\times SU(2)\times U(1)$
    as a function of the total number of gauge-group factors in the
    string model.  Although off-scale and therefore not shown on this plot, 
     the probability of realizing $G_{\rm SM}$ actually hits 1 for $f=21$.
    }  
\label{Fig6} 
\end{figure}

For the purposes of realizing the Standard Model from heterotic strings, 
we may also be interested in the relative probabilities
of achieving $SU(3)$, $SU(2)$, and $U(1)$ 
gauge-group factors individually.
This information is shown in Fig.~\ref{Fig6}(a).
Moreover, in Fig.~\ref{Fig6}(b), we show the  
joint probability of {\it simultaneously}\/ 
obtaining at least one of each of these 
factors within a given string model,
producing the combined factor $G_{\rm SM}\equiv SU(3)\times SU(2)\times U(1)$. 
Indeed, averaging across all our heterotic string models,
we find that
\begin{itemize}
\item  {\bf 10.64\%} of our heterotic string models
contain $SU(3)$ factors;  amongst these models,
the average number of such factors is $\approx 1.88$.
\item  {\bf 95.06\%} of our heterotic string models
contain $SU(2)$ factors;  amongst these models,
the average number of such factors is $\approx 6.85$.
\item  {\bf 90.80\%} of our heterotic string models
contain $U(1)$ factors;  amongst these models,
the average number of such factors is $\approx 4.40$.
\end{itemize}

%================== FIGURE ============================================
%================== END OF INSERTED FIGURE ============================

Note that these overall probabilities for $SU(3)$ and $SU(2)$ factors
are consistent with those shown in Fig.~\ref{Fig5}.
By contrast, across the set of allowed {\it gauge groups}\/,
we find that
\begin{itemize}
\item  {\bf 23.98\%} of our heterotic gauge groups
contain $SU(3)$ factors;  amongst these groups,
the average number of such factors is $\approx 2.05$.
\item  {\bf 73.87\%} of our heterotic gauge groups
contain $SU(2)$ factors;  amongst these groups,
the average number of such factors is $\approx 5.66$.
\item  {\bf 91.47\%} of our heterotic gauge groups
contain $U(1)$ factors;  amongst these groups,
the average number of such factors is $\approx 5.10$.
\end{itemize}
We see that the biggest relative change occurs for $SU(3)$ gauge-group
factors:  although such factors appear within almost $24\%$ of
the allowed gauge groups, these gauge groups emerge from underlying
string models only half as frequently as we would have expected.
This is why only $10\%$ of our distinct heterotic
string models contain $SU(3)$ gauge-group factors.

At first glance, it may seem that these results for $SU(2)$ and $U(1)$
factors conflict with the results in Fig.~\ref{Fig6}(a).
However, the total number of 
models containing at least one gauge-group factor of a given
type is dependent not only on the average rank contributed by 
a given class of gauge group as a function of $f$ [as shown in Fig.~\ref{Fig6}(a)],
but also on the overall number of models as a function of $f$
[as shown in Fig.~\ref{Fig1}]. 
Thus, these plots provide independent information
corresponding to different ways of correlating 
and presenting statistics for the same data set.

As we see from Fig.~\ref{Fig6}(a),
$SU(3)$ gauge factors do not statistically appear in our heterotic
string models until the overall gauge group has been ``shattered''
into at least eight irreducible factors.  
Moreover, as we have seen, the net probabilities of $SU(2)$ and $U(1)$ factors
peak only when there are relatively large numbers of factors.
Consequently, we observe from the joint probabilities in Fig.~\ref{Fig6}(b)
that the entire Standard-Model gauge group does not 
statistically appear until our overall gauge group has been shattered
into at least $10$ gauge-group factors.
This precludes the appearance of gauge groups such as $G_{\rm SM}\times SO(36)$,
$G_{\rm SM}\times E_6\times SO(24)$, and so forth --- all of which would
have been allowed on the basis of rank and central charge constraints.
Once again, it is the constraint of the self-duality of the complete charge lattice
--- {\it i.e.}\/, the modular invariance of the underlying string model --- which is ultimately
the origin of such correlations.
These results also agree with what has been found in several
explicit (supersymmetric) semi-realistic perturbative heterotic 
string models with Standard-Model gauge groups~\cite{Faraggimodels}.

Note from Fig.~\ref{Fig6}(b)
that the probability of obtaining the Standard-Model gauge group 
actually hits $100\%$ for $f=21$, and drops to zero for $f=22$.
Both features are easy to explain in terms of correlations we have already
seen.
For $f=21$, our gauge groups are {\it required}\/ to contain an $SU(3)$
factor since there are no simply-laced irreducible 
rank-two groups other than $SU(3)$.
[This is also why the $SU(3)$ factors always contribute
exactly two units of rank to the overall rank for $f=21$, as indicated in Fig.~\ref{Fig6}(a).]
For $f=22$, by contrast, no $SU(3)$ factors can possibly appear.

Another important issue for string model-building
concerns cross-correlations between {\it different}\/ gauge groups ---
\ie, the joint probabilities that two different gauge groups appear
simultaneously within a single heterotic string model.
For example, while one gauge-group factor may correspond to our observable
sector, the other factor may correspond to a hidden sector. 
Likewise, for model-building purposes, we might also be interested
in probabilities
that involve the entire Standard-Model
   gauge group $G_{\rm SM}\equiv SU(3)\times SU(2)\times U(1)$
or the entire Pati-Salam gauge group
    $G_{\rm PS}\equiv SU(4)\times SU(2)^2$.

%================================
\begin{table}
\centerline{
   \begin{tabular}{||r|| r|r|r|r|r|r|r|r|r|r||r|r||}
   \hline
   \hline
   ~ & $U_1$~ & $SU_2$ & $SU_3$ & $SU_4$ & $SU_5$ & $SU_{>5}$ & $SO_8$
            & $SO_{10}$ & $SO_{>10}$ & $E_{6,7,8}$ & SM~ & PS~ \\
   \hline
   \hline
    $U_1$ & 87.13& 86.56& 10.64& 65.83&  2.41&  8.20& 32.17& 14.72&  8.90&  0.35& 10.05& 61.48 \\
\hline
     $SU_2$ & ~ & 94.05& 10.05& 62.80&  2.14&  7.75& 37.29& 13.33& 12.80&  0.47&  9.81& 54.31 \\
\hline
     $SU_3$ & ~ &  ~ &  7.75&  5.61&  0.89&  0.28&  1.44&  0.35&  0.06&  $10^{-5}$ &  7.19&  5.04 \\
\hline
     $SU_4$ & ~ &  ~ &  ~ & 35.94&  1.43&  5.82& 24.41& 11.15&  6.53&  0.22&  5.18& 33.29 \\
\hline
     $SU_5$ & ~ &  ~ &  ~ &  ~ &  0.28&  0.09&  0.46&  0.14&  0.02&  0 &  0.73&  1.21 \\
\hline
     $SU_{>5}$ & ~ &  ~ &  ~ &  ~ &  ~ &  0.59&  3.30&  1.65&  1.03&  0.06&  0.25&  4.87 \\
\hline
     $SO_8$ & ~ &  ~ &  ~ &  ~ &  ~ &  ~ & 12.68&  6.43&  8.66&  0.30&  1.19& 22.02 \\
\hline
     $SO_{10}$ & ~ &  ~ &  ~ &  ~ &  ~ &  ~ &  ~ &  2.04&  2.57&  0.13&  0.25&  9.44 \\
\hline
     $SO_{>10}$ & ~ &  ~ &  ~ &  ~ &  ~ &  ~ &  ~ &  ~ &  3.03&  0.25&  0.03&  5.25 \\
\hline
     $E_{6,7,8}$ & ~ &  ~ &  ~ &  ~ &  ~ &  ~ &  ~ &  ~ &  ~ &  0.01&  0&  0.13 \\
\hline
\hline
     SM & ~ &  ~ &  ~ &  ~ &  ~ &  ~ &  ~ &  ~ &  ~ &  ~ &  7.12&  3.86 \\
\hline
     PS & ~ &  ~ &  ~ &  ~ &  ~ &  ~ &  ~ &  ~ &  ~ &  ~ &  ~ & 26.86 \\
   \hline
   \hline
    total:& 90.80& 95.06& 10.64& 66.53&  2.41&  8.20& 40.17& 15.17& 14.94&  0.57& 10.05& 62.05\\
   \hline
   \hline
   \end{tabular}
}
\caption{Percentages of four-dimensional heterotic string models which exhibit
   various combinations of gauge groups.
   Columns/rows labeled as $SU_{>5}$, $SO_{>10}$, and $E_{6,7,8}$ indicate
   {\it any}\/ gauge group in those respective categories [\eg, $SU_{>5}$ indicates
   any gauge group $SU(n)$ with $n>5$].
   Off-diagonal entries show the percentage of models whose gauge groups simultaneously 
   contain the factors associated with the corresponding rows/columns,
   while diagonal entries show the percentage of models which meet the corresponding
   criteria at least {\it twice}\/.
   For example, $7.75\%$ of models contain an $SU(2)\times SU(n)$
   factor with any $n>5$, while
   $35.94\%$ of models contain at least two $SU(4)$ factors.
   `SM' and `PS' respectively indicate the Standard-Model
   gauge group $G_{\rm SM}\equiv SU(3)\times SU(2)\times U(1)$
   and the Pati-Salam gauge group
    $G_{\rm PS}\equiv SU(4)\times SU(2)^2$.  Thus,
   only $0.13\%$ of models contain $G_{PS}$ together
   with an exceptional group, while 
   $33.29\%$ of models contain at least 
   $G_{\rm PS}\times SU(2)=SU(4)^2\times SU(2)^2$ and
   $26.86\%$ of models contain at least
   $G_{\rm PS}^2= SU(4)^2\times SU(2)^4$.
   A zero entry indicates that
  no string model with the required properties was found, whereas
  the entry $10^{-5}$ indicates
  the existence of a single string model with the
  given properties.
  Entries along the `total' row indicate the total percentages of models which have
   the corresponding gauge-group factor, regardless of what other 
   gauge groups may appear;  note that this is {\it not}\/ merely the
   sum of the joint probabilities along a row/column since these 
   joint probabilities are generally not exclusive.  For example, 
   although $86.56\%$ of models have both an $SU(2)$ factor
   and a $U(1)$ factor and $94.05\%$ of models contain
   $SU(2)^2$, the total percentage of models containing
   an $SU(2)$ factor is only slightly higher at $95.06\%$,
   as claimed earlier.
   Note that nearly every string model
   which contains an $SU(n\geq 3)$ gauge-group factor for $n\not=4$ also contains a $U(1)$
   gauge-group factor, as discussed earlier;  thus, the joint and individual 
   probabilities are essentially equal in this case.  Also note
   that only $10.05\%$ of heterotic string models contain
   the Standard-Model gauge group, regardless of other 
   correlations;  moreover, when this gauge group appears, it
   almost always comes with an additional $SU(2)$, $SU(3)$,
   or $SU(4)$ factor. }
\label{table1}
\end{table}
%================================

This information is collected in Table~\ref{table1}.
It is easy to read a wealth of information from this table.
For example, this table provides further confirmation
of our previous claim that nearly all heterotic string
models which contain an $SU(n\geq 3)$ factor for $n\not=4$
also contain a corresponding $U(1)$ factor.
[Recall that $SU(4)$ is a special case:  since $SU(4)\sim SO(6)$, the roots of $SU(4)$
have integer coordinates in a standard lattice embedding.]

Likewise, we see from this table that 
\begin{itemize}
\item  {\bf The total probability of obtaining the Standard-Model 
    gauge group across our entire sample set is only 10.05\%, regardless of what
    other gauge-group factors are present.}
\end{itemize}
This is similar to what is found for Type~I strings~\cite{blumenhagen},
and agrees with the sum of the data points shown in Fig.~\ref{Fig6}(b)
after they are weighted by the results shown in Fig.~\ref{Fig1}. 
Since we have seen that only $10.64\%$ of our heterotic string models
contain at least one $SU(3)$ gauge factor
[as compared with $95.06\%$ for $SU(2)$ and $90.80\%$ for
$U(1)$], 
we conclude that {\it the relative scarcity of $SU(3)$ factors
is the dominant feature involved
in suppressing the net probability of appearance 
of the Standard-Model gauge group}.
Indeed, the relative scarcity of $SU(3)$ gauge-group factors is amply
illustrated in this table.

It is also apparent that among the most popular GUT groups
$SU(5)$, $SO(10)$, and $E_6$, the choice $SO(10)$ is most
often realized across this sample set.
This again reflects the relative difficulty of realizing
`SU' and `E' groups.
Indeed,
\begin{itemize}
\item  {\bf The total probability of obtaining a potential GUT group of the 
    form $SO(2n\geq 10)$ across our entire sample set is 24.5\%,
    regardless of what other gauge-group factors are present. 
    For $SU(n\geq 5)$ and $E_{6,7,8}$ GUT groups, by contrast, these probabilities
    fall to 7.7\% and 0.2\% respectively.}  
\end{itemize}
Once again, we point out that these numbers are independent of those
in Table~\ref{table1}, since the sets of models exhibiting $SO(10)$ 
versus $SO(2n>10)$ factors
[or exhibiting $SU(5)$ versus $SU(n>5)$ factors] 
are not mutually exclusive.

Note that for the purposes of this tally, we are considering the $SO(4n\geq 12)$ groups
as potential GUT groups even though they do not have complex representations;
after all, these groups may dynamically break at lower energies to
the smaller $SO(4n-2\geq 10)$ groups which do.
Moreover, although we are referring to such groups 
as ``GUT'' groups, we do not mean to imply that string models
realizing these groups
are necessarily suitable candidates for realizing various grand-unification 
scenarios.
We only mean to indicate that these gauge groups are those that
are found to have unification possibilities in terms of the quantum
numbers of their irreducible representations.
In order for a string model to realize a complete unification scenario,
it must also give rise to the GUT Higgs field which is necessary 
for breaking the GUT group down to that of the Standard Model.
For each of the potential GUT groups we are discussing, the smallest 
Higgs field that can accomplish this breaking 
must transform in the adjoint representation of the GUT
group, and string unitarity constraints imply that such Higgs fields 
cannot appear in the massless string spectrum
unless these GUT groups are realized
at affine level $k>1$.  
This can only occur in models which 
also exhibit rank-cutting~\cite{rankcutting,Prep},
and as discussed in Sect.~2, 
this is beyond 
the class of models we are examining.
Nevertheless, there are many ways in which such models may
incorporate alternative unification scenarios.  For example, such GUT
groups may be broken by 
field-theoretic
means ({\it e.g.}\/, through effective or composite Higgs fields which emerge
by running to lower energies and which therefore
do not correspond to elementary string states at the string scale).

Thus far, we have paid close attention to the {\it ranks}\/ of the gauge groups.
There is, however, another important property
of these groups, namely their {\it orders}\/ or dimensions ({\it i.e.}\/, the
number of gauge bosons in the massless spectrum of 
the corresponding string model).
This will be particularly relevant in Sect.~6 when we examine correlations between
gauge groups and one-loop vacuum amplitudes (cosmological constants).

Although we find $1301$ gauge groups for our $\sim 10^5$ models,
it turns out that these gauge groups have only $95$ distinct orders.
These stretch all the way from $22$ [for $U(1)^{22}$] to $946$ [for $SO(44)$].  
Note that the $22$ gauge bosons for $U(1)^{22}$ 
are nothing but the Cartan generators at the origin of our
22-dimensional charge lattice,
with all higher orders signalling the appearance of
additional non-Cartan generators which
enhance these generators to form larger, non-abelian Lie groups. 

\begin{figure}[ht]
\centerline{
   \epsfxsize 3.1 truein \epsfbox {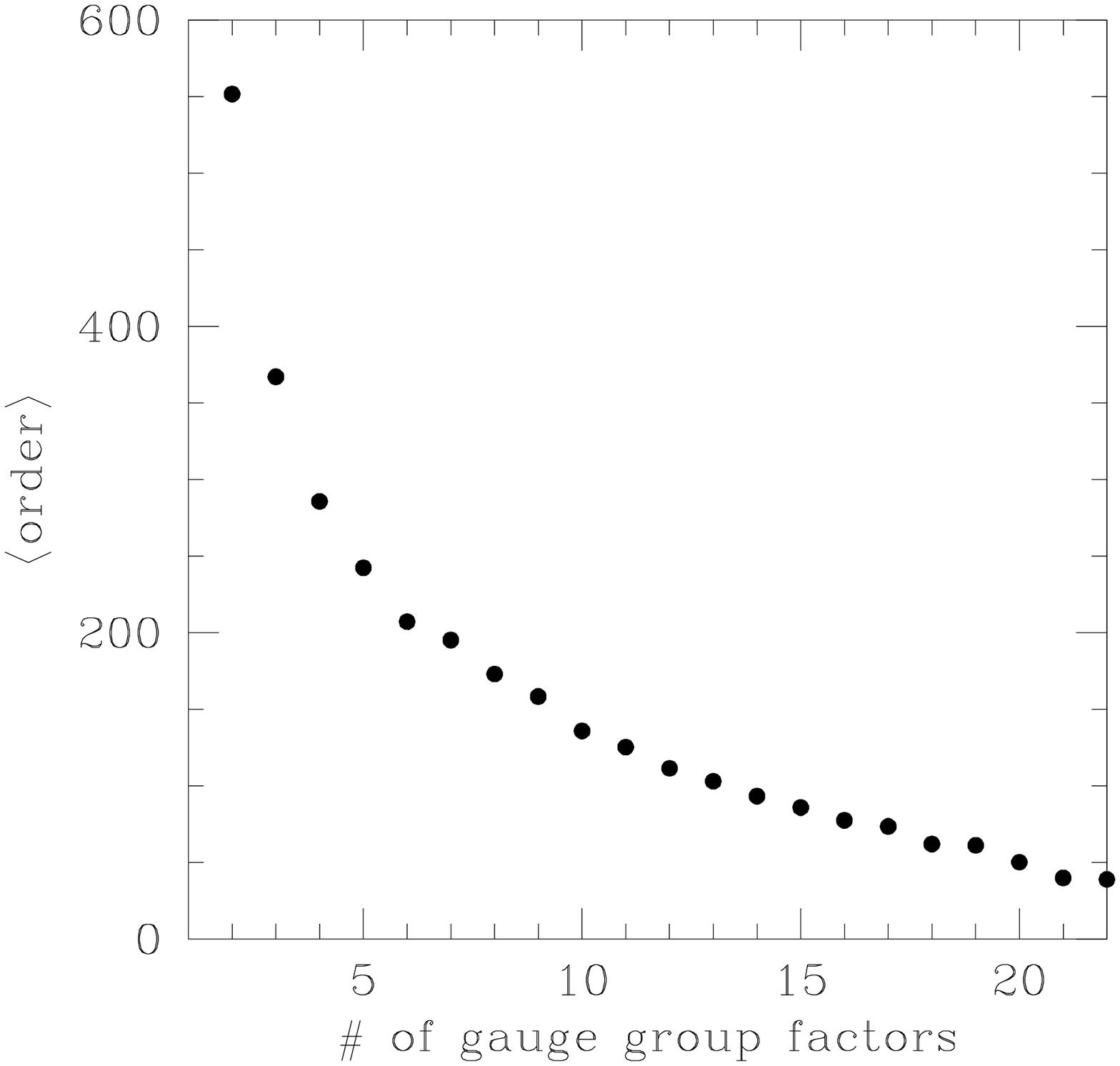}
   \epsfxsize 3.1 truein \epsfbox {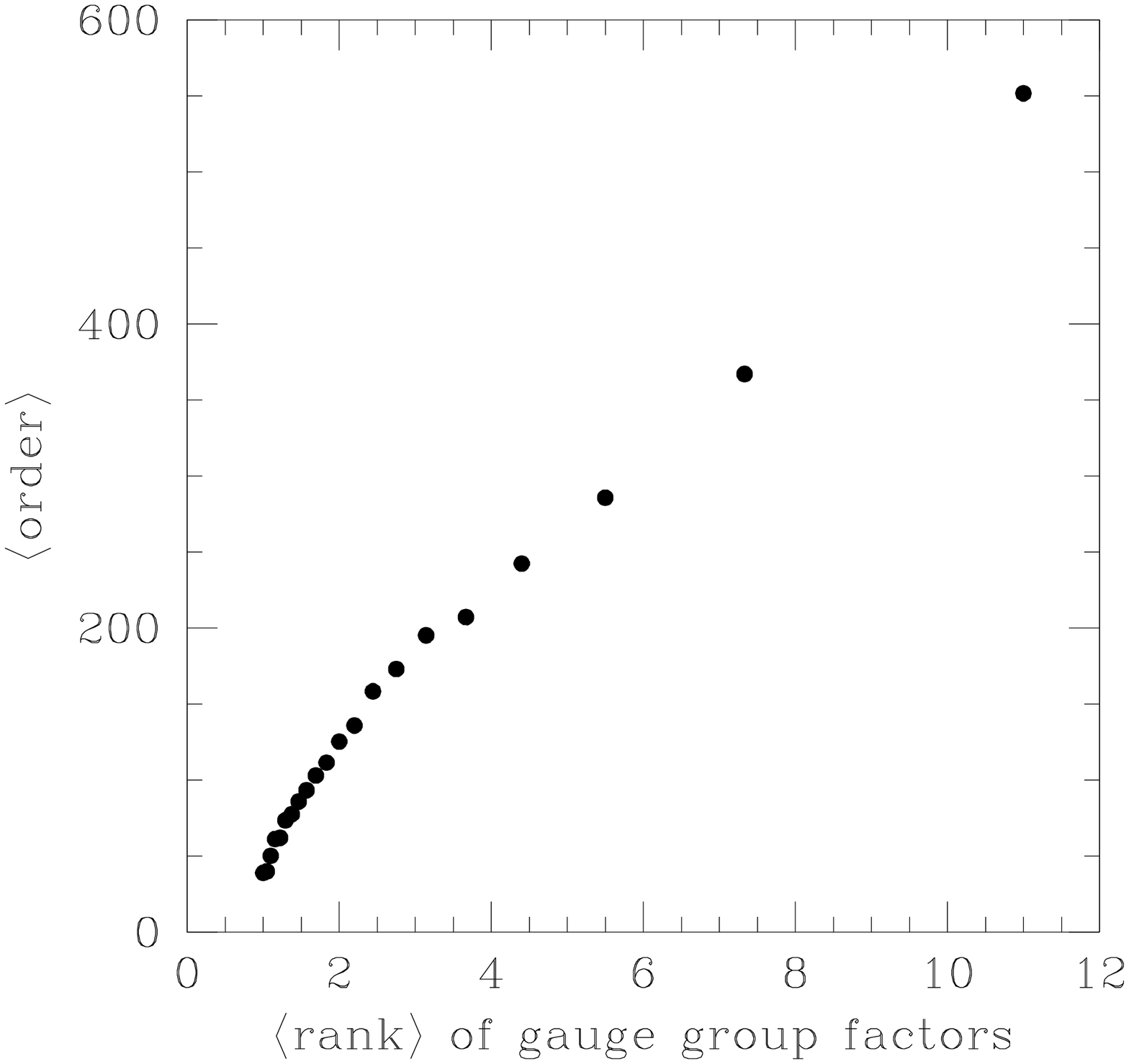}
    }
\caption{(a)  (left) 
    The orders (dimensions) of the gauge groups ({\it i.e.}\/, 
    the number of gauge bosons) averaged over all heterotic string
    models with a fixed number $f$ of gauge-group factors, plotted 
    as a function of $f$.  The monotonic shape of this curve 
    indicates that on average --- and despite the
    contributions from twisted sectors --- the net effect of 
    breaking gauge groups into smaller irreducible factors is to
    project non-Cartan gauge bosons out of the massless string spectrum.
    The $f=1$ point with order $946$ is off-scale and hence not shown.
     (b)  (right)  Same data, plotted versus the average rank
    per gauge group factor in such models, defined as $22/f$.
    On this plot, the extra point for $f=1$ with order $946$ would correspond to 
    $\langle {\rm rank}\rangle=22$ and thus continues the 
    asymptotically linear behavior.}
\label{Fig7} 
\end{figure}

In Fig.~\ref{Fig7}(a), we show the average orders of our string gauge groups 
as a function of the number $f$ of gauge-group factors, where the average
is calculated over all string models sharing a fixed $f$.
Within the set of models with fixed $f$,
the orders of the corresponding gauge groups can vary wildly.
However, the average exhibits a relatively smooth behavior,
as seen in Fig.~\ref{Fig7}(a).

%================== FIGURE ============================================
%================== END OF INSERTED FIGURE ============================

It is easy to understand the shape of this curve.
Ordinarily, we expect the order of a Lie gauge-group factor to scale 
as the square of its rank $r$:
\beq
                {\rm order}~\sim~ p \, r^2~~~~~~~~ {\rm for}~~r\gg 1~
\label{orderrank}
\eeq
where the proportionality constant for large $r$ is
\beq
       p~=~\cases{ 
             1 & for $SU(r+1)$\cr
             2 & for $SO(2r)$\cr
             \approx 2.17 & for $E_6$\cr
             \approx 2.71 & for $E_7$\cr
             \approx 3.88 & for $E_8$~.\cr}
\label{pvalues}
\eeq
(For the $E$ groups, these values of $p$ are merely the corresponding
orders divided by the corresponding ranks.)
Thus, for the total gauge group of a given string model, we expect
the total order to scale as 
\beq
           {\rm order}~\sim~ \langle p \rangle \cdot \langle r^2\rangle \cdot \langle \#~{\rm of~factors}\rangle  
\label{orderrank2}
\eeq
However, letting $f=\langle  \#~{\rm of~factors}\rangle$,
we see that for our heterotic string models,
$\langle r\rangle = 22/f$.
We thus find that 
\beq
           {\rm order}~\sim~ (22)^2\,\langle p \rangle \,{1\over  \langle \#~{\rm of~factors}\rangle}  
                      ~\sim~ 22 \,\langle p\rangle\, \langle {\rm rank}\rangle~
\label{orderrank3}
\eeq
where we are neglecting all terms which are subleading in the average rank.

In Fig.~\ref{Fig7}(b), we have plotted the same data
as in Fig.~\ref{Fig7}(a), but
as a function of $\langle {\rm rank}\rangle \equiv 22/f$. 
We see that our expectations of roughly linear behavior are indeed realized
for large values of $\langle {\rm rank}\rangle$,
with an approximate numerical value for $\langle p \rangle$ very close to $2$. 
Given Eq.~(\ref{pvalues}), this value for
$\langle p \rangle$ reflects the dominance of the `SO' groups,
with the contributions from `SU' groups tending to cancel the contributions
of the larger but rarer `E' groups.  
For smaller values of $\langle {\rm rank}\rangle$,
however, we observe a definite curvature to the plot in
Fig.~\ref{Fig7}(b).  This reflects the contributions of the subleading terms  
that we have omitted from Eq.~(\ref{orderrank}) and from our implicit identification
of $\langle r^2\rangle\sim\langle r \rangle^2$ in passing from 
Eq.~(\ref{orderrank2}) 
to Eq.~(\ref{orderrank3}). 

Fig.~\ref{Fig7}(a) demonstrates
that as we ``shatter'' our gauge groups (\eg, through orbifold twists),
the net effect is 
to project non-Cartan gauge bosons out of the string spectrum.
While this is to be expected, we emphasize that this need not always
happen in a string context.  Because of the constraints coming from
modular invariance and anomaly cancellation, performing an orbifold
projection in one sector requires that we introduce a corresponding
``twisted'' sector which can potentially give rise to new non-Cartan 
gauge bosons that replace the previous ones.
The most famous example of this phenomenon occurs in ten dimensions:
starting from the supersymmetric $SO(32)$ heterotic string 
(with $496$ gauge bosons),
we might attempt an orbifold twist to project down to the gauge
group $SO(16)\times SO(16)$ (which would have had only $240$ gauge bosons).  
However, if we also wish to preserve spacetime supersymmetry,
we are forced to introduce a twisted sector which provides
exactly $256$ extra gauge bosons to replace those that were lost, 
thereby enhancing the resulting
$SO(16)\times SO(16)$ gauge group 
back up to $E_8\times E_8$ (which again has $496$ gauge bosons).
As evident from Fig.~\ref{Fig7}(a), there are several
places on the curve at which increasing the number of 
gauge-group factors by one unit
does not appear to significantly decrease the average order; 
indeed, this
phenomenon of extra gauge bosons emerging from twisted
sectors is extremely common across our entire set of heterotic
string models.
However, we see from Fig.~\ref{Fig7}(a) that {\it on average}\/,
more gauge-group factors implies a diminished order,
as anticipated in Eq.~(\ref{orderrank3}).

\begin{figure}[ht]
\centerline{
   \epsfxsize 4.0 truein \epsfbox {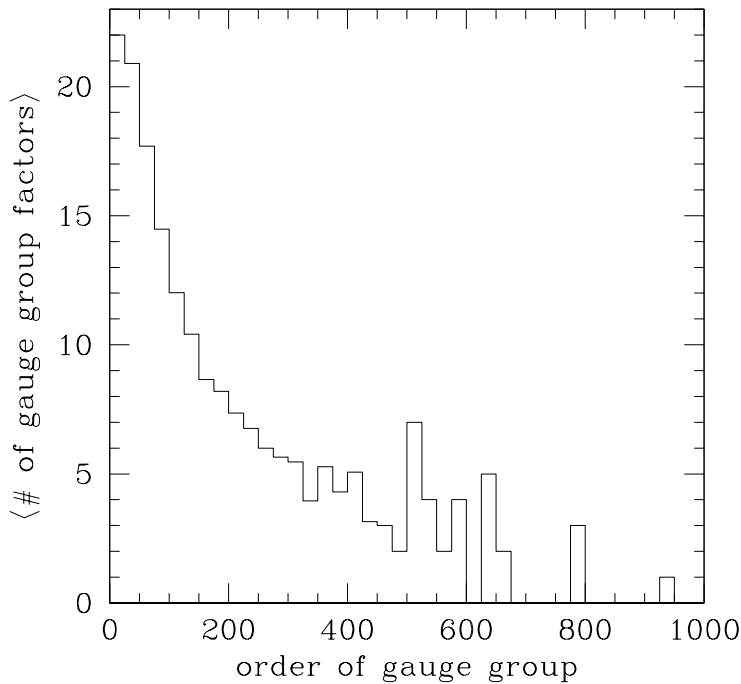}
    }
\caption{Histogram illustrating the ``inverse'' of Fig.~\protect\ref{Fig7}.  
    This plot shows the number of gauge-group factors, averaged over 
    all heterotic string models with a given gauge-group order (dimension).}  
\label{Fig8} 
\end{figure}

For later purposes,
it will also be useful for us to evaluate the ``inverse'' map
which gives the average number of gauge-group factors as function
of the total order. 
Since our models give rise to $95$ distinct orders, it is 
more effective to provide this map in the form of a histogram.
The result is shown in Fig.~\ref{Fig8}.
Note that because this ``inverse'' function is binned 
according to orders rather than averaged ranks, 
models are distributed differently across the data set
and thus Fig.~\ref{Fig8} actually
contains independent information relative to Fig.~\ref{Fig7}.
However, the shape of the resulting curve
in Fig.~\ref{Fig8} is indeed independent of 
bin {\it size}\/, as necessary for 
statistical significance.

%================== FIGURE ============================================
%================== END OF INSERTED FIGURE ============================

Once again, many of the features in Fig.~\ref{Fig8} 
can be directly traced back to the properties of
our original model ``tree''.
At the right end of the histogram, for example, we see the
contribution from the $SO(44)$ model
(for which $f=1$), with order $946$.
This is the model with the highest order.
After this, with order $786$, are two models with
gauge group $SO(40)\times SU(2)^2$, followed by 
12 models with gauge group $SO(36)\times SO(8)$
at order $658$.
This is why the 
histogram respectively shows exactly $\langle f\rangle=3$ and 
$\langle f\rangle =2$ 
at these orders.
This pattern continues as the orders descend,
except that we start having
multiple gauge groups contributing with the same
order.  For example, at order $466$,
there are twelve models with three distinct rank-22 
simply-laced gauge groups:
five models with 
gauge group $SO(24)\times SO(20)$,
two models with gauge group
$E_8\times SO(20)\times SO(8)$,
and five models with gauge group
$SO(30)\times SU(4)^2\times U(1)$.
Combined, this yields $\langle f\rangle=3$, as shown
in Fig.~\ref{Fig8} for order $466$.
Finally, at the extreme left edge of Fig.~\ref{Fig8},
we see the contributions from models with
$f=22$, which necessarily have orders $\leq 66$.

%================== FIGURE ============================================
\begin{figure}[ht]
\centerline{
   \epsfxsize 4.0 truein \epsfbox {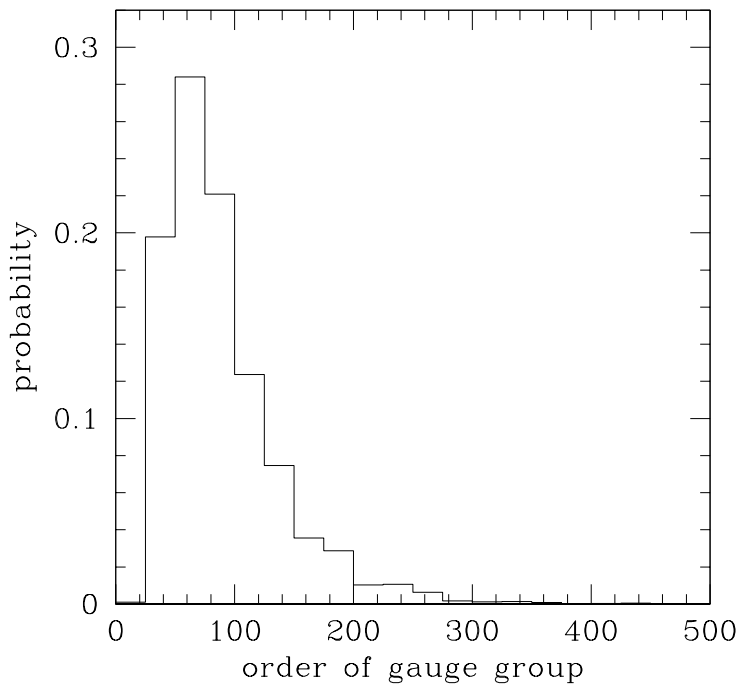}
    }
\caption{Histogram illustrating 
    the absolute probabilities of obtaining distinct four-dimensional heterotic
    string models
    as a function of the orders of their gauge groups.
    The total probability from all bins (the ``area under the curve'') is 1,
    with models having orders exceeding $300$ relatively rare.}  
\label{Fig9} 
\end{figure}
%================== END OF INSERTED FIGURE ============================

We can also plot the 
absolute probabilities of obtaining distinct four-dimensional string models
as a function of the orders of their gauge groups.
This would be the analogue of Fig.~\ref{Fig1}, but with probabilities
distributed as functions of orders rather 
than numbers of gauge-group factors.
The result is shown in Fig.~\ref{Fig9}.
As we see from Fig.~\ref{Fig9}, models having orders exceeding $200$ 
are relatively rare.  

%================== FIGURE ============================================
\begin{figure}[ht]
\centerline{
   \epsfxsize 4.0 truein \epsfbox {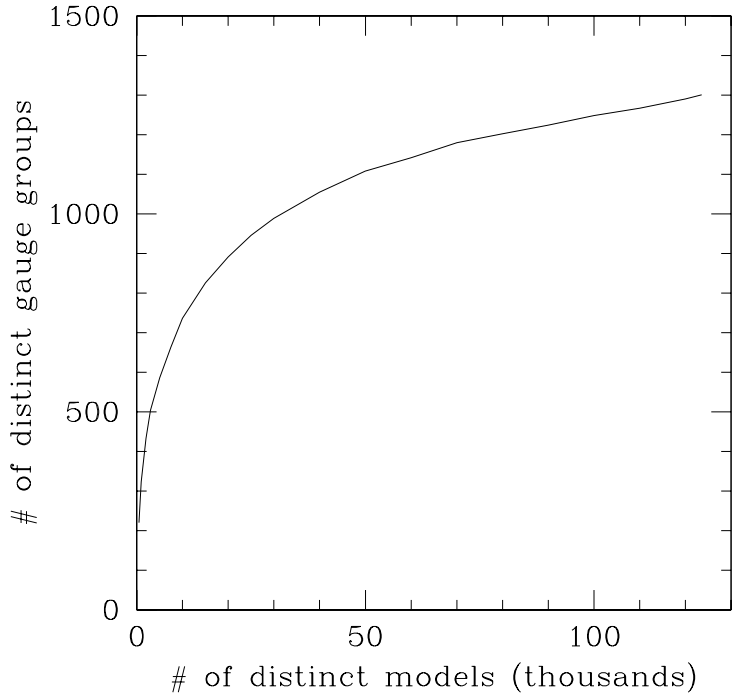}
    }
\caption{The number of gauge groups obtained as a function of the
    number of distinct heterotic string models examined.
   While the total number of models examined is insufficient
   to calculate a precise shape for this curve, one possibility is that
   this curve will eventually saturate at a 
   maximum number of possible gauge groups.} 
\label{Fig10} 
\end{figure}
%================== END OF INSERTED FIGURE ============================

As a final topic,
we have already noticed that our data set of $\gsim 10^5$ distinct models
has yielded only $1301$ different gauge groups. 
There is, therefore, a huge redundancy of gauge groups
as far as our models are concerned, with it becoming increasingly
difficult to find new heterotic string models exhibiting
gauge groups which have not been seen before.
In Fig.~\ref{Fig10}, we show the number
of gauge groups that we obtained as a function of the
number of distinct heterotic string models we examined.
Clearly, although the total number of models examined is insufficient
to calculate a precise shape for this curve, one possibility is that
this curve will eventually saturate at a 
maximum number of possible gauge groups which is relatively small. 
This illustrates the tightness of the modular-invariance
constraints in restricting the set of possible allowed gauge groups.

%=============================================================================
\section{Cosmological constants:  Statistical results} 
\setcounter{footnote}{0}

We now turn to calculations of the vacuum energy densities
(or cosmological constants)
corresponding to these heterotic string vacua.
Since the tree-level contributions to these cosmological constants
 all vanish as a result of
conformal invariance, we shall focus exclusively on their one-loop contributions.
These one-loop cosmological constants $\lambda$ may be expressed in terms of the
one-loop zero-point functions $\Lambda$, defined as
\beq
                 \Lambda ~\equiv~
     \int_{\cal F} {d^2 \tau\over ({\rm Im} \,\tau)^2}
             \, Z(\tau)~.
\label{Lambdadef}
\eeq
Here $Z(\tau)$ is the one-loop partition function of the tree-level
string spectrum of the model in question (after GSO projections have been implemented);
$\tau\equiv \tau_1+i \tau_2$ is the one-loop toroidal complex parameter,
with $\tau_i\in\IR$;
 and ${\cal F}\equiv \lbrace \tau:  |{\rm Re}\,\tau|\leq \half,
 {\rm Im}\,\tau>0, |\tau|\geq 1\rbrace$ is the fundamental domain
of the modular group.
Because the string models under consideration are non-supersymmetric
but tachyon-free, $\Lambda$ is guaranteed to be finite and in principle
non-zero.
The corresponding one-loop vacuum energy density (cosmological constant) $\lambda$ 
is then defined as $\lambda\equiv -\half \calM^4\Lambda$,
where ${\cal M}\equiv M_{\rm string}/(2\pi)$ is the reduced string scale.
Although $\Lambda$ and $\lambda$ have opposite signs, with $\Lambda$ being dimensionless,
we shall occasionally refer to $\Lambda$ as the cosmological constant
in cases where the overall sign of $\Lambda$ is not important.
             
Of course, just as with the ten-dimensional $SO(16)\times SO(16)$ string,
the presence of a non-zero $\Lambda$ indicates
that these string models are unstable beyond tree level.
Thus, as discussed in the Introduction,
these vacua are generically not situated at local minima within
our ``landscape'', and can be expected to become unstable as the string coupling
is turned on.
Nevertheless, we shall investigate the values of these amplitudes
for a number of reasons.  First, the amplitude defined in Eq.~(\ref{Lambdadef})
represents possibly the simplest one-loop amplitude that can be calculated for
such models;  as such, it represents a generic quantity whose behavior
might hold lessons for more complicated amplitudes.  For example, more
general $n$-point amplitudes are related to this amplitude through 
differentiations;  a well-known example of this is provided by
string threshold corrections~\cite{Kaplunovsky}, which are described by a similar 
modular integration with a slightly altered (differentiated) integrand.
Second, by evaluating and analyzing such string-theoretic expressions,
we can gain insight into the extent to which results from effective
supergravity calculations might hold in a full string context.
Indeed, we shall be able to judge exactly how significant a role
the massive string states might play in altering our field-theoretic
expectations based on considerations of only the massless states.
Third, when we eventually combine this information with our gauge-group
statistics in Sect.~6, we shall be
able to determine the extent to which gauge groups and the magnitudes of
such scattering amplitudes might be correlated in string theory.
But finally and most importantly, we shall investigate this amplitude because it 
relates directly back to fundamental questions of supersymmetry breaking
and vacuum stability.  Indeed, if we can find models for which $\Lambda$
is nearly zero, we will have found good approximations to stable vacua with
broken supersymmetry.  We shall also discover other interesting
features, such as
unexpected one-loop degeneracies in the space of non-supersymmetric
string models.  All of this may represent important information concerning
the properties of the landscape of {\it non}\/-supersymmetric strings.   

In general, the one-loop partition function $Z(\tau)$ which appears 
in Eq.~(\ref{Lambdadef})
is defined as the trace over the full Fock space of string states:
\beq
      Z(\tau)~\equiv~ {\rm Tr}\,
          (-1)^F\, \overline{q}^{H_R}\, q^{H_L}~.
\label{Zdef}
\eeq
Here $F$ is the spacetime fermion number,
$(H_R,H_L)$ are the right- and left-moving worldsheet
Hamiltonians, and $q\equiv \exp(2\pi i\tau)$.
Thus spacetime bosonic states contribute positively to $Z(\tau)$, while
fermionic states contribute negatively.
In general, the trace in Eq.~(\ref{Zdef}) may be evaluated in terms of the
characters $\chi_i$ and $\overline{\chi}_j$ of the left- and right-moving
conformal field theories on the string worldsheet,
\beq
      Z(\tau) ~=~ \tau_2^{-1}\, \sum_{i,j}\,
          \overline{\chi}_j (\overline{\tau}) \, N_{j i} \, \chi_i(\tau)~,
\label{Zchi}
\eeq
where the coefficients $N_{ij}$ describe the manner in which the left- and right-moving
CFT's are stitched together and thereby reflect the particular GSO
projections inherent in the given string model.
The $\tau_2^{-1}$ prefactor in Eq.~(\ref{Zchi}) represents the contribution to the 
trace in Eq.~(\ref{Zdef}) from the continuous spectrum of 
states corresponding to the uncompactified spacetime dimensions.

Since the partition function $Z(\tau)$ represents a trace over the string Fock space
as in Eq.~(\ref{Zdef}), it encodes the information about the net degeneracies
of string states at each mass level in the theory.  Specifically, 
expanding $Z(\tau)$ as a double-power series in $(q,\qbar)$,  
we obtain an expression of the form
\beq
 Z(\tau) ~=~  \tau_2^{-1} \,
       \sum_{mn}\, b_{m n} \,\qbar^{m} \,q^{n}~
\label{bare}
\eeq
where $(m,n)$ represent the possible eigenvalues of the right- and left-moving worldsheet
Hamiltonians $(H_R,H_L)$, and where $b_{mn}$ represents the net number of bosonic
minus fermionic states (spacetime degrees of freedom)
which actually have those eigenvalues and satisfy the GSO constraints.
Modular invariance requires that $m-n\in \IZ$ for all $b_{mn}\not=0$;
a state is said to be ``on-shell'' or ``level-matched'' if $m=n$, and corresponds
to a spacetime state with mass ${\cal M}_n =  2\sqrt{n}M_{\rm string}$.
Thus, states for which $m+n\geq 0 $ are massive and/or massless,
while states with $m+n <0$ are tachyonic.
By contrast, states with $m-n\in\IZ\not=0$ are considered to be ``off-shell'':
they contribute to one-loop amplitudes such as $\Lambda$ with a dependence on $|m-n|$,
but do not correspond to physical states in spacetime.

Substituting Eq.~(\ref{bare}) into Eq.~(\ref{Lambdadef}), we have
\beqn
   \Lambda &=& \sum_{m,n} \, b_{mn} \,\int_{\cal F} 
      {d^2\tau\over \tau_2^2}\,\tau_2^{-1} \,\qbar^m q^n\nonumber\\
         &=& \sum_{m,n} \, b_{mn} \,\int_{\cal F} 
      {d^2\tau\over \tau_2^3}\, \exp\left\lbrack -2\pi(m+n) \tau_2 \right\rbrack \,
               \cos\lbrack 2\pi(m-n) \tau_1 \rbrack ~.
\label{format}
\eeqn
(Note that since ${\calF}$ is symmetric under $\tau_1\to-\tau_1$,
only the cosine term survives in passing to the second line.)
Thus, we see that 
the contributions from different regions of $\cal F$ 
will depend critically on the values of $m$ and $n$.
The contribution to $\Lambda$ from 
the $\tau_2<1$ region is always finite and non-zero
for all values of $m$ and $n$.  However, given that $m-n\in\IZ$,
we see that the contribution from the
$\tau_2>1$ region is zero if $m\not= n$,
non-zero and finite if $m=n>0$,
and infinite if $m=n<0$.

For heterotic strings, our worldsheet vacuum energies are bounded by
$m\geq -1/2$ and $n\geq -1$.  
Moreover, all of the models we are considering in this paper
are tachyon-free, with $b_{mn}=0$ for all $m=n<0$.
As a result, in such models, each contribution to $\Lambda$ 
is finite and non-zero.
By contrast, note that a supersymmetric string model 
would have $b_{mn}=0$ for all $(m,n)$, leading to
$\Lambda=0$.  
Non-zero values of $\Lambda$ are thus a signature for 
broken spacetime supersymmetry.

For simplicity, we can change variables from $(m,n)$ to $(s\equiv m+n,d\equiv |n-m|)$.
Indeed, since $\Lambda\in \IR$, we can always take $d\geq 0$ and define
\beq
             a_{sd} ~\equiv~ b_{(s-d)/2 , (s+d)/2} ~+~ b_{(s+d)/2 , (s-d)/2} ~. 
\label{adef}
\eeq
We can thus rewrite Eq.~(\ref{format}) in the form
\beq
        \Lambda ~=~ \sum_{s,d}  \,a_{sd}\, I_{sd}~~~~~~~
     {\rm where}~~~~
     I_{sd}~\equiv~ 
         \int_{\cal F} 
      {d^2\tau\over \tau_2^3}\, \exp( -2\pi s \tau_2 ) \,
               \cos( 2\pi d \tau_1 ) ~.
\label{Idef}
\eeq
For heterotic strings, this summation is over all $s\geq -1$
and $|d|=0,1,...,[s]+2$ where $[x]$ signifies the greatest integer $\leq x$.  
Of course, only those states with $d=0$ are on-shell, 
but $\Lambda$ receives contributions from off-shell states 
with non-zero $d$ as well.
In general, the values of $s$ which appear in a given string  
model depend on a number of factors, most notably the conformal dimensions of
the worldsheet fields (which in turn depend on various internal compactification radii);  
however, for the class of models considered in this paper,
we have $s\in\IZ/2$.
The numerical values of $I_{sd}$ 
    from the lowest-lying string states with $s\leq 1.5$
are listed in Table~\ref{integraltable}.

%================================
\begin{table}[ht]
\centerline{
   \begin{tabular}{||r|r|r||r|r|r||}
   \hline
   \hline
    $s$~  &  $d$ &  $I_{sd}$~~~~~ &  $s$~  &  $d$ &  $I_{sd}$~~~~~ \\
   \hline
   \hline
 $-1.0$  &$  1 $&$  {\tt    -12.192319 }  $
&  $ 1.0  $&$ 0  $&$ {\tt    0.000330   }$\\
$ -0.5  $&$ 1  $&$ {\tt   -0.617138  } $
&  $ 1.0  $&$ 1  $&$ {\tt   -0.000085  } $\\
$ 0.0  $&$ 0  $&$  {\tt  0.549306   }$
&  $ 1.0  $&$ 2  $&$ {\tt    0.000035   }$\\
$ 0.0  $&$ 1  $&$  {\tt  -0.031524  } $
&  $ 1.0  $&$ 3  $&$ {\tt   -0.000018   }$\\
$ 0.0  $&$ 2  $&$  {\tt   0.009896   }$
   &  $ 1.5  $&$ 0  $&$    {\tt  0.000013}  $\\
$ 0.5  $&$ 0  $&$ {\tt    0.009997   }$
   &  $ 1.5  $&$ 1  $&$    {\tt -0.000004}  $\\
$ 0.5  $&$ 1  $&$ {\tt   -0.001626   }$
   &  $ 1.5  $&$ 2  $&$     {\tt 0.000002}  $\\
$ 0.5  $&$ 2  $&$ {\tt    0.000587   }$
   &  $ 1.5  $&$ 3  $&$    {\tt -0.000001}  $\\
\hline
\hline
\end{tabular}
 }
\caption{Contributions to the one-loop string vacuum amplitude $I_{sd}$ 
         defined in Eq.~(\protect\ref{Idef})
    from the lowest-lying string states with $s\leq 1.5$.}
\label{integraltable}
\end{table}
%======================================

We immediately observe several important features.
First, although the
coefficients $a_{sd}$ tend to experience an exponential (Hagedorn) growth
$a_{sd}\sim e^{\sqrt{s}}$, 
we see that the values of $I_{sd}$ generally decrease exponentially with $s$,
\ie, $I_{sd}\sim e^{-s}$.
This is then sufficient to overcome the Hagedorn growth
in the numbers of states, 
leading to a convergent value of $\Lambda$.
However, because of the balancing between these two effects,
it is generally necessary
to determine the complete particle spectrum of each vacuum state up
to a relatively high (\eg, fifth or sixth) mass level, with $s\approx 5$ or $6$,
in order to accurately 
calculate the full cosmological constant $\Lambda$ for each string model. 
This has been done for all results quoted below.

Second, we observe that the contributions $I_{s,0}$ from all on-shell
states are positive, \ie, $I_{s,0}>0$ for all $s$.
Thus, as anticipated on field-theoretic grounds, 
on-shell bosonic states contribute positively to $\Lambda$,
while on-shell fermionic states contribute negatively.
(Note that this is reversed for the actual cosmological constant $\lambda$,
which differs from $\Lambda$ by an overall sign.)
However, we more generally observe that $I_{sd}>0$ $(<0)$ for even (odd) $d$.
Thus, the first group of off-shell states with $d=1$ tend to 
contribute oppositely to the corresponding on-shell states with $d=0$,
with this behavior partially compensated by the second group of off-shell
states with $d=2$, and so forth.
This, along with the fact~\cite{missusy} that  
the coefficients $a_{sd}$ necessarily exhibit a regular oscillation in their 
overall signs as a function of $s$,
also aids in the convergence properties of $\Lambda$.

Finally, we observe that by far the single largest contributions
to the vacuum amplitude $\Lambda$ actually come from states
with $(s,d)=(-1,1)$, or $(m,n)=(0,-1)$.
These states are off-shell tachyons.  
At first glance one might suspect that it would be possible to project
such states out of the spectrum (just as one does with the on-shell tachyons),
but it turns out that this is impossible:  {\it all
heterotic string models necessarily have off-shell tachyons with $(m,n)=(0,-1)$}.
These are ``proto-graviton'' states emerging 
in the Neveu-Schwarz sector:
\beq
        \hbox{proto-graviton:}~~~~~~~~~~~~
              \tilde b_{-1/2}^\mu |0\rangle_R ~\otimes~ ~|0\rangle_L~
\label{protograviton}
\eeq
where $\tilde b_{-1/2}^\mu$ represents the excitation
of the right-moving worldsheet Neveu-Schwarz
fermion $\tilde \psi^\mu$.
Since the Neveu-Schwarz heterotic string ground state
has vacuum energies $(H_R,H_L)=(-1/2,-1)$,
we see that the ``proto-graviton'' state in Eq.~(\ref{protograviton}) has
worldsheet energies 
$(H_R,H_L)=(m,n)=(0,-1)$;  indeed, this is nothing but the graviton state 
without its left-moving oscillator excitation.
However, note that regardless of the particular GSO projections, the graviton state
must always appear in the string spectrum.  Since GSO projections are insensitive
to the oscillator excitations, this implies that the proto-graviton must also
necessarily appear in the string spectrum.

By itself, of course, this argument does not prove that we must necessarily have
$a_{-1,1}\not=0$.
However, it is easy to see that the only state which could possibly cancel
the contribution from the (bosonic) proto-graviton in the Neveu-Schwarz sector
is a (fermionic) proto-gravitino in the Ramond sector: 
\beq
        \hbox{proto-gravitino:}~~~~~~~~~~~~
         \lbrace \tilde b_{0}\rbrace^\alpha  |0\rangle_R ~\otimes~ ~ |0\rangle_L~.
\label{protogravitino}
\eeq
Here  $\lbrace \tilde b_{0}\rbrace^\alpha$ schematically indicates the Ramond zero-mode
combinations which collectively give rise to the spacetime Lorentz spinor index $\alpha$.
However, if the gravitino itself is projected out of the spectrum (producing
the non-supersymmetric string model), then that same GSO projection must simultaneously 
project out the proto-gravitino state.  In other words, while all heterotic strings
contain a proto-graviton state, only those with spacetime supersymmetry will
contain a compensating proto-gravitino state. 
Thus, all non-supersymmetric heterotic string
models must have an uncancelled $a_{-1,1}> 0$.

This fact has important implications for the overall sign of $\Lambda$. 
 {\it A priori}\/, we might have expected that whether $\Lambda$ is positive
or negative would be decided primarily by the net numbers of massless, on-shell
bosonic and fermionic states in the string model --- \ie,  by the sign
of $a_{00}$.
However, we now see that because $a_{-1,1}>0$ and $I_{-1,1}<0$,  
there is already a built-in bias towards negative values of $\Lambda$ for heterotic strings.
Indeed, each off-shell tachyon can be viewed as providing an initial negative 
offset for $\Lambda$ of magnitude $I_{-1,1}\approx -12.19$,
so that there is an approximate critical value for $a_{00}$, given by 
\beq
        a_{00} \biggl |_{\rm critical} ~\approx~-{I_{-1,1}\over I_{0,0}}~\approx~ 22.2~,
\eeq
which is needed just to balance each off-shell tachyon and
obtain a vanishing $\Lambda$.
Of course, even this estimate is low, as it ignores
the contributions of off-shell {\it massless}\/ states which, like the off-shell
tachyon, again provide negative contributions for bosons and 
positive contributions for fermions.

The lesson from this discussion, then, is clear:
\begin{itemize}
\item  {\bf In string theory, contributions from the infinite towers 
       of string states, both on-shell and off-shell,
       are critical for determining not only the magnitude but also the overall sign
       of the one-loop cosmological constant.}  Examination of the massless string 
       spectrum (\eg, through a low-energy effective 
       field-theory analysis) is insufficient.     
\end{itemize}

\begin{figure}[ht]
\centerline{
   \epsfxsize 4.0 truein \epsfbox {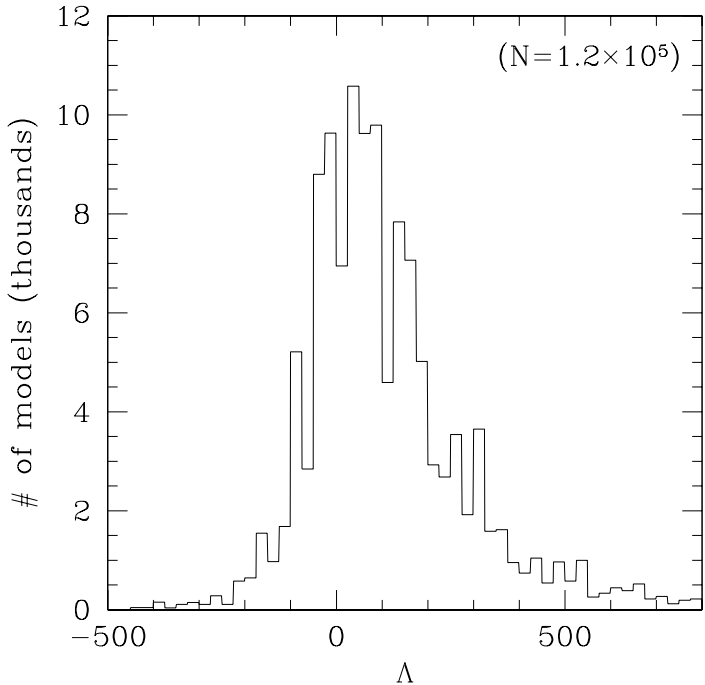}
    }
\caption{Histogram showing calculated values of the one-loop amplitude $\Lambda$
   defined in Eq.~(\protect\ref{Lambdadef})
   across our sample of $ N \gsim 10^5$ tachyon-free perturbative 
    heterotic string vacua with string-scale
   supersymmetry breaking.  Both positive and negative values of $\Lambda$ are obtained,
    with over $73\%$ of models having positive values.
    The smallest $|\Lambda|$-value found is $\Lambda\approx 0.0187$, which appears
        for eight distinct models.
   (This figure adapted from Ref.~\cite{thesis}.)}
\label{Fig11} 
\end{figure}

We now turn to the values of $\Lambda$ that are obtained across our
set of heterotic string models.
For each model, we evaluated the on- and off-shell degeneracies $a_{sd}$ 
to at least the fifth level ($s\approx 5$), 
and then tabulated the corresponding value of $\Lambda$ as in Eq.~(\ref{Idef}).
A histogram illustrating the resulting distribution
of cosmological constants is shown in Fig.~\ref{Fig11}.

%================== FIGURE ============================================
%================== END OF INSERTED FIGURE ============================

As can be seen, both positive and negative values of $\Lambda$ are
obtained, and indeed the bulk of these models have
values of $\Lambda$ centered near zero.  
In fact, despite the contributions of the off-shell tachyons,
it is evident from Fig.~\ref{Fig11} that 
there is a preference
for positive values of $\Lambda$, with just over $73\%$ of our
models having $\Lambda>0$. 
However, we obtained no model with $\Lambda=0$; 
indeed, the closest value we obtained for any model
is $\Lambda \approx 0.0187$, which appeared for
eight distinct models.

Given that we examined more than $10^5$ distinct heterotic string models,
it is natural to wonder why no smaller values of $\Lambda$ were found.
This question becomes all the more pressing in light
of recent expectations~\cite{BP} that the set of 
cosmological constant values should be approximately randomly distributed,
with relative spacings and a smallest overall value that 
diminish as additional models are considered.

It is easy to see why this does not happen, however:
 {\it just as for gauge groups, it turns out that there is a tremendous
degeneracy in the space of string models,
with many distinct heterotic string models sharing exactly the same value of
$\Lambda$.}
Again, we stress that these are distinct models with distinct
gauge groups and particle content.  Nevertheless, such models 
may give rise to exactly the same one-loop cosmological constant!

The primary means by which two models can have the same 
cosmological constant is by having 
the same set of state degeneracies $\lbrace a_{sd}\rbrace$.
This can happen in a number of ways.
Recall that these degeneracies represent only the net numbers
of bosonic minus fermionic degrees of freedom;  thus
it is possible for two models to have different numbers of
bosons and fermions separately, but to have a common difference 
between these numbers.
Secondly, it is possible for two models to have partition functions
$Z_1(\tau)$ and $Z_2(\tau)$ 
which differ by a purely imaginary function of $\tau$;
in this case, such models they will once again share a common set 
of state degeneracies $\lbrace a_{sd}\rbrace$
although their values of $b_{mn}$ will differ. 
Finally, it is possible for $Z_1(\tau)$ and $Z_2(\tau)$ 
to differ when expressed in terms of conformal field-theory characters
(or in terms of Jacobi theta functions $\vartheta_i$), but with this difference
proportional to the vanishing Jacobi factor
\beq
             J ~\equiv~  {1\over \eta^4} \left(\vartheta_3^4 - \vartheta_4^4 - \vartheta_2^4\right) ~=~0~
\eeq
where $\eta$ and $\vartheta_i$ respectively
represent the Dedekind eta-function and 
Jacobi theta-functions, defined as
\beqn
    \eta(q)  ~\equiv&  q^{1/24}~ \displaystyle\prod_{n=1}^\infty ~(1-q^n)&=~
                \sum_{n=-\infty}^\infty ~(-1)^n\, q^{3(n-1/6)^2/2}\nonumber\\
    \vartheta_2(q)~\equiv&  2 q^{1/8} \displaystyle\prod_{n=1}^\infty (1+q^n)^2 (1-q^n)&=~
                 2\sum_{n=0}^\infty q^{(n+1/2)^2/2} \nonumber\\
    \vartheta_3(q)~\equiv&  \displaystyle\prod_{n=1}^\infty (1+q^{n-1/2})^2 (1-q^n) &=~
                1+ 2\sum_{n=1}^\infty q^{n^2/2} \nonumber\\
    \vartheta_4(q) ~\equiv& \displaystyle\prod_{n=1}^\infty (1-q^{n-1/2})^2 (1-q^n) &=~
                1+ 2\sum_{n=1}^\infty (-1)^n q^{n^2/2} ~.
\label{thetadefs}
\eeqn
Such partition functions $Z_1(\tau)$ and $Z_2(\tau)$ differing by $J$ will then
have identical $(q,\qbar)$ power-series expansions, once again
leading to identical degeneracies $\lbrace a_{sd}\rbrace$ and identical values of $\Lambda$.

\begin{figure}[ht]
\centerline{
   \epsfxsize 3.1 truein \epsfbox {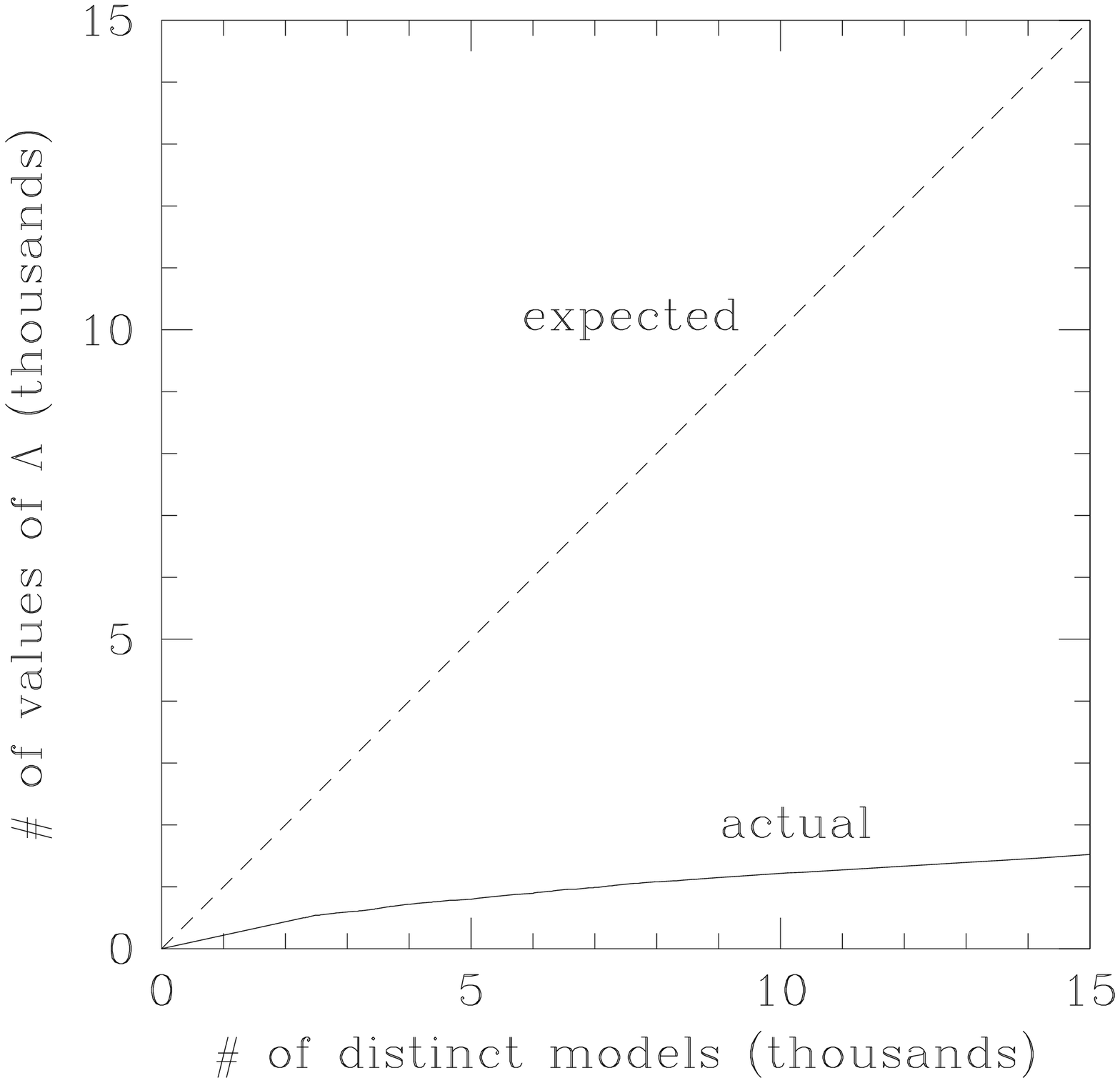}
   \epsfxsize 3.1 truein \epsfbox {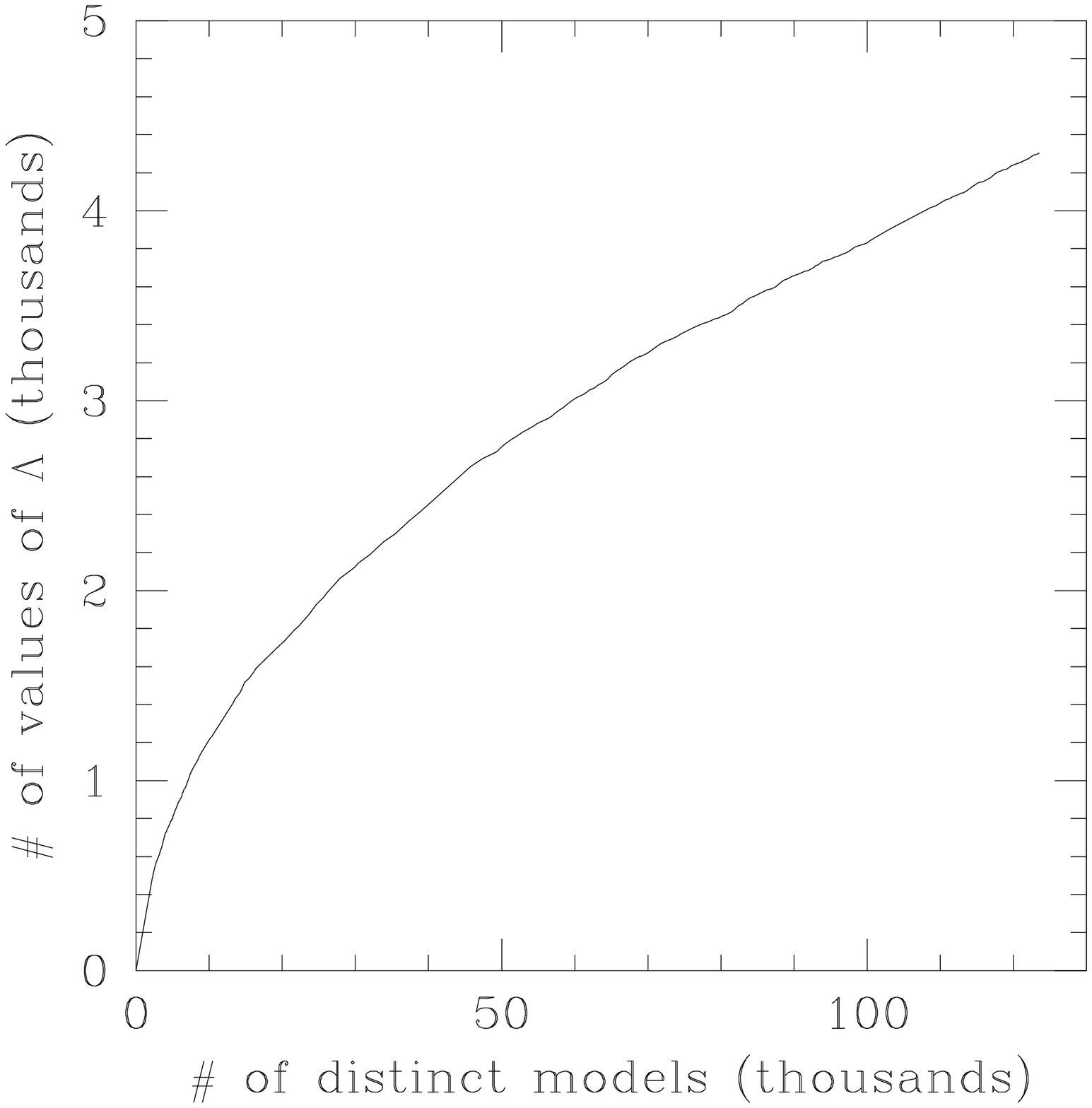}
    }
\caption{
    Unexpected degeneracies in the space of non-supersymmetric
    string vacua.  As evident from these
    figures, there is a tremendous degeneracy according to which many
    distinct non-supersymmetric heterotic string models with different gauge
    groups and particle contents nevertheless exhibit exactly the {\it same}\/ 
    numbers of bosonic and fermionic states
    and therefore have identical one-loop
    cosmological constants. 
   (a) (left)  Expected versus actual numbers of cosmological
    constants obtained for the first fifteen thousand models.    
   (b) (right)  Continuation of this plot as more models are examined.
   While the number of models examined is insufficient
   to calculate a precise shape for this curve, one possibility is that
   this curve will eventually saturate at a 
   maximum number of allowed cosmological constants, 
   as discussed in the text.   (Right figure adapted from Ref.~\cite{thesis}.)}
\label{Fig12} 
\end{figure}

In Fig.~\ref{Fig12}, we have plotted the actual numbers of 
distinct degeneracy sets $\lbrace a_{sd}\rbrace$ found
(and therefore the number of distinct cosmological constants $\Lambda$ obtained)
as a function of the number of distinct models examined.
It is clear from these results that this 
cosmological-constant redundancy is quite severe, with only
$4303$ different values of $\Lambda$ emerging from 
over $1.2\times 10^5$ models!
This represents a redundancy factor of approximately $28$,
and it is clear from Fig.~\ref{Fig12} that this factor 
tends to grow larger and larger as
the number of examined models increases. 
Thus, we see that 
\begin{itemize}
\item  {\bf More string models does not necessarily imply more values of $\Lambda$.}
  Indeed, many different string models with entirely different spacetime phenomenologies
  (different gauge groups, matter representations, hidden sectors, and
   so forth) exhibit identical values of $\Lambda$.
\end{itemize}

%================== FIGURE ============================================
%================== END OF INSERTED FIGURE ============================

In fact, the shape of
the curve in Fig.~\ref{Fig12}(b) might lead us to conclude that
there
may be a finite and relatively small number of self-consistent
matrices $\lbrace a_{sd}\rbrace$ which our models may be capable of
exhibiting.  If this were the case, then we would expect
the number of such matrices $\lbrace a_{sd}\rbrace$ already seen, $\Sigma$, to 
have a dependence on the total number of models examined, $t$, of the form
\beq
    \Sigma(t) ~=~ N_0\,\left( 1~-~ e^{-t/t_0}\right)~,
\label{namenlosetwo}
\eeq
where $N_0$ is this total number of matrices $\lbrace a_{sd}\rbrace$ and $t_0$, the ``time constant'',
is a parameter characterizing the scale of the redundancy.  Fitting the
curve in Fig.~\ref{Fig12}(b) to Eq.~(\ref{namenlosetwo}), we find that values of
$N_0\sim 5500$ and $t_0 \sim 70\,000$ seem to be indicated.  (One cannot
be more precise, since we have clearly not examined
a sufficient number of models to observe saturation.)  Of course,
this sort of analysis assumes that our models uniformly span the space
of allowed $\lbrace a_{sd}\rbrace $ matrices (and also that our model set 
uniformly spans the space of models).

As if this redundancy were not enough, it turns out 
that there is a further redundancy beyond that
illustrated in Fig.~\ref{Fig12}.
In Fig.~\ref{Fig12}, note that we are actually plotting the numbers
of distinct sets of degeneracy matrices $\lbrace a_{sd}\rbrace$,
since identical matrices necessarily imply identical values of $\Lambda$.
However, it turns out that 
there are situations in which even {\it different}\/ values of 
$\lbrace a_{sd}\rbrace$ can lead to identical values of $\Lambda$!~
Clearly, such an occurrence 
would be highly non-trivial, requiring that two sets 
of integers $\lbrace a^{(1)}_{sd}\rbrace$
and
$\lbrace a^{(2)}_{sd}\rbrace$
differ by non-zero integer coefficients 
$c_{sd}\equiv a^{(1)}_{sd}-a^{(2)}_{sd}$
for which $\sum_{sd} c_{sd} I_{sd} =0$.

At first glance, 
given the values of $I_{sd}$ tabulated in Table~\ref{integraltable},
it may seem that no such integer
coefficients $c_{sd}$ could possibly exist.
Remarkably, however, it was shown in Ref.~\cite{PRL}
that there exists a function
\beqn
      Q &\equiv &  {1\over 128 \,\tau_2}\, {1\over \overline{\eta}^{12}\, \eta^{24}} \,
       \sum_{\scriptstyle i,j,k=2  \atop \scriptstyle i\not= j\not= k}^4
      \,  |\thetai|^4 \,  \Biggl\lbrace  \,
 \thetai^4 \thetaj^4 \thetak^4
      \,\biggl\lbrack \,2\, |\thetaj \thetak|^8  -
     \thetaj^8 \thetakbar^{8} - \thetajbar^{8} \thetak^8  \biggr\rbrack
                    \nonumber\\
        &&~~~~~~~~~~~~~~ +~ \thetai^{12} \,\biggl[
      \, 4 \,\thetai^8  \thetajbar^{4} \thetakbar^{4} +
               (-1)^i~13 \,|\thetaj  \thetak|^8 \biggr] \, \Biggr\rbrace~
\label{Qdef}
\eeqn
which, although non-zero, has the property that
\beq
     \int_{\cal F} {d^2 \tau\over ({\rm Im} \,\tau)^2} \,Q~=~0~
\label{ident}
\eeq
as the result of an Atkin-Lehner symmetry~\cite{Moore}.
Power-expanding the expression $Q$ in Eq.~(\ref{Qdef}) using 
Eq.~(\ref{thetadefs}) then yields a set of
integer coefficients $c_{sd}$ for which $\sum_{sd} c_{sd}I_{sd}=0$
as a consequence of Eq.~(\ref{ident}). 
Thus, even though neither of the partition functions $Z_1(\tau)$ and $Z_2(\tau)$
of two randomly chosen models exhibits its own Atkin-Lehner symmetry
(consistent with an Atkin-Lehner ``no-go'' theorem~\cite{Balog}),
it is possible that their {\it difference}\/ might nevertheless exhibit such a symmetry. 
If so, then such models are ``twins'',
once again sharing the same value of $\Lambda$.

As originally reported in Ref.~\cite{PRL}, this additional 
type of twinning redundancy turns out to be pervasive throughout
the space of heterotic string models, leading to a further $\sim 15\%$ reduction
in the number of distinct values of $\Lambda$.
Indeed, we find not only twins, but also ``triplets'' and ``quadruplets'' ---
groups of models whose
degeneracies $a^{(i)}_{sd}$ differ sequentially 
through repeated additions of such coefficients $c_{sd}$.
Indeed, we find that our $4303$ different sets $\lbrace a_{sd}\rbrace$
which emerge from our $\sim 10^5$ models
can be categorized as
$3111$ ``singlets'',
$500$ groupings of ``twins'',
$60$ groupings of ``triplets'',
      and
$3$ groupings of ``quadruplets''.
[Note that indeed $3111+2(500)+3(60)+4(3)=4303.$]
Thus, the number of distinct cosmological constants emerging
from our $\sim 10^5$ models is not actually $4303$, but only
$3111+500+60+3= 3674$, which represents an additional $14.6\%$ reduction. 

At first glance, 
since there are relatively few groupings of twins,
triplets, and quadruplets,  
it might seem that this additional reduction 
is not overly severe. 
However, this fails to take into
account the fact that our previous redundancy may be (and in fact is)
statistically clustered around these sets.
Indeed, across our entire set of $10^5$ distinct string models, 
we find that
\begin{itemize}
\item  $30.7\%$ are ``singlets'';
   $48.2\%$ are members of a ``twin'' grouping;
   $21.0\%$ are members of a ``triplet'' grouping; and
   $0.1\%$ are members of a ``quadruplet'' grouping.
\end{itemize}
Thus, we see that this twinning phenomenon is responsible for a massive 
degeneracy across the space of non-supersymmetric heterotic string vacua.\footnote{
    Indeed, reviving the ``raindrop'' analogy introduced
    in the Introduction, we see that the rain falls mainly on the plane.}

Note that this degeneracy may be of considerable relevance for various solutions
of the cosmological-constant problem.  
For example, one proposal in Ref.~\cite{kane} imagines
a large set of degenerate vacua which combine to form a ``band'' of states
whose lowest state has a significantly suppressed vacuum energy.
However, the primary ingredient in this formulation is the existence of a large
set of degenerate, non-supersymmetric vacua.  This is not unlike what
we are seeing in this framework.  
Of course, there still remains the outstanding question concerning how
transitions between these vacua can arise, as would be needed in order
to generate the required ``band'' structure.

In all cases, modular invariance is the primary factor that underlies 
these degeneracies.
Despite the vast set of possible heterotic string spectra, there are only so
many modular-invariant functions $Z(\tau)$ which can serve as the partition
functions for self-consistent string models.  It is this modular-invariance constraint
which ultimately limits the possibilities for the degeneracy coefficients
$\lbrace a_{sd}\rbrace$, and likewise it is modular invariance (along with Atkin-Lehner
symmetries) which leads to identities such as Eq.~(\ref{ident}) which
only further enhance this tendency towards degeneracy. 
 
Needless to say, our analysis in this section has been limited to one-loop
cosmological constants.  It is therefore natural to ask whether such degeneracies
might be expected to persist at higher orders.
Of course, although modular invariance is a one-loop phenomenon,
there exist multi-loop generalizations of modular invariance;
likewise it has been speculated that there also exist multi-loop extensions
of Atkin-Lehner symmetry~\cite{Moore}.
Indeed, modular invariance is nothing but the reflection of the underlying
finiteness of the string amplitudes, and we expect this  
finiteness to persist to any order in the string perturbation expansion.
It is therefore possible that degeneracies such as these will persist to higher
orders as well.

In any case, this analysis dramatically illustrates that many of our na\"\i ve
expectations concerning the values of string amplitudes such as the cosmological
constant and the distributions of these values may turn out to be 
grossly inaccurate
when confronted with the results of explicit string calculations.
The fact that string theory not only provides infinite towers of states
but then tightly constrains the properties of these towers
through modular symmetries --- even when spacetime supersymmetry is broken --- 
clearly transcends our na\"\i ve field-theoretic expectations.
It is therefore natural to expect that such properties will continue to
play a significant
role in future statistical studies of the heterotic landscape,
even if/when stable non-supersymmetric heterotic string models 
are eventually constructed.

%=============================================================================
\section{Gauge groups and cosmological constants:\\  Statistical correlations}
\setcounter{footnote}{0}

We now turn to {\it correlations}\/ between our gauge groups
and cosmological constants.  To what extent does the gauge group
of a heterotic string model influence the magnitude of its 
cosmological constant, and vice versa?  Note that in field theory, these quantities are
completely unrelated --- the gauge group is intrinsically
non-gravitational, whereas the cosmological constant is of primarily
gravitational relevance.  However, in string theory, we can expect
correlations to occur.

To begin the discussion, let us again construct a ``tree'' 
according to which our string models are grouped according
to their gauge groups.
While in Sect.~4 we grouped our models on ``branches'' according to their numbers
of gauge-group factors,
in this section, for pedagogical purposes,
we shall instead
group our models into ``clusters'' according to their orders.
\begin{itemize}
\item {\sl order=946:}\/  As we know, there is only one distinct string model
    with this order, with gauge group $SO(44)$.  The corresponding
    cosmological constant is $\Lambda\approx 7800.08\equiv \Lambda_1$.
     This is the largest value of $\Lambda$ for any string model,
      and it appears for the $SO(44)$ string model only.
\item {\sl order=786:}\/   This cluster contains two distinct models, both of
    which have gauge group $SO(40)\times SU(2)^2$.
     However, while one of these models has $\Lambda\approx 3340.08$,
     the other has a cosmological constant exactly equal to $\Lambda_1/2$!
\item {\sl order=658:}\/  This cluster contains 12 distinct models, all with
       gauge group $SO(36)\times SO(8)$.  Remarkably, once again,
       all have cosmological constants $\Lambda=\Lambda_1/2$, even though
       they differ in their particle representations and content.
       This is a reflection of the huge cosmological-constant redundancies
       discussed in Sect.~5.
\item {\sl order=642:}\/  This cluster contains only one model, with gauge
       group $SO(36)\times SU(2)^4$ and cosmological constant
       $\Lambda\approx 3620.06\equiv \Lambda_2$.
\item {\sl order=578:}\/  Here we have one model with gauge group
       $SO(34)\times SU(4)\times U(1)^2$.  Remarkably, this model
       has cosmological constant given by $\Lambda=\Lambda_1/4$!
\end{itemize}
These kinds of redundancies and scaling patterns persist
as we continue to implement twists that project out gauge bosons from our string models.
For example, we find 
\begin{itemize}
\item {\sl order=530:}\/  
       Here we have 10 distinct string models:
       nine have gauge group $SO(32)\times SO(8)\times SU(2)^2$,
       while one has gauge group $E_8\times SO(24)\times SU(2)^2$.
      Two of the models in the first group have $\Lambda=\Lambda_2$,
     while the remaining eight models in this cluster exhibit five
     new values of $\Lambda$ between them. 
\item {\sl order=514:}\/  Here we have two distinct string models,
    both with gauge group $SO(32)\times SU(2)^6$:
    one has $\Lambda=\Lambda_1/2$, while the other has $\Lambda=\Lambda_1/4$.  
\item {\sl order=466:}\/  Here we have 12 distinct string models:
    five with gauge group $SO(24)\times SO(20)$,
    two with gauge group $E_8\times SO(20)\times SO(8)$,
    and five with gauge group $SO(30)\times SU(4)^2\times U(1)$.
    All of the models in the last group have $\Lambda=\Lambda_1/4$,
    while two of the five in the first group have $\Lambda=\Lambda_1/2$.
    This is also the first cluster in which models with $\Lambda<0$ appear. 
\end{itemize}
Indeed, as we proceed towards models with smaller orders,
we generate not only new values of the cosmological constant,
but also cosmological-constant values
which are merely rescaled from previous values by factors of 2, 4, 8, and so forth. 
For example, the original maximum value $\Lambda_1$ which appears
only for the $SO(44)$ string model has many ``descendents'':
there are $21$ distinct models with $\Lambda=\Lambda_1/2$;
$61$ distinct models with $\Lambda=\Lambda_1/4$;
$106$ distinct models with $\Lambda=\Lambda_1/8$;
and so forth.

Ultimately, these rescalings are related to the fact that our 
models are constructed through successive $\IZ_2$ twists.
Although there are only limited numbers of modular-invariant partition
functions $Z(\tau)$, these functions may be rescaled 
without breaking the modular invariance.  
However, we emphasize that {\it this rescaling 
of the partition function does not represent 
a trivial overall rescaling of the associated particle spectrum}.
In each model, for example, there can only be one distinct gravity multiplet;  likewise,
the string vacuum states are necessarily unique.
Thus, it is somewhat remarkable that two models with completely different
particle spectra can nevertheless give rise to rescaled versions of the 
same partition function and cosmological constant.

\begin{figure}%[ht]
\centerline{
   \epsfxsize 3.1 truein \epsfbox {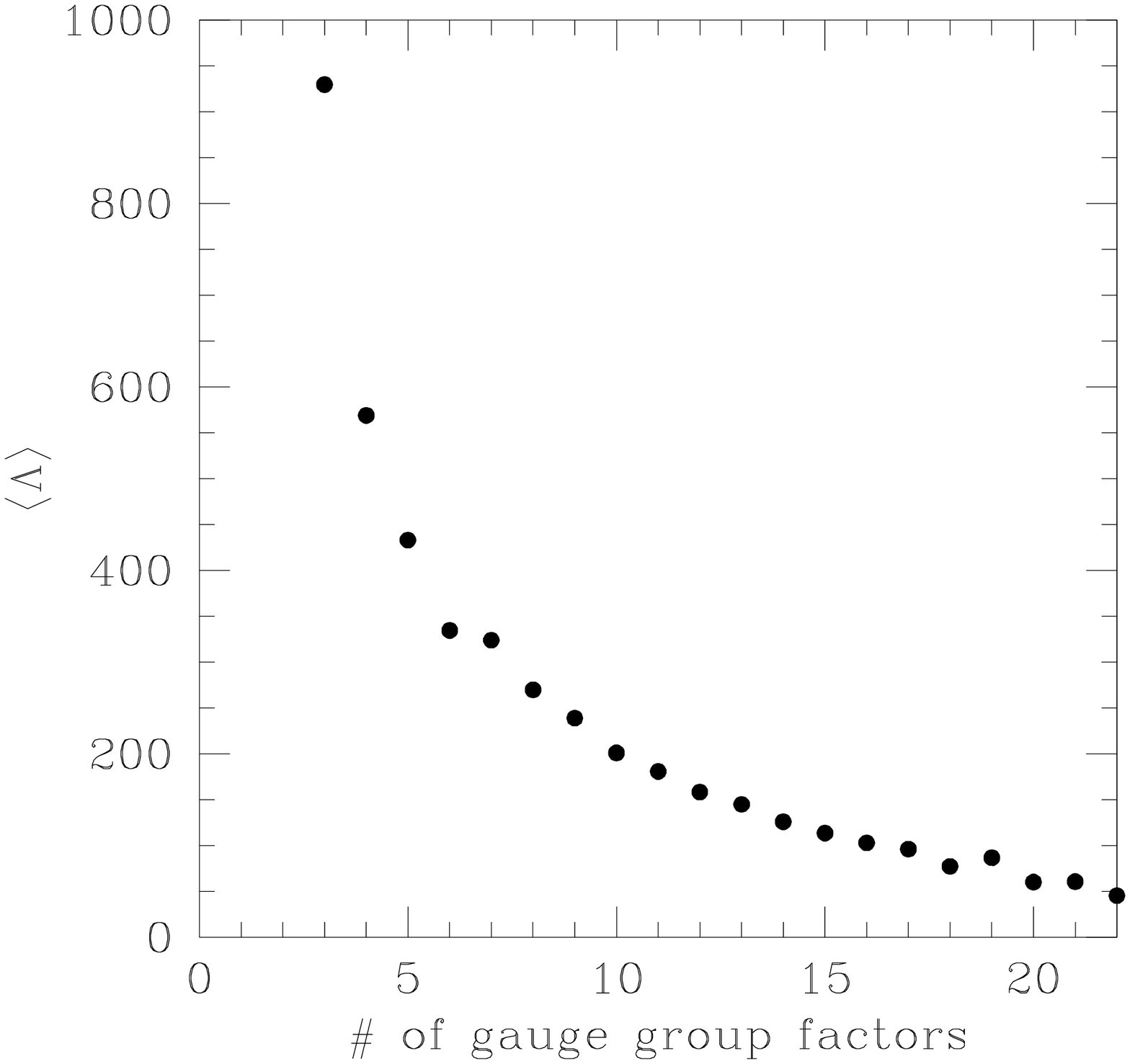}
   \epsfxsize 3.1 truein \epsfbox {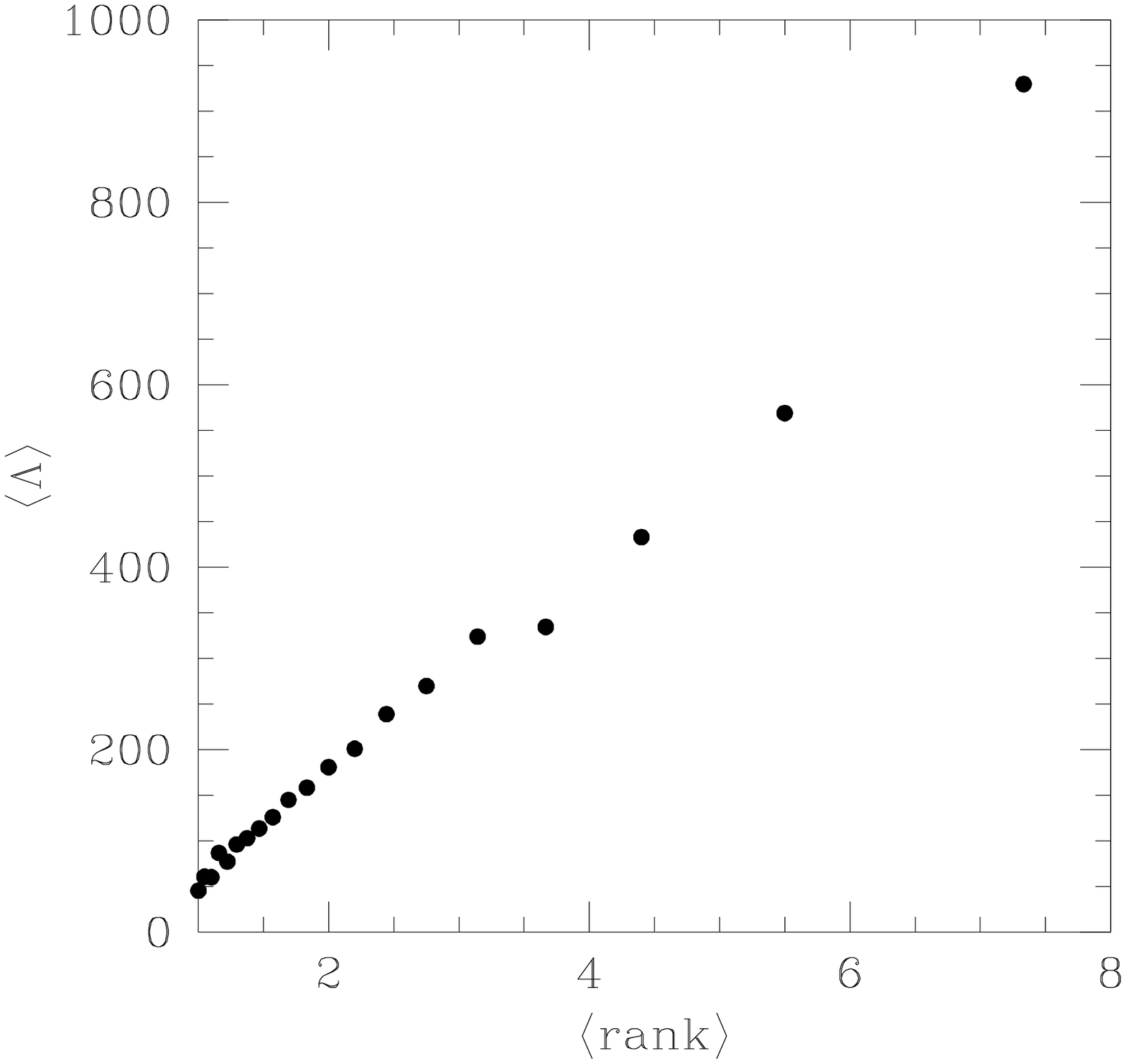}
    }
\centerline{
   \epsfxsize 3.4 truein \epsfbox {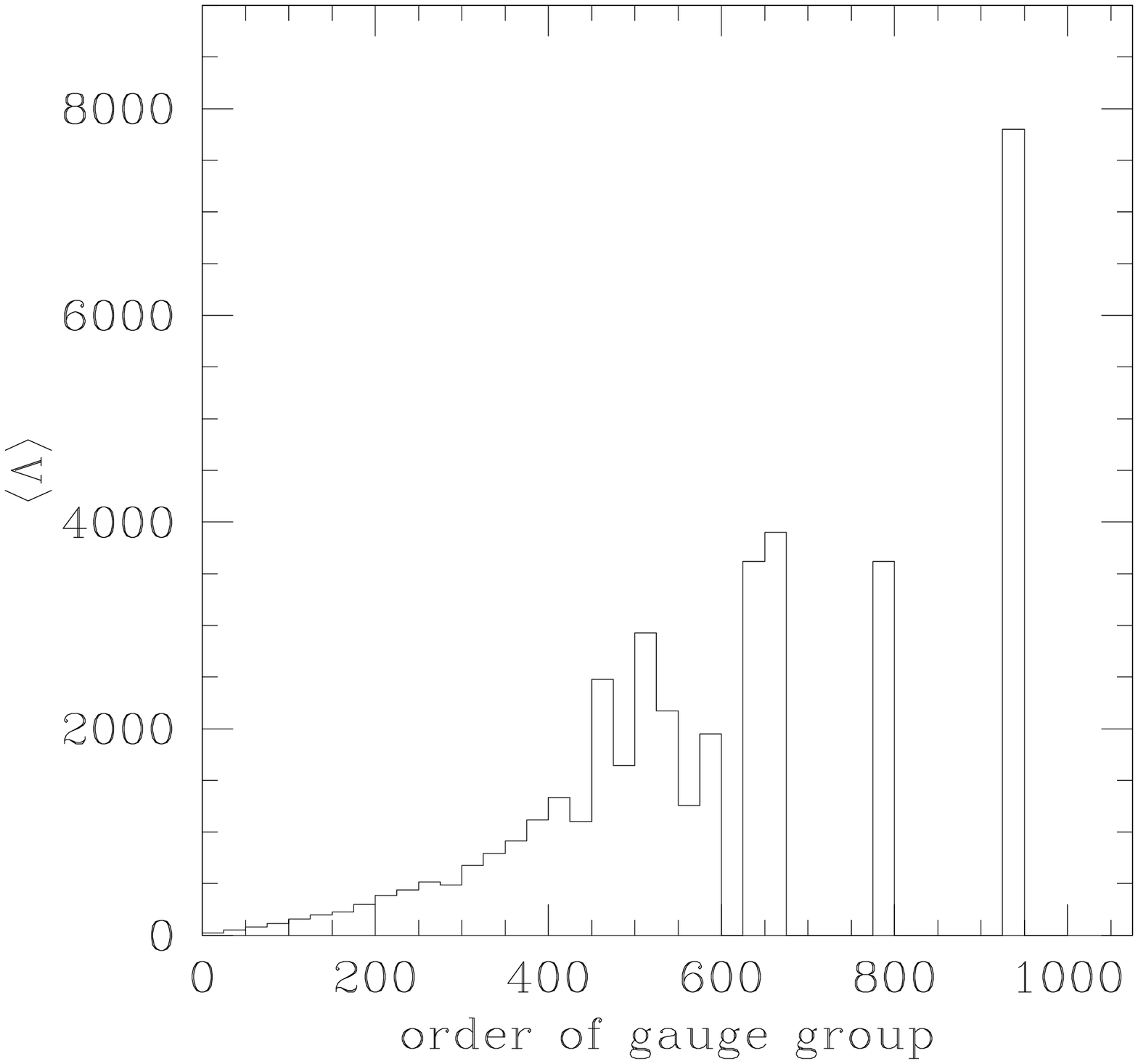}
    }
\caption{Correlations between cosmological constants and gauge groups.
    (a) (upper left) Average values of the cosmological constants obtained
    as a function of the number $f$ of gauge-group factors in the corresponding
    string models.  Note that all cosmological constant averages are positive
    even though approximately 1/4 of individual string models have $\Lambda<0$,
    as shown in Fig.~\protect\ref{Fig11}.
    For plotting purposes, we show data only for values $f\geq 3$.
    (b) (upper right)  Same data plotted versus the average
    rank per gauge group factor, defined as $22/f$.
    (c) (lower figure)  Histogram showing the average values of $\Lambda$
   as a function of the order (dimension) of corresponding gauge group.
   Note that in every populated bin, we find $\langle \Lambda\rangle >0$ even
   though most bins contain at least some string models with $\Lambda<0$.
   Thus, in this figure, bins with $\langle \Lambda\rangle=0$ are to be
   interpreted as empty rather than as bins for which $\Lambda>0$ models
   exactly balance against $\Lambda<0$ models.}
\label{Fig13} 
\end{figure}

Having described the characteristics of our cosmological-constant tree,
let us now turn to its overall statistics and correlations.
For each branch of the tree, we can investigate the values
of the corresponding cosmological constants, averaged over 
all models on that branch.
If we organize our branches according to the numbers of gauge-group
factors as in Sect.~4, 
we then find the results shown in Figs.~\ref{Fig13}(a) and \ref{Fig13}(b).
Alternatively, we can also cluster our models according to
the orders of their gauge groups, as described above, and calculate average
cosmological constants as a function of these orders.
We then find the results shown in Fig.~\ref{Fig13}(c).

%================== FIGURE ============================================
%================== END OF INSERTED FIGURE ============================

Clearly, we see from Fig.~\ref{Fig13} that 
there is a strong and dramatic correlation between gauge groups
and cosmological constants:  
\begin{itemize}
\item  {\bf Models with highly shattered 
gauge groups and many irreducible gauge-group factors
tend to have smaller cosmological constants, while those
with larger non-abelian gauge groups tend to have larger
cosmological constants.} 
Indeed, we see from Fig.~\ref{Fig13}(b) that
the average cosmological constant grows approximately 
linearly with the average rank of the gauge-group factors
in the corresponding model.
\end{itemize}

It is easy to understand this result.
As we shatter our gauge groups into smaller irreducible factors,
the average size of each {\it representation}\/ of the gauge group also
becomes smaller.  (For example, we have already seen this 
behavior in Fig.~\ref{Fig8} for gauge
bosons in the adjoint representation.)
Therefore, we expect the individual tallies of bosonic
and fermionic states 
at each string mass level
to become smaller as the total gauge group
is increasingly shattered.
If these individual numbers of bosons and fermions become smaller,
then we statistically expect the magnitudes of their differences $\lbrace a_{sd}\rbrace$ 
in Eq.~(\ref{Idef}) to become smaller as well.
We therefore expect the cosmological constant to be correlated
with the degree to which the gauge group of the heterotic 
string model is shattered, as found in Fig.~\ref{Fig13}.
The fact that the average cosmological grows approximately linearly 
with the average rank of the gauge-group factors  
[as shown in Fig.~\ref{Fig13}(b)]
then suggests that on average, 
this cosmological constant scaling is dominated
by vector representations of the gauge groups (whose dimensions 
grow linearly with the rank of the gauge group).
Such representations 
indeed tend to dominate the string spectrum at the massless level,
since larger representations are often excluded on the basis of unitarity
grounds for gauge groups with affine levels $k=1$.

%================== FIGURE ============================================
\begin{figure}[ht]
\centerline{
   \epsfxsize 4.0 truein \epsfbox {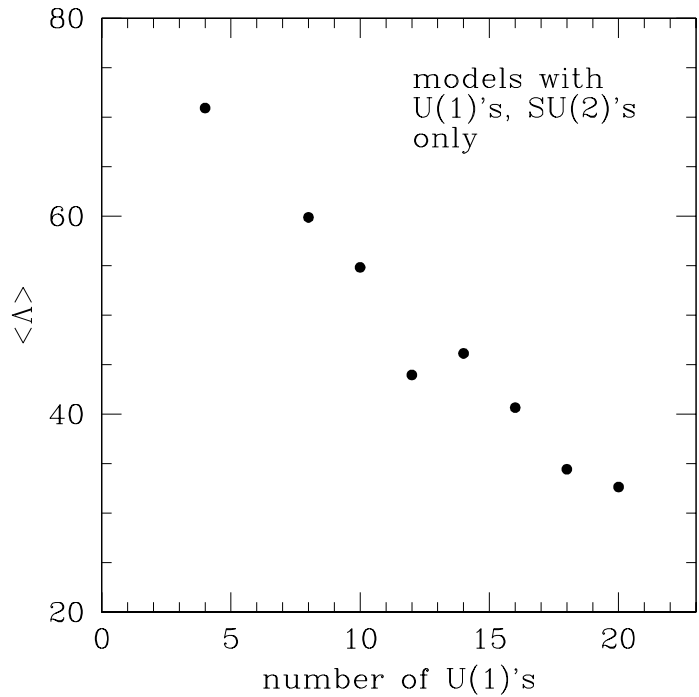}
    }
\caption{The effects of non-abelianity:  Average values of $\Lambda$
  for 12$\,$000 heterotic string models with gauge groups of the form
   $U(1)^n\times SU(2)^{22-n}$, plotted as a function of $n$.
   Since varying $n$ does not change the average rank of each gauge-group
   factor, this correlation is statistically independent 
    of the correlations shown in Fig.~\protect\ref{Fig13}.}
\label{Fig14} 
\end{figure}
%================== END OF INSERTED FIGURE ============================

Even {\it within}\/ a fixed degree of shattering (\ie, even within
a fixed average rank per gauge group), one may ask whether the degree to
which the gauge group is abelian or non-abelian 
may play a role.
In order to pose this question in a way that is independent of the
correlations shown in Fig.~\ref{Fig13}, we can restrict our attention
to those 
heterotic string models in our sample set
for which $f=22$ [\ie, models with
gauge groups of the form $U(1)^n\times SU(2)^{22-n}$].
It turns out that there are approximately $12\,000$ models in this class.
Varying $n$ does not change the average rank of each gauge-group
factor, and thus any correlation between the cosmological
constant and $n$ is statistically independent of the correlations shown 
in Fig.~\ref{Fig13}.
The results are shown in Fig.~\ref{Fig14}, averaged over the $12\,000$ relevant
string models.
Once again, we see that ``bigger'' (in this case, non-abelian) groups
lead to larger average values of the cosmological constant.  The roots
of this behavior are the same as those sketched above.

We can also investigate how $\Lambda$ statistically depends 
on cross-correlations of gauge groups.  Recall that in Table~\ref{table1}, we
indicated
the percentages of four-dimensional heterotic string models which exhibit
various combinations of gauge groups.
In Table~\ref{tablelam}, we indicate the average values of $\Lambda$
for those corresponding sets of models.
We see, once again, that the correlation between average values of $\Lambda$
and the ``size'' of our gauge groups is striking.
For example, looking along the diagonal in Table~\ref{tablelam},
we observe that the average values of $\Lambda$ for models containing gauge
groups of the form $G\times G$ always monotonically increase as
$G$ changes from $SU(3)$ to $SU(4)$ to $SU(5)$ to $SU(n>5)$;
likewise, the same behavior is observed as $G$ varies
from $SO(8)$ to $SO(10)$ to $SO(2n>10)$.
Indeed, as a statistical collection,
we see from Table~\ref{tablelam} that 
the models with the largest average values of $\Lambda$
are those with at least two exceptional factors.

%================================
\begin{table}
\centerline{
   \begin{tabular}{||r|| r|r|r|r|r|r|r|r|r|r||r|r||}
   \hline
   \hline
   ~ & $U_1$~ & $SU_2$ & $SU_3$ & $SU_4$ & $SU_5$ & $SU_{>5}$ & $SO_8$
            & $SO_{10}$ & $SO_{>10}$ & $E_{6,7,8}$ & SM~ & PS~ \\
   \hline
   \hline
$U_1$ &   104.6&   104.6&    83.2&   112.9&   110.7&   162.3&   131.2&   172.1&   238.8&   342.2&    80.8&   107.6 \\ \hline
$SU_2$ &     ~&   120.7&    80.8&   109.1&   106.6&   157.1&   155.5&   167.9&   282.6&   442.5&    80.4&   103.9 \\ \hline
$SU_3$ &     ~&     ~&    85.9&    90.9&   113.3&   136.1&   117.6&   162.8&   193.5&   220.2&    83.0&    88.3 \\ \hline
$SU_4$ &     ~&     ~&     ~&   115.2&   115.0&   150.9&   129.1&   166.7&   235.3&   314.2&    88.9&   110.5 \\ \hline
$SU_5$ &     ~&     ~&     ~&     ~&   135.9&   156.3&   128.1&   191.6&   199.2&     ---~~ &   107.7&   110.3 \\ \hline
$SU_{>5}$ &     ~&     ~&     ~&     ~&     ~&   200.9&   156.4&   203.2&   274.5&   370.7&   133.5&   142.8 \\ \hline
$SO_8$ &     ~&     ~&     ~&     ~&     ~&     ~&   192.7&   167.5&   301.6&   442.8&   115.3&   123.3 \\ \hline
$S0_{10}$ &     ~&     ~&     ~&     ~&     ~&     ~&     ~&   207.8&   253.4&   289.3&   166.0&   159.6 \\ \hline
$SO_{>10}$ &     ~&     ~&     ~&     ~&     ~&     ~&     ~&     ~&   417.4&   582.8&   190.8&   220.0 \\ \hline
$E_{6,7,8}$ &     ~&     ~&     ~&     ~&     ~&     ~&     ~&     ~&     ~&  1165.9&   220.2&   272.3 \\ \hline
\hline
SM &     ~&     ~&     ~&     ~&     ~&     ~&     ~&     ~&     ~&     ~&    82.5&    85.5 \\ \hline
PS &     ~&     ~&     ~&     ~&     ~&     ~&     ~&     ~&     ~&     ~&     ~&   104.9 \\ \hline
\hline
total: &   108.8&   121.4&    83.2&   113.8&   110.7&   162.2&   163.0&   173.0&   298.5&   440.2&    80.8&   108.3 \\ 
   \hline
   \hline
   \end{tabular}
}
\caption{Average values of $\Lambda$ for the four-dimensional heterotic string models which 
    exhibit various combinations of gauge groups.  This table follows the same organization
    and notational conventions as Table~\protect\ref{table1}.
    Interestingly, scanning across the bottom row of this table, we see that
    those string models which contain at least the 
     Standard-Model gauge group have the smallest average values of $\Lambda$.  }
\label{tablelam}
\end{table}
%================================

Conversely, scanning across the bottom row of Table~\ref{tablelam},
we observe that the average value of $\Lambda$ is minimized
for models which contain at least the Standard-Model gauge group
$SU(3)\times SU(2)\times U(1)$ among their factors. 
Given our previous observations about the correlation between
the average value of $\Lambda$ and the size of the corresponding
gauge groups, it may seem surprising at first glance that models in which 
we demand only
a single $U(1)$ or $SU(2)$ gauge group do not have an even smaller
average value of $\Lambda$.  However,
models for which we require only a single $U(1)$ or $SU(2)$ factor  
have room for potentially larger gauge-group factors in their complete
gauge groups 
than do models which simultaneously exhibit
the factors $U(1)\times SU(2)\times SU(3)\equiv G_{\rm SM}$.
Thus, on average, demanding the appearance of the entire 
Standard-Model gauge group is more effective in minimizing
the resulting average value of $\Lambda$ than demanding a single
$U(1)$ or $SU(2)$ factor alone. 
In fact, we see from Table~\ref{tablelam} that
demanding $G_{\rm SM}\times U(1)$ is even more effective
in reducing the average size of $\Lambda$, while
demanding a completely shattered gauge group of the form
$U(1)^n\times SU(2)^{22-n}$ produces averages which are
even lower, as shown in Fig.~\ref{Fig14}.

\begin{figure}[t]
\centerline{
   \epsfxsize 4.0 truein \epsfbox {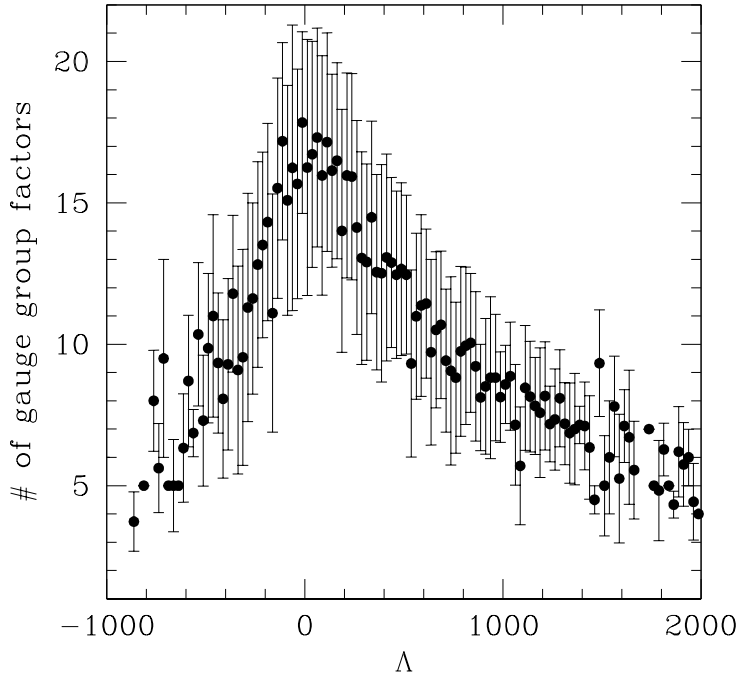}
    }
\caption{The ``inverse'' of Fig.~\protect\ref{Fig13}(a):
  Here we have binned our heterotic string models according
   to their values of $\Lambda$ and then plotted $\langle f \rangle$,
   the average value of the number of gauge-group factors,
    for the models in each bin.
   The error bars delimit the range
   $\langle f\rangle \pm \sigma$ where $\sigma$
    are the corresponding standard deviations.
   We see that while a particular value of $\Lambda$ restricts $f$
    to a fairly narrow range, a particular value of $f$ only focuses
    $\Lambda$ to lie within two separate ranges of different central
   magnitudes $|\Lambda|$ and opposite signs.}
\label{Fig15}
\end{figure}

In this discussion, it is important to stress that we are dealing
with statistical {\it averages}\/:  individual
gauge groups and cosmological constants can vary significantly
from model to model.
In other words, even though we have been plotting average
values of $\Lambda$ in Figs.~\ref{Fig13} and \ref{Fig14},
there may be significant standard deviations in these plots.
In order to understand the origin of these standard deviations,
let us consider the ``inverse'' of Fig.~\ref{Fig13}(a)
which can be constructed
by binning our heterotic string models according
to their values of $\Lambda$ and then plotting the average value
of $f$, the number of gauge-group factors, for the models in each bin.
The result is shown in Fig.~\ref{Fig15}, where we have plotted not
only the average values of $f$ but also the corresponding standard deviations.
Once again, we see that smaller values of $|\Lambda|$ are clearly correlated
with increasingly shattered gauge groups.
However, while a particular value of $\Lambda$ is directly correlated
with a contiguous, relatively small range for $f$,
the reverse is not true:
a particular value of $f$ is correlated with {\it two}\/ distinct
ranges for $\Lambda$ of opposite signs.
Indeed, even the central magnitudes for $\Lambda$ in these two ranges
are unequal because of the
asymmetry of the data
in Fig.~\ref{Fig15}, with the ``ascending'' portion of the curve
having a steeper slope than the descending portion.
As a result of this asymmetry, 
and as a result of the different numbers of string models which populate
these two regions,
the total value
$\langle \Lambda\rangle$  averaged across
these two regions does not cancel, but instead follows
the curve in Fig.~\ref{Fig13}(a).
Thus, while the curves in Figs.~\ref{Fig13} and \ref{Fig14} technically
have large standard deviations, we have not shown these standard deviations
because they do not reflect
large uncertainties in the corresponding allowed values of $\Lambda$.
Rather, they merely reflect the fact that the allowed values of $\Lambda$
come from two disjoint but relatively well-focused regions.

%================== FIGURE ============================================
%================== END OF INSERTED FIGURE ============================

Of course, as we approach the ``top'' of the curve
in Fig.~\ref{Fig15} near $|\Lambda|\approx 0$, 
these two distinct regions merge together.
However, even in this limit,
it turns out that the sizes of the standard deviations depend
on which physical quantity in the comparison is
treated as the independent variable and held fixed.
For example, if we restrict our attention to heterotic string models
containing a gauge group of the form $U(1)^n\times SU(2)^{22-n}$
(essentially holding $f$, the number of gauge-group factors, fixed
at $f=22$), we still find corresponding values of $\Lambda$ populating
the rather wide range $-400\lsim \Lambda\lsim 500$.
In other words, holding $f$ fixed does relatively little to focus $\Lambda$.
By contrast,
we have already remarked in Sect.~5 that across our entire
sample of $\sim 10^5$ models, the smallest value of $|\Lambda|$
that we find is $\Lambda\approx 0.0187$.
This value emerges for nine models, eight of which share
the same state degeneracies $\lbrace a_{sd}\rbrace$ and one
of which is their ``twin'' (as defined at the end of Sect.~5).
If we take $\Lambda$ as the independent variable and hold 
$\Lambda\approx 0.0187$ fixed (which represents only one very
narrow slice within the bins shown in Fig.~\ref{Fig15}),
we then find that essentially {\it all}\/ of
the corresponding gauge groups are extremely ``small'':  four are of
the form $U(1)^n\times SU(2)^{22-n}$ with $n=12, 13, 15,17$,
and the only two others are
$SU(3)\times SU(2)^4\times U(1)^{16}$
and
$SU(3)^2\times SU(2)^3\times U(1)^{15}$.
In other words, models with $\Lambda\approx 0.0187$
have $\langle f\rangle \approx 21.67$,
with only a very small standard deviation.
Indeed,
amongst all distinct models with $|\Lambda|\leq 0.04$,
we find none with
gauge-group factors of rank exceeding $5$;
only $8.2\%$ of such models have an individual gauge-group factor of rank exceeding $3$
and only $1.6\%$ have an individual gauge-group factor of rank exceeding $4$.
None were found that exhibited any larger gauge-group factors.
Thus, we see that keeping $\Lambda$ small goes a long way towards keeping
the corresponding gauge-group factors small.

It is, of course, dangerous to extrapolate from these observations
of $\sim 10^5$ models in order to make claims concerning a full string landscape
which may consist of $\sim 10^{500}$ models or more.
Nevertheless, as we discussed at the end of Sect~2,
we have verified that all of these statistical correlations appear
to have reached the ``continuum limit'' (meaning that they appear
to be numerically stable
as more and more string models
are added to our sample set).
Indeed, although the precise minimum value of $|\Lambda|$
is likely to continue to decline as more and more models
are examined,
the correlation between small values of $|\Lambda|$
and small gauge groups is likely to persist.
Needless to say, it is impossible to estimate how large our
set of heterotic string models
must become before
we randomly find a model with $|\Lambda|\approx 10^{-120}$;
indeed, if the curve in Fig.~\ref{Fig12} truly saturates
at a finite value, such models may not even exist.
However, if we assume (as in Ref.~\cite{BP}) that such models
exist --- providing what would truly be a stable string ``vacuum''  ---
then it seems overwhelmingly likely that
\begin{itemize}
\item  {\bf Perturbative heterotic string vacua with 
     observationally acceptable cosmological constants
     can be expected to have extremely small gauge-group factors,
     with $U(1)$, $SU(2)$, and $SU(3)$ overwhelmingly
     favored relative to larger groups such as $SU(5)$,
     $SO(10)$, or $E_{6,7,8}$.  Thus, for such string vacua,
     the Standard-Model gauge group is much more likely 
     to be realized
     at the string scale than any of its grand-unified
     extensions.}
\end{itemize}
As always, such a claim is subject to a number of 
additional assumptions:  we are limited to perturbative
heterotic string vacua, we are examining only one-loop
string amplitudes, and so forth.
Nevertheless, we find this type of correlation to
be extremely intriguing, and feel that it is likely
to hold for higher-loop contributions to the vacuum
amplitude as well.

Note, in particular, that the critical ingredient
in this claim is the assumption of small cosmological constant.
Otherwise, statistically weighting all of our string models
equally without regard for their cosmological constants,
we already found in Sect.~4 that the Standard Model is relatively {\it disfavored}\/,
appearing only $10\%$ of the time, while
the $SO(2n\geq 10)$ GUT groups appear with the much greater
frequency $24.5\%$.
Thus, it is the requirement of a small cosmological constant 
which is responsible for redistributing these probabilities
in such a dramatic fashion.   

There is, however, another possible way to interpret
this correlation between the magnitudes of $\Lambda$ 
and the sizes of gauge groups.  As we have seen, smaller
values of $\Lambda$ tend to emerge as the gauge group becomes
increasingly shattered.  However, as we know, there is a
fundamental limit to how shattered our gauge groups can
become:  $f$ simply cannot exceed $22$, \ie,
there is no possible gauge-group factor with rank
less than 1.  Thus, the correlation between $\Lambda$ and
average gauge-group rank may imply that  
there is likewise a minimum possible value for $\Lambda$.
If so, it is extremely unlikely that perturbative heterotic string
models will be found in which $\Lambda$ is orders of magnitude less 
than the values we have already seen.

%================== FIGURE ============================================
\begin{figure}[ht]
\centerline{
   \epsfxsize 4.0 truein \epsfbox {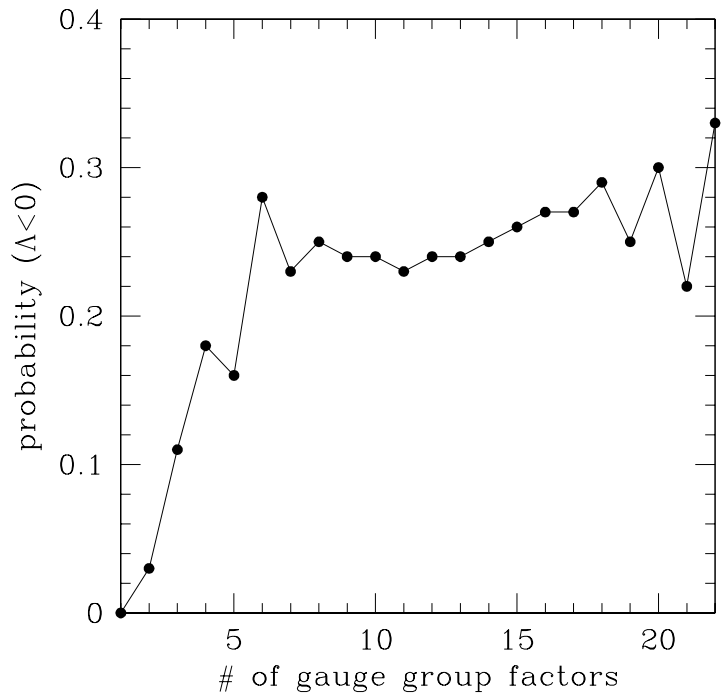}
    }
\caption{The probability that a randomly chosen heterotic string
   model has a negative value of $\Lambda$ (\protect\ie, a positive value of the
   vacuum energy density $\lambda$), plotted as a function
  of the number of gauge-group factors in the total gauge group of the model.
    We see that we do not have a significant probability of 
    obtaining models with $\Lambda<0$ until our gauge group is
    ``shattered'' into at least four or five factors;  this
   probability then remains roughly independent of the number of factors
    as further shattering occurs.}
\label{Fig16} 
\end{figure}
%================== END OF INSERTED FIGURE ============================

Another important characteristic of such string models
is the {\it sign}\/ of $\Lambda$.  For example, whether
the vacuum energy $\lambda= -\half \calM^4 \Lambda$ is positive
or negative can determine whether the corresponding spacetime is   
de Sitter (dS) or anti-de Sitter (AdS).~
In Fig.~\ref{Fig16}, we show 
the probability that a randomly chosen heterotic string
   model has a negative value of $\Lambda$, plotted as a function
  of the number of gauge-group factors in the total gauge group of the model.
For small numbers of factors, the corresponding models all tend to have
very large positive values of $\Lambda$.  
Indeed, as indicated in Fig.~\ref{Fig16},
we do not accrue a significant probability of 
obtaining models with $\Lambda<0$ until our gauge group is
``shattered'' into at least four or five factors.
The probability of obtaining negative values of $\Lambda$ then
saturates near $\approx 1/4$, remaining
roughly independent of the number of gauge-group factors
as further shattering occurs.
Thus, we see that regardless of the value of $f$,
the ``ascending'' portion of Fig.~\ref{Fig15} 
is populated by only a quarter as many models as
populate the ``descending'' portion.
Since the overwhelming majority of models have relatively large numbers of
gauge-group factors (as indicated in Fig.~\ref{Fig1}),
we see that on average, approximately 1/4 of our models have
negative values of $\Lambda$.
This is consistent with the histogram in Fig.~\ref{Fig11}. 

%================== FIGURE ============================================
\begin{figure}[h]
\centerline{
   \epsfxsize 4.0 truein \epsfbox {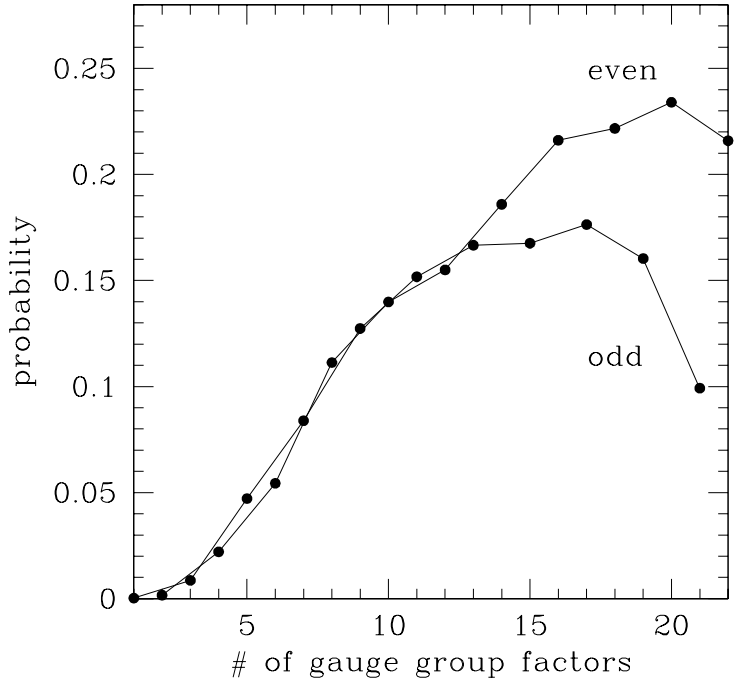}
    }
\caption{The degeneracy of values of the cosmological constant relative 
   to gauge groups.  We plot the probability that a given value of 
   $\Lambda$ (chosen from amongst the total set of obtained values) 
   will emerge from a model
   with $f$ gauge-group factors, as a function of $f$ for even and odd
   values of $f$.
   This probability is equivalently defined as the number of distinct
   $\Lambda$ values obtained from models with $f$ gauge-group factors,
   divided by the total number of $\Lambda$ values found across all values of $f$.
   The resemblance of these curves to those in Fig.~\protect\ref{Fig1}
   indicates that the vast degeneracy of cosmological-constant values
   is spread approximately uniformly 
   across models with different numbers of gauge-group factors.}
\label{Fig17} 
\end{figure}
%================== END OF INSERTED FIGURE ============================

Finally, we can investigate how the vast redundancy in the values of the
cosmological constant 
is correlated with the corresponding gauge groups  and the degree to which
they are shattered.
In Fig.~\ref{Fig17}, we plot  the number of distinct
$\Lambda$ values obtained from models with $f$ gauge-group factors,
divided by the total number of $\Lambda$ values found across all values of $f$.
Note that 
the sum of the probabilities 
plotted in Fig.~\ref{Fig17}
exceeds one.  This is because 
a given value of $\Lambda$ may emerge from models with many different values of $f$ ---
\ie, the sets of values of $\Lambda$ for each value of $f$ are not exclusive. 
It turns out that this is a huge effect, especially as $f$ becomes relatively large.

The fact that the curves in Fig.~\ref{Fig1} and Fig.~\ref{Fig17} 
have similar shapes
as functions of $f$
implies that the vast degeneracy of cosmological-constant values
is spread approximately uniformly 
across models with different numbers of gauge-group factors.
Moreover, as indicated in Fig.~\ref{Fig12}(b), this degeneracy factor 
itself tends to decrease as more and more models are
examined, leading to a possible saturation of distinct
$\Lambda$ values, as discussed above.

%=============================================================================
\section{Discussion}
\setcounter{footnote}{0}

In this paper, we have investigated the statistical properties 
of a fairly large class of perturbative, four-dimensional, non-supersymmetric,
tachyon-free heterotic string models.
We focused on their gauge groups, their one-loop cosmological constants, and the
statistical correlations that emerge between these otherwise disconnected quantities.

Clearly, as stated in the Introduction, much more work remains to be done,
even within this class of models.
For example, it would be of immediate interest to examine other
aspects of the full particle spectra of these models
and obtain statistical information
concerning Standard-Model embeddings, spacetime chirality, numbers of 
generations, and $U(1)$ hypercharge assignments, as well as cross-correlations
between these quantities.
This would be analogous to what has been done for Type~I orientifold
models in Refs.~\cite{blumenhagen,schellekens}.
One could also imagine looking at the gauge couplings and their runnings,
along with their threshold corrections~\cite{Kaplunovsky},
to see whether it is likely that unification occurs given low-energy precision
data~\cite{Prep,faraggi}.
It would also be interesting to examine the properties of the cosmological
constant {\it beyond}\/ one-loop order, with an eye towards understanding to what
extent the unexpected degeneracies we have found persist.  
An analysis of other string amplitudes and correlation functions is
clearly also of interest, particularly as they relate to Yukawa couplings
and other phenomenological features.
Indeed, a more sophisticated analysis examining all of these features
with a significantly larger sample set of models is currently underway~\cite{next}. 

Needless to say, another option is to expand the class of heterotic
string models under investigation.
While the models examined in this paper are all non-supersymmetric,
it is also important to repeat much of this work for 
four-dimensional heterotic models with $\calN=1$, $\calN=2$, and $\calN=4$ supersymmetry.
There are two distinct reasons why this is an important undertaking.
First, because they are supersymmetric, such models are commonly believed
to be more relevant to particle physics in addressing issues of gauge coupling
unification and the gauge hierarchy problem.
But secondly, at a more mathematical level,
such models have increased stability properties relative to the
non-supersymmetric models we have been examining here.  Thus,
by examining the statistical properties of such models and comparing
them with the statistical properties of supersymmetric models,
we can determine the extent to which supersymmetry has an effect
on these other phenomenological features.
Such analyses are also underway~\cite{next}.
Indeed, in many cases these perturbative supersymmetric heterotic strings are dual
to other strings (\eg, Type~I orientifold models) whose statistical
properties are also being analyzed.  Thus, analysis of the perturbative
heterotic landscape, both supersymmetric and non-supersymmetric, will
enable {\it statistical}\/ tests of duality symmetries across 
the entire string landscape.

Beyond this, of course, there are many broader classes of closed string
models which may be examined --- some of these are discussed
in Sect.~2.   Indeed, of great interest are completely
 {\it stable}\/ non-supersymmetric models.  As discussed in 
Ref.~\cite{heretic}, such models could potentially provide
 {\it non}\/-supersymmetric
solutions not only for the cosmological-constant problem,
but also for the apparent gauge hierarchy problem.
Such models could therefore provide a framework for 
an alternative, non-supersymmetric approach towards string phenomenology~\cite{heretic}.
However, given that no entirely stable non-supersymmetric perturbative heterotic
string models have yet been constructed, 
the analysis of this paper represents the current ``state of the art'' as far
as non-supersymmetric perturbative heterotic string model-building
is concerned.

As mentioned in the Introduction, this work may be viewed as part
of a larger ``string vacuum project'' whose goal is to map
out the properties of the landscape of string vacua. 
It therefore seems appropriate to close with two warnings 
concerning the uses and abuses of such large-scale statistical studies
as a method of learning about the properties of the landscape.

The first warning concerns what may be called ``lamppost'' effect --- 
the danger of restricting one's attention to only those portions
of the landscape where one has control over calculational 
techniques.
(This has been compared to searching for a small object in the darkness of night:
the missing object may be elsewhere, 
but the region under the lamppost may represent the
only location where the search can be conducted at all.)
For example, our analysis
in this paper has been restricted to string models exploiting
``free-field'' constructions (such as string constructions using bosonic lattices or
free-fermionic formalisms).  While this class of string models
is very broad and lends itself naturally to a computer-automated 
search and analysis, it is entirely possible that the models with the
most interesting phenomenologies are beyond such a class.
Indeed, it is very easy to imagine that different constructions
will have different strengths and weaknesses as far as their
low-energy phenomenologies are concerned, and that one type of
construction may easily realize features that another cannot
accommodate.

By contrast, the second danger can be called the ``G\"odel effect'' ---
the danger that no matter how many conditions (or input ``priors'')
one demands for a phenomenologically realistic string model,
there will always be another observable 
for which the set of realistic models will make differing predictions.
Therefore, such an observable will remain beyond our statistical
ability to predict.
(This is reminiscent of the ``G\"odel incompleteness theorem'' 
which states that in any axiomatic system, there is always
another statement which, although true, cannot be deduced purely from the axioms.) 
Given that the full string landscape is very large, consisting of perhaps
$10^{500}$ distinct models or more,
the G\"odel effect may represent a very real danger.
Thus, since one can never be truly sure of having examined
a sufficiently sizable portion of the landscape, it is likewise never 
absolutely clear whether we can be truly free of such G\"odel-type ambiguities
when attempting to make string predictions. 

Of course, implicit in each of these effects is the belief that 
one actually knows what one is looking for --- that we know {\it which}\/ theory 
of particle physics should be embedded
directly into the string framework and viewed as emerging from a particular 
string vacuum.
However, it is possible that nature might pass through many layers
of effective field theories at higher and higher energy scales before reaching an 
ultimate string-theory embedding.  In such cases, the potential constraints
on a viable string vacuum are undoubtedly weaker.

Nevertheless, we believe that there are many valid purposes 
for such statistical studies of actual string models.
First, as we have seen at various points in this paper,
it is only by examining actual string models --- and not effective
supergravity solutions --- that many surprising features come to light.
Indeed, one overwhelming lesson that might 
be taken from the analysis in this paper is that the string landscape 
is a very rich place, full of unanticipated properties and characteristics 
that emerge only from 
direct analysis of concrete string models.

Second, through their direct enumeration,
we gain valuable experience in the construction and analysis 
of phenomenologically viable models. 
This is, in some sense, a direct test of string theory as a phenomenological
theory of physics. 
For example, it is clear from the results of this paper that
obtaining the Standard-Model gauge group is a fairly non-trivial task
within free-field constructions based on $\IZ_2$ periodic/antiperiodic
orbifold twists;  as we have seen in Fig.~\ref{Fig6}(b), one must induce 
a significant amount of gauge-group shattering before a sizable population 
of models with the Standard-Model gauge group emerges.
This could not have been anticipated on the basis of low-energy
effective field theories alone, and is ultimately a reflection
of worldsheet model-building constraints.
Such knowledge and experience are extremely valuable for
string model-builders, and can serve as useful guideposts.

Third, as string phenomenologists, we must ultimately come to terms
with the landscape.  Given that such large numbers of string vacua
exist, it is imperative that string theorists learn about these
vacua and the space of resulting possibilities.
Indeed, the first step in any scientific examination of a large data
set is that of enumeration and classification;  this has been
true in branches of science ranging from astrophysics and botany to zoology.  
It is no different here.

But finally, we are justified 
in interpreting observed statistical correlations
as general landscape features to the extent that 
we can attribute such correlations to
the existence of underlying string-theoretic consistency constraints.
Indeed, when the constraint operates only within a single class
of strings, then the corresponding statistical correlation
is likely to hold across only that restricted portion of the landscape.  
For example, in cases where we were   
able to interpret our statistical correlations 
about gauge groups and cosmological constants
as resulting from deeper constraints such as conformal and
modular invariance, we expect these correlations to 
hold across the entire space of perturbative closed-string vacua.
As such, we may then claim to have extracted true phenomenological
predictions from string theory.  This is especially  
true when the string-correlated quantities 
would have been completely disconnected in quantum field theory.

Thus, it is our belief that such statistical landscape studies have their
place, particularly when the results of such studies are interpreted correctly
and in the proper context.
As such, we hope that this initial study of the perturbative
heterotic landscape may represent one small step in this direction.

%=========================================================================== 
\section*{Acknowledgments}
\setcounter{footnote}{0}

Portions of this work were originally presented at 
the String Phenomenology Workshop at the Perimeter Institute in April 2005,
at the Munich String Phenonomenology Conference in August 2005,
and at the Ohio State String Landscape Conference in October 2005.
This work is supported in part by the National Science Foundation
under Grant PHY/0301998, by the Department of Energy under Grant
Grant~DE-FG02-04ER-41298, and by a Research Innovation Award from 
Research Corporation.  I am happy to thank 
R.~Blumenhagen, M.~Douglas, S.~Giddings, J.~Kumar,
G.~Shiu, S.~Thomas, H.~Tye, 
and especially M.~Lennek for discussions.
I am also particularly grateful to D.~S\'en\'echal for
use of the computer programs~\cite{Senechal} which
were employed fifteen years ago~\cite{PRL}  to generate these string models
and to determine their gauge groups.
While some of this data was briefly mentioned in Ref.~\cite{PRL},
the bulk of the data remained untouched.
Because of its prohibitively huge size (over 4 megabytes in 1990!), this data
was safely offloaded for posterity onto a standard nine-track magnetic computer tape.
I am therefore also extremely grateful to M.~Eklund and P.~Goisman
for their heroic efforts in 2005 to locate the only remaining tape drive
within hundreds of miles capable of reading
such a fifteen-year-old computer artifact, and to S.~Sorenson
for having maintained such a tape reader in his 
electronic antiquities collection and
using it to resurrect the data 
on which this statistical analysis was based.
%  Technology marches on, but string models are eternal.

\bigskip
\vfill\eject

%=================================================================================
\bibliographystyle{unsrt}

\end{document}